\documentclass[
reprint,
runinaddress,
preprintnumbers,
nofootinbib,
 amsmath,amssymb,
 aps,
prd,
]{revtex4-1}
\usepackage{lipsum}
\usepackage{dcolumn}
\usepackage{bm}
\usepackage{color}
\usepackage{graphicx,amsmath,amssymb}
\usepackage{tikz}
\usetikzlibrary{matrix}
\usepackage{slashed}
\usepackage{fontenc}

\newcommand{\eqnsplit}[1]{\begin{align}\begin{split}#1\end{split}\end{align}}
\newcommand{\eqn}[1]{\begin{align}#1\end{align}}
\newcommand{\eqns}[1]{\begin{align*}#1\end{align*}}

\definecolor{awesome}{rgb}{1.0, 0.13, 0.32}
\definecolor{darkpastelgreen}{rgb}{0.01, 0.75, 0.24}
\definecolor{azuree}{rgb}{0.0, 0.5, 1.0}
\definecolor{guppiegreen}{rgb}{0.0, 1.0, 0.5}
\definecolor{mediumspringgreen}{rgb}{0.0, 0.98, 0.6}
\definecolor{richelectricblue}{rgb}{0.03, 0.57, 0.82}
\definecolor{mediumcandyapplered}{rgb}{0.89, 0.02, 0.17}
\definecolor{scarlet}{rgb}{1.0, 0.13, 0.0}
\definecolor{blue-violet}{rgb}{0.54, 0.17, 0.89}
\definecolor{carnationpink}{rgb}{1.0, 0.65, 0.79}
\definecolor{fluorescentpink}{rgb}{1.0, 0.08, 0.58}
\definecolor{bittersweet}{rgb}{1.0, 0.44, 0.37}
\definecolor{electricgreen}{rgb}{0.0, 1.0, 0.0}
\definecolor{chartreuse(web)}{rgb}{0.5, 1.0, 0.0}

\definecolor{regionA}{RGB}{164,42,23}
\definecolor{regionB}{RGB}{32,76,124}
\definecolor{cyanRT}{RGB}{113,227,207}
\definecolor{purpleRT}{RGB}{185,59,121}
\definecolor{redRT}{RGB}{234,109,87}
\definecolor{greenInter}{RGB}{168,246,106}

\newcommand{\lccc}{\left\{ }
\newcommand{\rccc}{\right\} }

\newcommand{\sref}[1]{$\S$~\ref{#1}}
\newcommand{\aref}[1]{appendix~\ref{#1}}
\newcommand{\fref}[1]{figure~\ref{#1}}

\usepackage{hyperref}
\hypersetup{
    bookmarks=true,			
    unicode=false,          
    pdftoolbar=true,        
    pdfmenubar=true,        
    pdffitwindow=false,     
    pdfstartview={FitH},    
    pdftitle={Universality, intertwiners and black hole information},    
    pdfauthor={A. Akal},     
    pdfnewwindow=true,      
    colorlinks=true,       
    linkcolor=richelectricblue,          
    citecolor=blue-violet,        
    filecolor=blue,      
    urlcolor=guppiegreen           
}

\begin{document}
\preprint{YITP-20-132}
\title{Universality, intertwiners and black hole information}
\author{I. Akal}
\email[Current e-mail: ]{a.akal@uu.nl}
\affiliation{Center for Gravitational Physics\\
Yukawa Institute for Theoretical Physics\\
Kyoto University, Kyoto 606-8502, Japan\\}
\date{\today}

\begin{abstract}
The central question in this article is how information does leak out from black holes. Relying on algebraic arguments and the concept of superselection sectors, we propose the existence of certain operators whose correlations extend across the black hole atmosphere and range into the interior. Contained in the full algebra, these black hole intertwiners will not belong to the subalgebra describing semiclassical bulk physics. We study this proposal in the context of operator reconstructions for code spaces containing a large number of microstates. As long as the atmosphere is excluded from a particular subsystem, the global state seen under the action of the associated algebra is maximally mixed and therefore described by a single classical background. Once the relevant correlations are encoded, i.e. if the algebra is sufficiently enlarged, perfect state distinguishability becomes possible. We arrive at this by computing the von Neumann entropy which may explain the result obtained by applying the quantum extremal surface prescription to the mixed state. We then examine these insights in the context of black hole evaporation and argue that information is transferred to the radiation via black hole intertwiners. We derive the Page curve. The mechanism above suggests that black hole information is topologically protected. An infalling observer would experience no drama. This may resolve the unitarity problem without running into any firewall or state puzzle, the latter being evident in generalized entropy computations. We also examine the question of how certain wormhole topologies may be understood given these findings. We argue that their occurrence in gravity replica computations may be related to the maximal correlation between radiation and atmosphere surrounding the old black hole. This may suggest a connection between topology change and near horizon quantum gravitational effects.
\end{abstract}

\maketitle

{\footnotesize 
\tableofcontents
}

\section{Introduction}
\label{sec:intro}

\subsection{Holography and black holes}
\label{subsec:holo-BHs}

Understanding the quantum properties of black holes will provide deep insights into how our universe came into existence. 

Originated within the framework of string theory, the anti-de Sitter (AdS)/conformal field theory (CFT) correspondence is a nonperturbative formulation of quantum gravitational physics in AdS space in terms of a single quantum field theory (QFT) on the asymptotic conformal boundary \cite{Maldacena:1997re,Gubser:1998bc,Witten:1998qj} which serves as an explicit top down realization of the holographic principle \cite{tHooft:1993dmi,Susskind:1994vu}.

In recent years, several concepts from quantum information have helped to better understand the underlying mechanisms behind holography. 
An influential insight has been obtained by Ryu and Takayanagi \cite{Ryu:2006bv,*Ryu:2006ef}, see \cite{Hubeny:2007xt} for the covariant generalization.
The Ryu--Takayanagi (RT) proposal states that the entanglement entropy of a given boundary subregion is determined by the area of a minimal surface in the bulk that is homologous to the subregion \cite{Ryu:2006bv,*Ryu:2006ef}. 

The RT prescription has been confirmed via bulk computations based on a certain saddle point approximation to the gravitational path integral \cite{Lewkowycz:2013nqa}. It has also been extended to more general cases including higher order quantum corrections \cite{Faulkner:2013ana,Engelhardt:2014gca}. Since then, such findings have triggered many investigations at the intersection of quantum gravity and information theory.

One particular example is the proposed connection between the AdS/CFT correspondence and quantum error correction (QEC) codes \cite{Almheiri:2014lwa,Harlow:2016vwg}. These can be seen as the isometries from bulk subspaces to the dual conformal boundary. The proposal provides interesting insights into how the duality especially operates in its semiclassical limit. 

Even though QEC in holography is usually assumed to be exact, recent discussions have pointed out that certain key aspects of the duality can only be satisfied if the correction is approximate, especially, in situations where the code space contains a large number of black hole microstates \cite{Hayden:2018khn}.\footnote{This basically depends on what is chosen to be the code space, which in the context of AdS/CFT is ambiguous. Originally, the former has been proposed to be the low energy semiclassical subspace properly included in the physical CFT Hilbert space. However, one may even extend the code space, for instance, by adding black hole microstates. The way how QEC operates in holography rather depends on what is assumed to be contained in the code space.
\label{footnote:cpsace}
} 

This explicitly happens in form of very small uncorrectable errors nonperturbatively suppressed in the semiclassical limit. The connection to QEC has particularly helped understanding how bulk operators may be reconstructed from the boundary. 

A powerful prediction of holography flourishes in the context of the black hole information paradox. 
Semiclassically, Hawking has shown that the radiation released by black holes is described by a thermal state \cite{hawking1975particle}, means that the whole process would end up in information loss after the black hole has entirely evaporated via Hawking radiation \cite{Hawking:1976ra}. 

Such a conclusion clearly contradicts with the notion of unitarity and, thus, ends up in a clear conflict between quantum mechanics and general relativity.
However, at least according to the way how the AdS/CFT correspondence operates, it is clear that the information paradox should be resolved, since the lower dimensional boundary QFT dual to the gravitational system in the bulk is unitary.

Unitarity would be manifest in the evolution of the fine grained entropy of the radiation. If the evaporation follows the rules of quantum mechanics, then the entanglement entropy of the radiation should first increase. This is of course identical to Hawking's result. However, at a certain time, where the entropy almost equals half of the initial coarse grained entropy of the black hole, the latter being described by the Bekenstein--Hawking (BH) formula \cite{hawking1975particle,Bekenstein:1973ur}
\eqn{
S_\text{BH} = \frac{\mathcal{A}_\text{hor}}{4 G_N},
\label{eq:S-BH}
}
where $\mathcal{A}_\text{hor}$ denotes the area of the horizon and $G_N$ is the Newton constant, the entropy should start decreasing and become zero when the evaporation is completed. This is known as the Page curve \cite{Page:1993wv,Page:2013dx}. 

At the semiclassical level, there is no such a turning point at the so called Page time. Instead, the fine grained entropy of the radiation would keep growing and possibly stop when it equals the coarse grained entropy of the initial black hole. 

Of course, this is contradictory, since the entanglement entropy of the radiation should never be able to exceed the coarse grained entropy of the evaporating black hole, at any time. This behavior clearly deviates from the Page curve and therefore indicates information loss.

Since AdS/CFT already suggests that information must be preserved, one should be able to directly verify the Page curve in holography. This has recently been achieved in \cite{Penington:2019npb,Almheiri:2019psf}. Based on certain toy models, where the evaporation of the black hole in the bulk is realized via absorbing boundary conditions, it has been shown that the entanglement entropy of the radiation initially increases. 

At the Page time, there occurs a phase transition and the entropy of the radiation equals the coarse grained entropy of the black hole. The latter would decrease, since the area of the horizon would continue shrinking.

This result is obtained despite considering a thermal mixed state. The novelty that has been the game changer is the idea of using the quantum RT or quantum extremal surface (QES) prescription \cite{Faulkner:2013ana,Engelhardt:2014gca}, respectively, to semiclassically compute the fine grained entropy of the radiation instead of using the standard von Neumann entropy formula.
 
This is of course very different from Hawking's semiclassical computation \cite{hawking1975particle}. There, the main quantity is the state of the radiation whose von Neumann entropy gives rise to a steadily increasing curve.

The phase transition happens due to the fact that after the Page time the entanglement wedge\footnote{The entanglement wedge of a given region on the boundary is defined to be the bulk region which can be reconstructed from the former. It is bounded by the RT surface associated with the boundary region \cite{Czech:2012bh,Wall:2012uf,Headrick:2014cta,Dong:2016eik,Jafferis:2015del}. 
It is expected that the latter operates as a QEC code. This is one possible explanation for how the boundary would reconstruct the degrees of freedom supported in the bulk subspace.
}
of the radiation extends to the interior of the black hole. This goes back to the minimality condition in computing the generalized entropy. 
The bulk entropy associated with the initially empty minimal surface of the radiation becomes larger than the one associated with the new nonempty minimal surface.

In certain sense, the additional region in the interior purifies the bulk entropy and hinders its further increase after the Page time.\footnote{
The region which belongs to the entanglement wedge of the radiation and extends to the interior of the black hole has been called island \cite{Almheiri:2019hni}. Its appearance directly follows from the minimization procedure intrinsic to the prescription of computing the generalized entanglement entropy \cite{Engelhardt:2014gca}. 
}

These findings have also been obtained via gravity computations. The RT formula has been verified by employing a certain saddle point approximation to the gravity path integral \cite{Lewkowycz:2013nqa}. In the bulk, this means that one has to perform the analytic continuation of the $n$-th order Renyi entropy. This can be derived from the gravitational path integral by applying the replica trick. 

Computing the generalized entropy for the radiation in this way shows that, after the Page time, the dominating saddle is associated with wormholes connecting the replica manifolds \cite{Almheiri:2019qdq,Penington:2019kki}. In the gravity computation, these replica wormholes therefore turn out to be responsible for the expected unitary behavior around the Page time.\footnote{It turns out that the quantum RT prescription also works beyond AdS/CFT, as recently demonstrated in the case of asymptotically flat space \cite{Anegawa:2020ezn,Hashimoto:2020cas,Hartman:2020swn}. Therefore, given some quantum system coupled to gravity, it is suggestive to consider the prescription as a general proposal for computing the fine grained entropy semiclassically.

Based on these observations, one might argue that the gravity path integral already knows about all correlations beyond the semiclassical one. However, this might be misleading. What really turns out to be powerful in deriving the radiation entropy for old black holes is the generalized RT prescription including higher order quantum corrections. Although highly nontransparent, and in some sense in an uncontrolled way, it seems to succeed in extracting the information about the quantum gravity correlations from the saddle point approximation to the path integral. 

The latter, under certain assumptions, is consistent with the RT proposal \cite{Lewkowycz:2013nqa}. It is important to note that in the gravitational path integral approach, there is the implicit assumption that the gravity under consideration is holographic having a dual field theory on its boundary. In particular, one uses the AdS/CFT dictionary to compare between the gravity replica computation and the field theory result. From that perspective, the gravitational path integral should be expected computing the entropy of a density matrix that is a hypothetical density matrix of the underlying quantum gravity. Of course, the latter should account for all correlations including those which do not only rely on the semiclassical degrees of freedom.
}

In the generalized entropy computations described above, the Page transition is obtained in the presence of a smooth horizon. There seems to be no firewall \cite{Almheiri:2012rt,Almheiri:2013hfa}. The firewall proposal builds on the earlier idea of black hole complementarity \cite{Susskind:1993if,Susskind:1993mu,Lowe:1995ac}, which further supported the expectation that black holes should be subject to the standard rules of quantum mechanics \cite{tHooft:1984kcu}, 
and argues that any local QFT in the bulk together with the assumption of global unitarity is inconsistent with the presence of a smooth horizon for old black holes, namely, those evolving after the Page time. 

The argument goes as follows. In the presence of a smooth horizon, outgoing modes should be quantum entangled with the interior modes. It is assumed that all the Hawking quanta sharply result from the horizon. At the Page time, the black hole is supposed to be almost maximally entangled with the early radiation. However, due to unitarity, the late time radiation has to be entangled with the early radiation. This is in contradiction with strong subadditivity \cite{lieb1973proof}. 
It is not possible for a single system to be highly quantum entangled with two different systems. 

In the present case, the latter are the interior mode of the Hawking quantum and the early radiation. As a resolution, it has been suggested that  slightly before or right at the Page time, a firewall consisting of high energy quanta spontaneously forms at the horizon which breaks the entanglement between the outgoing Hawking quantum and its interior mode. 

However, this obviously leads to a drama for an observer falling into the black hole and therefore contradicts with the equivalence principle in general relativity. We end up with another paradox.

The firewall argument does not seem to apply to the generalized entropy computations, since after the Page time, the entanglement wedge of the early radiation extends to the interior of the black hole which is described by a classical background. The two causally separate regions, where the second one is the radiation outside the smooth horizon, do not belong to two different Hilbert spaces. Therefore, one would not run into some contradiction with strong subadditivity.\footnote{
We would like to note that this observation already suggests a weakening of one of the original postulates underlying black hole complementarity which led to the firewall proposal. We comment on this point below. 
}

This closely resembles the resolution of the firewall paradox in the case of the two sided eternal black hole which is expected to have a smooth horizon as well and is dual to the thermofield double (TFD) state \cite{Maldacena:2001kr}
\eqn{
| \text{TFD} \rangle = \sum_i \sqrt{p_i}\ | i \rangle_L | i \rangle_R
\label{eq:TFD}
}
with 
\eqn{
p_i = \frac{e^{-\beta E_i}}{Z(\beta)},\quad Z(\beta) = \sum_i e^{-\beta E_i},
\label{eq:pi-Z}
}
where $Z(\beta)$ denotes the canonical partition function.
Indeed, the gravitational replica computations \cite{Almheiri:2019qdq,Penington:2019kki} turn out to be supporting the ER=EPR argument \cite{Maldacena:2013xja}, see also \cite{VanRaamsdonk:2009ar} for the possible link between quantum correlations and spacetime connectedness. At the Page time, there forms a wormhole connecting to the interior in higher dimensional space of the conformal radiation \cite{Almheiri:2019hni}. 

However, we should emphasize that the TFD state is rather an artificial construction. It does not describe a black hole locally observed from outside and it is certainly not evaporating. The TFD is a pure state, where the noninteracting CFTs on the two sides are maximally entangled with each other. Of course, in order to make the two sided eternal black hole evaporating, one may embed it into some larger space where the radiation on both sides is transferred to two external regions which may or may not be coupled with each other \cite{Verlinde:2020upt}.

Since AdS/CFT is expected to solve the information problem to the full extent, it should not only be possible to reproduce the correct unitary evolution for the fine grained entropy of the radiation around the Page time. In particular, we need to understand the mechanism of how quantum information is transferred to the outside. This surely necessitates to understand the evolution of the state describing the radiation and not only its entropy. Even though the quantum RT prescription proves powerful in reproducing the Page transition, it does not resolve the latter problem. The thermal state which lies at the heart of the information puzzle still occurs as in Hawking's semiclassical computation.

According to the quantum RT prescription the bulk entropy in the presence of the thermal state stops to contribute and the fine grained entropy of the radiation becomes the coarse grained entropy of the black hole at this stage. Obviously, this is not consistent with the semiclassical expectation. Viewing the formation and evaporation process as a scattering event, where a pure initial state must result in a pure final state, there should not appear two different values for the entropy. Otherwise, we encounter some ambiguity regarding the state of the radiation. This has recently been referred to as the state paradox \cite{Bousso:2019ykv,Bousso:2020kmy}. It turns out that the absence of a firewall as realized in \cite{Penington:2019npb,Almheiri:2019psf} results in an apparent contradiction of that type.

One possibility to explain the state problem would be arguing that the quantum RT surface somehow knows about the full quantum gravity correlations and therefore computes the fine grained entropy of the radiation consistent with unitarity. However, this would not say anything about the details of the resolution.\footnote{For doing so, it does not use the proper density matrix of the radiation system. If the latter would be given, it should be sufficient to compute its von Neumann entropy in order to obtain the unitary behavior around the Page time. 

Nevertheless, at least intuitively, we may get an understanding of how the quantum RT formula manages to reproduce the unitarity-consistent fine grained entropy even in the presence of the highly mixed state.

Namely, by definition, the classical minimal surface is a codimension two object which attaches to two distinct points on the boundary that enclose a certain subregion. It is a macroscopic object extending across the bulk. Any correlation, not necessarily restricted to the effective field theory description, would be taken into account by the minimal surface. 

On the other hand, the quantum RT surface is obtained in such a way that after adding the bulk entropy of local quantum fields, it has to give the minimal entropy. The RT surface defines the entanglement wedge of the subsystem and therefore determines the von Neumann entropy of local quantum fields in the entanglement wedge. Because of the mentioned minimization procedure, it can happen that the latter vanishes due to purification.

This is basically what happens around the Page time. 
The relevant quantum correlations traverse the quantum RT surface. The latter may change the entanglement wedge so that the von Neumann entropy of bulk quantum fields may change as well.
}

A second option might be the possibility that the quantum RT formula computes a certain ensemble of entropies in different quantum mechanical theories \cite{Bousso:2019ykv}. 
This perspective has received further support by findings suggesting
that averaging over certain two dimensional CFTs exhibits properties of some three dimensional gravity theory \cite{Afkhami-Jeddi:2020ezh,Maloney:2020nni}. Starting with three dimensional pure gravity, a putative ensemble understanding 
has also been explored \cite{Cotler:2020ugk,Maxfield:2020ale}.

Recall that the derivation of the RT formula from the gravitational path integral is based on certain assumptions \cite{Lewkowycz:2013nqa}. 
Most importantly, in doing so, the gravity under consideration has been assumed to be dual to a quantum mechanical system without gravity on the boundary. The gravitational path integral defined in an appropriate way would therefore compute the entropy of a density matrix that is a hypothetical density matrix of the underlying quantum gravity.

There have been earlier arguments for the averaging picture, which are particularly supported by the replica computations in \cite{Penington:2019kki} in the presence of Jackiw--Teitelboim (JT) gravity \cite{Teitelboim:1983ux,Jackiw:1984je}. 
Since the latter is proposed to be dual to a random matrix ensemble \cite{Saad:2018bqo,Saad:2019lba,Stanford:2019vob}, it may somewhat be
expected to encounter a certain gravity/ensemble correspondence.
According to the assumption underlying the gravitational derivation of the classical RT formula, such a property should then be reflected in the computation of the generalized entropy.

Even though such findings suggest an interesting new direction, there is so far no clear understanding of what quantity the ensemble averaging has to be performed over.
How this picture goes through in AdS/CFT is even more unclear. Indeed, averaging over different theories is not how holography seems to work in its best understood formulation. Namely, type IIB supergravity on $\text{AdS}_5 \times  S^5$ is dual to a single CFT. As far as it is known, there is no other boundary theory different from $\mathcal{N}=4$ super Yang--Mills that gives rise to the same semiclassical bulk description.

Let us turn to the proposal in \cite{Bousso:2020kmy}. The general expectation is that before the Page time, the entropy of the radiation is given by the entropy associated with the ensemble average of the corresponding density matrix. The latter is assumed to be determined by the thermal state semiclassically predicted by Hawking, i.e.
\eqn{
\overline{\rho_\text{rad}} = \rho_\text{semi}.
\label{eq:Hawkingstate}
} 
Computing its von Neumann entropy
\eqn{
S(\rho_\text{semi}) = -\mathrm{Tr} \left[ \rho_\text{semi} \log \rho_\text{semi} \right]
\label{eq:averS}
}
would lead to an increasing curve after the Page time.
However, it is argued that computing the generalized entropy by following the quantum RT prescription results in an averaged entropy, where the averaging is realized over all entropies belonging to different quantum mechanical theories. 

In order to make this more precise, let us assume an initial pure state describing the situation before the black hole has formed. Furthermore, each of the quantum theories shall evolve to different pure final states when the associated black hole has fully evaporated. The argument is that the generalized entanglement entropy of the radiation is assumed to be  equal to the following averaged entropy \cite{Bousso:2019ykv}
\eqn{
\overline{S(\rho)} \equiv \int d\nu\ c(\nu) S\left(  \rho^{(\nu)} \right) \simeq \sum_i c_i S( \rho_i ),
\label{eq:Srad-ensemble}
}
where the former is computed in the presence of the thermal mixed state predicted semiclassically. In the case of \eqref{eq:Srad-ensemble}, the entropy would start decreasing after the Page time and become zero in the final stage, even though the states may be different from each other.
This might explain the state paradox. However, this only works if we assume the equality between the generalized entropy and \eqref{eq:Srad-ensemble}. So far, it is not known how the details of this equality work out.

Understanding the mechanism which is responsible for the unitarization of the radiation should provide a resolution of the state puzzle. We view this as an important criterion in seeking for a possible way out of the unitarity problem. 

The question we should ask is, in which specific situations one does not run into a conflict between unitarity and the presence of a smooth horizon. 
Prior to the firewall proposal \citep{Almheiri:2012rt}, certain possibilities to achieve the unitarization of the radiation have been suggested. 
One specific example is the idea of an extended framework which minimally departs from local QFT \cite{Giddings:2012bm,Giddings:2012dh}. 

Certain toy models incorporating nonlocal physics, as seen from the semiclassical perspective, have been discussed \cite{Giddings:2012gc}. These suggest that quantum information might be transferred to relatively soft modes 
in a certain region which extends around the black hole, namely, in a controlled way without running into contradiction with a smooth horizon.

On the other hand, one may ask, where the Hawking quanta do exactly originate from. Expecting that the process of pair creation takes place at the horizon is a basic assumption underlying the firewall proposal \cite{Almheiri:2012rt}. However, based on earlier numerical computations of emission rates from black holes \cite{Page:1976df}, it has been argued that the quanta should originate from an extended near horizon region \cite{Giddings:2015uzr}. Such a quantum region has been called black hole atmosphere. 

Let us note that the typical wavelength of the radiation quanta is of the same scale as the size of the black hole. This does not necessarily lead to a contradiction with the semiclassical particle picture. Namely, a local observer approaching the horizon from infinity will measure a wavelength which is highly blue shifted.

Following the discussion above, we see that there appear two possible ways for avoiding a firewall forming at the horizon \cite{Almheiri:2012rt}. 

One of them seems to weaken the original postulates in the context of black hole complementarity \cite{Susskind:1993if,Susskind:1993mu,Lowe:1995ac}. This precisely appears to be the case when the fine grained entropy of the radiation is computed by following the quantum RT prescription \cite{Penington:2019npb,Almheiri:2019psf,Almheiri:2019hni,Almheiri:2019qdq,Penington:2019kki}. The old black hole, as viewed by a distant observer, cannot be described by the same degrees of freedom which describe the black hole interior, cf. \cite{Almheiri:2020cfm}. 

After the Page time, a substantial part of the interior belongs to the entanglement wedge of the radiation. Thus, the original black hole degrees of freedom describe only a small portion of the interior. 
However, according to the original complementarity arguments, the opposite is assumed to be true during black hole formation and evaporation.

The second situation is related to the presence of an extended near horizon region as a source of the radiation. If we accept a long distance origin of the Hawking quanta, then any mechanism responsible for the unitarization process must be operational at a larger scale, rather than in the microscopic vicinity of the black hole horizon (i.e. stretched horizon \cite{Susskind:1993if}). 

We may then conclude that any mechanism explaining the unitary behavior for the fine grained entropy should not contradict with the two scenarios described above. 

Of course, assuming a long distance origin for the radiation, there would be no reason for expecting the firewall forming at the horizon. Thus, any of the postulates underlying black hole complementarity may be weakened or may not apply at all. These are based on the assumption that the Hawking pairs are produced exactly at the horizon, where one particle falls into the black hole and the other escapes. The described generalized entropy computations, however, suggest that one specific postulate has to be weakened. 

Therefore, the presence of a region extending to the interior and belonging to the entanglement wedge of the radiation after the Page time should imply the long distance origin of the radiation and vice versa. Any resolution of the unitarization problem is expected to align with these observations.

In this paper, we mainly focus on the one sided black hole in AdS/CFT. We argue that there is exactly such a mechanism responsible for the unitarization of the radiation. From the perspective of a low energy observer, who by definition has only access to semiclassical degrees of freedom, there exist certain objects whose hidden quantum correlations (encoded in the structure of the emergent bulk spacetime) extend across the near horizon region that may be identified as the black hole atmosphere. 

This goes back to the presence of a large number of superselection sectors associated with the interior degrees of freedom.
We will see that such operators, dubbed black hole intertwiners, 
play a substantial role in the unitarization process. Before summarizing the details in \sref{sec:summ}, we would like to comment on certain aspects regarding the two sided eternal black hole and reconstructions of interior operators.

\subsection{Interior operators}
\label{subsec:intOps}

Formally, the two sided black hole is the purification of the one sided thermal black hole which can be obtained by tracing out either the left or right side of the density matrix associated with the TFD
\eqn{
\rho_\text{therm} = \mathrm{Tr}_{\{L,R\}} \left[ | \text{TFD} \rangle \langle \text{TFD} | \right]
\label{eq:thermFromTFD}
}
proposed to be dual to the Hartle--Hawking state above the the Hawking--Page transition \cite{Maldacena:2001kr}. The von Neumann entropy of the black hole is given by the BH entropy.
For the two sided eternal black hole, there is no information problem as well as no firewall formation. It is a pure state where both sides are maximally entangled with each other in the presence of a smooth horizon. 

Seen from the perspective of the left (right) CFT, the exterior mode is entangled with the interior mode behind the horizon. However, by construction, it is entangled with the right (left) CFT. What naively seems to contradict strong subadditivity is actually not doing so. The interior mode is not an independent system. It corresponds to the exterior mode of the right (left) CFT, and is therefore encoded in the entanglement wedge of the latter.

The two sided eternal black hole has played an important role in understanding how bulk operators behind horizons can be reconstructed in terms of boundary operators.
Consider, for instance, the AdS vacuum. In this case, there is a well known perturbative procedure for reconstructing bulk operators in terms of nonlocal boundary operators \cite{Banks:1998dd,*Bena:1999jv,*Hamilton:2005ju,*Hamilton:2006az,*Hamilton:2007wj,*Heemskerk:2012mn,*Kabat:2015swa}. 

However, this task becomes more tricky if the bulk operator is located inside the black hole. Interestingly, if the black hole state is known, there is a clear procedure for how to realize the boundary reconstruction \cite{Papadodimas:2012aq,*Papadodimas:2013wnh,*Papadodimas:2013jku} which requires certain mirror operators depending on unmeasured details of the associated microstate that belong to the fine grained sector of the Hilbert space. 

A similar state dependence has been found for operator reconstructions in the two sided eternal black hole by considering certain time shifted states associated with the same background geometry \cite{Papadodimas:2015xma}. Doing so for the left side, for instance, the different time shifted states may be viewed as some sort of microstates from the perspective of the other side. It turns out that the same operator cannot be used to describe the interior in the full range of such time shifted states.

An important question is how operator reconstructions are realized if the bulk is dressed with gauge symmetries. Due to the presence of the graviton, understanding this problem is for sure necessary for any holographic setup. This issue has been explored for the AdS vacuum based on the current method \cite{Kabat:2012hp,*Kabat:2012av}. The bulk objects are nonlocal Wilson lines ending on the boundary, or on charged local operators which have clear representations in the CFT.

It is then natural to ask how gauge invariant operators inside black holes can be reconstructed from the boundary. For the two sided eternal black hole in the presence of $U(1)$ gauge symmetry, reconstructing such objects in terms of CFT operators necessarily requires a new kind of nonlocal operator in the bulk; a Wilson line that stretches through the wormhole and attaches to both boundaries \cite{Harlow:2015lma,Guica:2015zpf}. In some sense, the existence of this gauge invariant operator can be seen as an indication of whether or not the two sides are connected through the bulk via the wormhole.

For instance, consider a charged scalar bulk operator within the future light cone in the two sided background. In order to make the operator gauge invariant, it shall be connected via a Wilson line to a given point on the right boundary. It turns out that the reconstruction in terms of CFT operators requires the (smeared and dressed) contribution coming from the left side to be multiplied by the wormhole threading Wilson line. The latter runs between an arbitrarily chosen point on the left boundary (depending on the dressing of the charged operator in the left CFT) and the point on the right boundary specified by constructing the gauge invariant bulk operator.

One might ask whether or not, and if yes, where the wormhole threading Wilson line can be split into two pieces. This is interesting, since, naively, it does not seem to be possible, even though the boundary Hilbert space is obviously factorizable. Indeed, the Wilson line can be constructed in terms of CFT operators. This provides a possible resolution of the factorization puzzle in AdS/CFT \cite{Harlow:2015lma,Guica:2015zpf}. 

The splitting can be viewed as a fusion of differently charged local operators framed to the left and right boundaries. In the vicinity of the TFD, the two charged operators would just be placed to the left and right of the bifurcation surface. The charged fields can be massive, expressed in terms of high dimension operators in the CFT. The line splitting may therefore indicate an interesting connection between bulk factorization and the RT surface. Interestingly, it shows that describing low energy bulk physics somehow requires high dimension operators in the CFT. 

In the context of the information paradox, such a behavior is indeed expected. The resolution is believed to rely on properties of short distance physics in the bulk. In certain sense, this is already indicated by the proposal in \cite{Papadodimas:2012aq,*Papadodimas:2013wnh,*Papadodimas:2013jku}, where reconstructions of local bulk operators inside black holes necessarily depend on mirror operators that belong to the fine grained sector of the underlying Hilbert space. Recall that the mirror operators are state dependent and originally constructed in the gravitational context.
However, the findings described above apply for $U(1)$ gauge symmetry.

Nevertheless, it has been argued that analogous objects should also exist for the gravitational version \cite{Harlow:2015lma}.
To some extent, this problem has already been explored in \cite{Guica:2015zpf}, where by comparing the $U(1)$ case with the gravitational setup, it is argued that the wormhole threading Wilson line can be related to the mirror operators appearing in the two sided eternal black hole.

More direct insights have been derived for a two sided bulk theory described in terms of JT gravity. At low energies, there turn our to be certain objects extending nonlocally between the two boundaries \cite{Harlow:2018tqv}. 
These can also be split via a pair of dynamical charges where the associated bulk fields are created by fermionic fields in the Sachdev--Ye--Kitaev model proposed to be dual to the bulk theory. 
This may be seen as the gravity counterpart of the factorization problem.

We see that, whenever there are gauge symmetries in the bulk, reconstructing low energy operators
can require heavy degrees of freedom which do not appear in the effective low energy limit. In other words, one necessarily needs certain operators that belong to the fine grained sector of the full theory. Such observations, therefore, provide further evidence that bulk gauge fields as well as gravity should be seen as emergent entities.

Let us now consider a single typical microstate. We may bring the pure state into a TFD like state \cite{Papadodimas:2012aq}. Any operator located inside the original black hole would still be an interior operator. However, as described, the latter may be defined in the manner as one usually does in the case of the TFD. The construction under consideration does not correspond to the standard two sided eternal black hole. 

In the TFD like case, seen from the left side assumed to be describing the coarse grained sector, such an operator would lie behind the horizon and would be in causal contact with the fine grained part. Boundary reconstructions of such operators are expressed in terms of state dependent mirror operators. 

The latter can be understood as operators which would originate from the second copy of the CFT, if we would consider the usual two sided eternal black hole. There, the mirror operators can namely be related to nonlocal, bulk threading objects by introducing certain time shifted states on one side \cite{Guica:2015zpf}. As seen from the nonshifted side, the states may be interpreted as a sort of microstates \cite{Papadodimas:2015xma}. For each of the states, there would be associated wormhole threading operators. 

The picture above would apply to typical microstates. We consider states which contribute to the density matrix of the one sided thermal black hole. Recall that those have energies concentrated in a sufficiently narrow energy band centered around the mass of the thermal black hole. For each microstate, we may construct the TFD like state.

Depending on the fine grained sector, where, for simplicity, we here do not assume any change with respect to the coarse grained part among the states, we would end up with state dependent operators running between the two hypothetical boundaries associated with the different sectors. This would be analogous to the situation in the presence of time shifted states.

Accordingly, for each microstate, it would be possible to reconstruct operators in the interior based on the corresponding nonlocal operators.
Since the single sided thermal black hole is obtained by tracing out either left or right side of the TFD, it is natural to expect that doing so in the latter case will cut through these bulk threading operators.

As a consequence, the degrees of freedom associated with the fine grained sector, i.e. the internal state, cannot be accessed. The reduced density matrix would be maximally mixed, thus, resulting in the thermal state.

In the following, we would like to summarize our findings and outline the structure of this article.

\section{Summary and structure}
\label{sec:summ}

In the present work, starting from algebraic arguments and the concept of superselection sectors, we propose the existence of certain operators, called black hole intertwiners, whose correlations extend across the near horizon region and range into the black hole interior. 

These operators belong tho the full algebra of the underlying quantum gravity dual to the CFT. The subalgebra describing semiclassical bulk physics is ignorant of them.
The mentioned near horizon region may be identified as the black hole atmosphere which can be viewed as the exterior bulk region where the unitarization of the radiation is operating during black hole evaporation.

We begin by studying this proposal in the context of boundary reconstructions for code spaces containing a large number of black hole microstates. 
In particular, by focusing on a specific distinguishability measure, we make use of the fact that the task of studying boundary reconstructions can be reformulated as a problem of microstate distinguishability under a finite number of measurements from a given boundary subregion.

Any global pure state, taken to be the superposition of microstates, appears to be maximally mixed, if the intertwiner correlations cannot be accessed from the boundary subregion. 
In other words, this happens to be the case if they are not covered by the corresponding entanglement wedge.
In such a case, the boundary subsystem is effectively described by the mentioned subalgebra. 

In particular, even though we globally consider a pure state by superimposing an exponentially large number of microstates determined by the BH entropy, each corresponding to distinct bulk quantum states, the state seen under the action of the subalgebra will be maximally mixed and described by a single classical black hole background. 

This basically explains why any semiclassical observer perceives the thermal black hole.

Of course, the effective thermalization due to the described operator restriction applies if the size of the region under consideration attains the size of the entire boundary Cauchy surface.\footnote{The role of the atmosphere (or zone) region in the thermalization process has, for instance, recently been discussed in \cite{Nomura:2019qps,Nomura:2019dlz}.} Under that condition, since the internal degrees of freedom cannot be accessed from the algebraically reduced state, we need to impose the usual homology constraint on the minimal surface as dictated by the RT proposal.

On the other hand, if the relevant correlations across the black hole atmosphere can be accessed from a given finite boundary subregion, that is if the associated algebra is sufficiently enlarged, it becomes possible to perfectly distinguish between microstates contributing to the thermal ensemble.

These observations totally align with semiclassical computations carried out by applying the quantum RT prescription to the highly mixed state. We arrive at the same result by computing the von Neumann entropy for the corresponding reduced density matrix. This may then explain what kind of quantum gravitational correlations the quantum RT prescription intrinsically takes into account.

We then apply these ideas to an evaporating black hole. Considering a simple toy model simulating a unitary evaporation process, we argue that information is transferred to the radiation via black hole intertwiners.

Initially, that is immediately after the collapse when no radiation has been released yet, the encoding of the information about the interior states already extends to the quantum atmosphere via hidden (as seen from the perspective of semiclassical observers) nonlocal correlations relying on the intertwiners.

As soon as the black hole starts to release its radiation across the near horizon region, the outgoing Hawking quanta become quantum correlated with the internal degrees of freedom. 

At the Page time, this correlation becomes maximal so that the radiation gets maximally correlated with the surrounding quantum atmosphere. 
As a consequence, late time radiation released in the near horizon region would not trigger an entropy increase anymore, this, invoking the expected Page transition.

The Page curve directly follows from computing the von Neumann entropy for the reduced density matrix associated with the radiation system.

Due to the intrinsic nonlocal nature of the intertwiner correlations, the described mechanism suggests that black hole information is topologically protected.
It is not possible to access the internal degrees of freedom by doing local measurements in the bulk.

Because of the same reasoning, an observer falling into the black hole would not experience any drama. There would be no firewall forming at the horizon of the old black hole.

It is worth emphasizing that the proposed mechanism provides a resolution of the unitarity problem without running into any firewall or state puzzle, the latter being apparent in computations of the fine grained entropy based on the QES prescription. 

We also examine the question of how replica wormhole topologies may be understood in the light of the observations summarized above. We argue that their occurrence may be related to the maximal correlation between the radiation system and the quantum atmosphere surrounding the old black hole.

Since we arrive at the same conclusion, as it has been worked out semiclassically by studying quantum RT surfaces, this might serve as further support for the expectation above.
This would then suggest that topology changing phenomena are related to such near horizon quantum gravitational effects. 

In particular, it may indicate that contributing wormholes should be comparable to the size of the black hole, which would be in line with suggestions provided by gravity computations of the generalized entropy.

The remaining part of the paper is organized in the following way.

In \sref{sec:alphas}, we review certain aspects about boundary reconstructions, discuss the connection to state distinguishability and comment on further details relevant for the arguments in the subsequent sections. 

In \sref{sec:alphas-intertwiners}, we introduce the notion of superselection sectors and intertwiners. We briefly recap certain observations with particular emphasis on relevant information theoretic quantities.

In \sref{sec:sss-bhs}, we present some of the main observations in this paper. We make the connection to black holes and discuss the concepts reviewed in the preceding section in the context of boundary reconstructions. We also comment on similarities to certain discrete formulations.

In \sref{sec:bh-evap}, we apply our findings to a simplified model describing an evaporating black hole and discuss how information is transferred to the radiation. 

In \sref{sec:wormholes}, we elaborate on possible connections to certain wormhole topologies appearing in gravity replica computations of the fine grained entropy.

In \sref{sec:fin}, we conclude with a few final remarks.

In appendix~\ref{subsec:equi}, we give a brief summary of the general canonical principle and appendix~\ref{subsec:sdecomp} deals with certain aspects related to Hilbert space decompositions.

\section{Boundary reconstructions}
\label{sec:alphas}

In the present section, we discuss boundary reconstructions for code spaces containing a large number of possible microstates. 
In \sref{subsec:statedep}, we begin by discussing certain aspects related to state dependence in holography. 
In \sref{subsec:largecode}, we briefly review certain observations that are related to the effective subspace structure at semiclassical level.
In \sref{subsec:bdy-mes}, we particularly discuss how a highly controlled, state independent reconstruction can be achieved by referring to computations based on the QES prescription. 
In the subsequent part, \sref{subsec:holevo}, we discuss the idea of state distinguishability from boundary subregions. We focus on a particular measure and expound how certain predictions are related to the one reviewed in \sref{subsec:bdy-mes}.

\subsection{State dependence}
\label{subsec:statedep}

\begin{figure}
  \centering
    \includegraphics[width=0.43\textwidth]{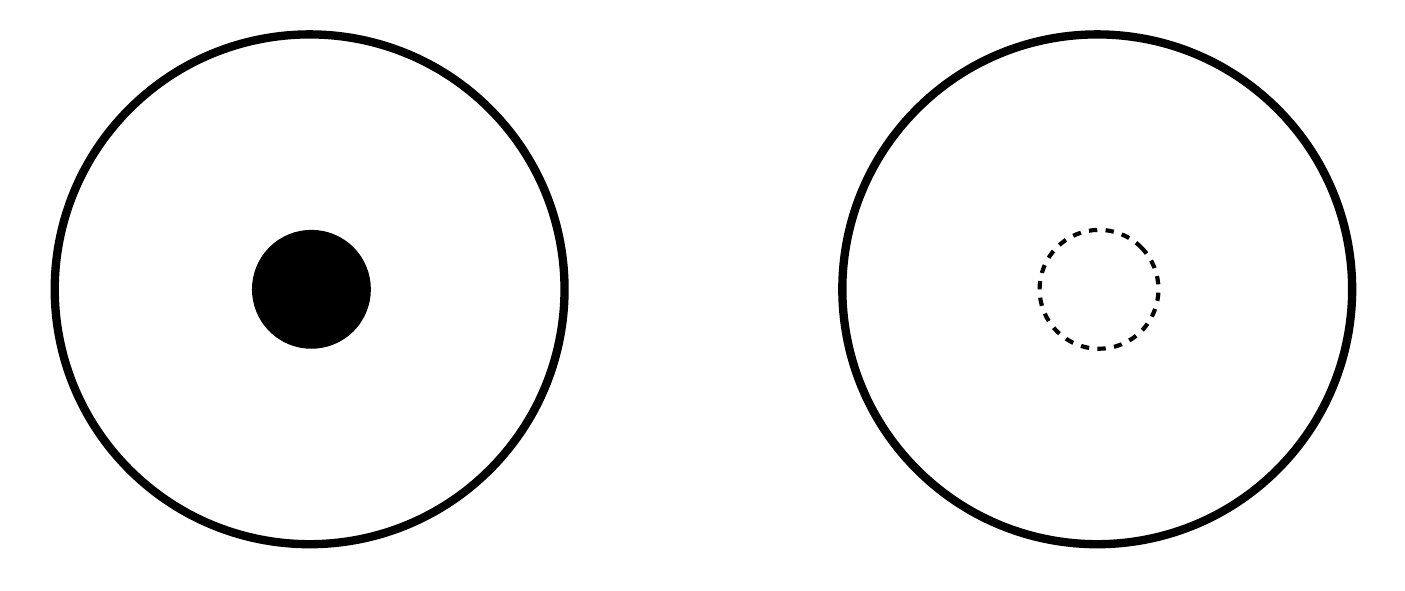}
    \caption{Thermal black hole in global AdS (left). The dual ensemble lives on the boundary. A typical microstate is dual to an excited heavy state in the CFT (right). In this case, there is no homology constraint on the RT surface associated with a finite boundary subregion. Its entanglement wedge may cover the entire black hole. To distinguish the two setups, we use a dashed circle indicating the black hole when the boundary state is pure.}
    \label{fig:bh-ads}
\end{figure}

Consider a black hole in AdS/CFT placed at the bulk center as depicted in \fref{fig:bh-ads}. Due to known properties of the duality, reconstructions of a given bulk operator from a finite boundary subregion will generally depend on the state of the black hole. 
To make this more concrete, let us focus on some boundary subregion $A$ that shall be slightly larger than half of the entire boundary Cauchy surface. 

For instance, the entire black hole
can be covered by the entanglement wedge of $A$, if it is in a pure state, see \fref{fig:bh-ads} (right).
The situation, when the state is unknown, can be modeled by the thermal ensemble as shown in the left panel. In the latter case, the black hole will always lie outside the entanglement wedge of $A$. In particular, any bulk operator acting on the black hole cannot be reconstructed from $A$.

As we will discuss later, this basically corresponds to the situation in which the radiation would look thermal, thus, resulting in information loss.

Nevertheless, if the black hole is covered by the entanglement wedge of $A$, which, again, is only true if the state is known, reconstructing bulk operators from $A$ would be possible. Such operators, of course, are well defined for all microstates contributing to the ensemble. 
However, this would not directly imply that there exists a single boundary reconstruction working for all microstates. Under such conditions, any set of measurements on a given finite boundary subregion would be state dependent.

Using the generalized RT formula including quantum corrections \cite{Faulkner:2013ana}, or more generally, the QES proposal \cite{Engelhardt:2014gca}, as well as the idea of entanglement wedge reconstruction \cite{Wall:2012uf,Headrick:2014cta,Jafferis:2015del} based on the concept of subregion-subregion duality \cite{Czech:2012bh}, it can be shown that a single boundary operator may even be used for highly mixed microstate subspaces \cite{Hayden:2018khn}.

Equivalently, the microstates can be viewed as being sufficiently entangled with some purifying reference system that has the same size as the large code space. In particular, the dimension of the latter can be exponentially large, namely, bounded from above by a constant fraction of the BH entropy.

This goes as follows. Let us first note that according to the standard arguments in entanglement wedge reconstruction, any bulk operator in the entanglement wedge of a finite boundary subregion $A$, can always be reconstructed on $A$, even when the state is not known.

However, suppose one intends to reconstruct some low energy operator acting in some bulk region bounded by two possible quantum RT surfaces that enclose the black hole which is sufficiently entangled with the external reference system. This shall even include bulk operators that act on the black hole. 

It turns out that the degrees of freedom which should at least be contained in the entanglement wedge of $A$ correspond to $\alpha S_\text{BH}$, where the dimensionless parameter $\alpha$ is the difference between the two QESs divided by the horizon area of the black hole.

Motivated by ideas on certain decoding channels having the property of universal subsystem QEC, the degrees of freedom associated with the fraction of the BH entropy have been identified as alpha bits \cite{Hayden:2018khn}. The latter are expected to be revealed through Hawking radiation during the evaporation process which we, of course, assume to be unitary. 

Such conclusions rely on observations obtained from simple qubit toy models with random unitary evolution that mimics the chaotic dynamics of microstates. Moreover, for being in accordance with the expectations from AdS/CFT, it has been argued that the process of QEC should operate approximately and not exact, as originally proposed in \cite{Almheiri:2014lwa,Harlow:2016vwg}, see also \cite{Dong:2016eik}.

We should note that from the quantum gravity point of view, the concept of alpha bits originating from information theory is less clear. Their connection to black holes
is particularly based on certain assumptions such as random unitary evolved states formed by a large number of qubits. So far, it is not clear how these concepts should be understood in the boundary QFT as well as from the gravity point of view. 

Moreover, for the conclusions described above, it is just sufficient to make use of the QES prescription \cite{Engelhardt:2014gca} and subregion-subregion duality \cite{Czech:2012bh,Jafferis:2015del,Dong:2016eik}. We may therefore expect that such observations should not necessarily be restricted to some specific QEC arguments. 

More importantly, although the role of the quantum RT surface is highly essential in all these considerations, we still do not know how to understand its entropy contributions within the framework of QEC. It is still unexplained how the geometric nature of bulk entropy arises from the latter.

The effects described above only work if the dimension of the code space is assumed to be exponentially large. 
In what follows, we first review the main arguments justifying this assumption. This will particularly be relevant when we refer to related aspects at a later stage of this paper.

For reasons of completeness, a general canonical principle based on the concept of equiprobable states is briefly reviewed in \aref{subsec:equi}. As already referred to in \cite{Hayden:2018khn}, the latter proves helpful, at least effectively under certain conditions, in arguing for the existence of such large spaces. 

We comment on related aspects in the following.

\subsection{Large code space}
\label{subsec:largecode}

Suppose there exists some factorized code space
\eqn{
\mathcal{H}_\text{CFT} \supset \mathcal{H}_\text{code} \simeq \mathcal{H}_\text{BH}\otimes \mathcal{H}_\text{ext},
\label{eq:subspace}
}
where $\mathcal{H}_\text{BH}$ describes the interior subspace and $\mathcal{H}_\text{ext}$ captures the exterior state outside the horizon. In addition, assume that
\eqn{
\lim_{G_N \rightarrow 0} 4 G_N \log d_{\text{code},\text{BH}} = \mathcal{A}_\text{hor}.
}
We note that due to eigenstate thermalization of low energy operators in AdS/CFT, an approximate linear subspace containing a large amount of microstates would not be an appropriate description.\footnote{According to the eigenstate thermalization hypothesis, in an isolated quantum system, the expectation value of an operator has a long time average which approximates the value predicted by the microcanonical ensemble if (i) the diagonal matrix elements are smoothly varying functions over small energy bands and (ii) the off diagonal elements are exponentially small in the system size \cite{deutsch1991quantum,srednicki1994chaos}. 

Equivalently, this would mean that the matrix elements can be written as $\bar O \delta_{ij} + \mathcal{O}(e^{-S/2})$ where $\bar O$ is the 
long time average of the operator expectation value in the microcanonical ensemble and $S$ denotes the (coarse grained, i.e. thermodynamic) entropy of the system.}

However, to leading order, this may at least approximately exist under certain conditions.

In the following, we will first be considering this simplification.

Take the microcanonical ensemble, where all microstates forming the maximally mixed state are assumed to be placed in a fixed, $G_N$ independent, small energy band $E \pm \delta E$ around the mass $m=E$ of the thermal black hole. Recall that this is a crucial criterion in the eigenstate thermalization hypothesis.\footnote{Eigenstate thermalization is a characteristic feature of quantum gravity which is intimately connected to the existence of a large number of invisible sectors at the semiclassical level. We will discuss that the presence of such superselection sectors plays a profound role in bulk operator reconstructions. 

More specifically, their existence turns out to be crucial for unitary evaporation of black holes, seen from the perspective of any local semiclassical observer, and have therefore direct implications for the resolution of the black hole information paradox.} 
The black hole mass is supposed to be scaling with the inverse of $G_N$.  

Given that the energy band is narrow enough, all microstates may be described by the same geometry at leading order, namely, the one which is associated with the average energy $E$. The dimension of this space is given by the exponential of the BH entropy.

\begin{figure}[b]
  \centering
    \includegraphics[width=0.45\textwidth]{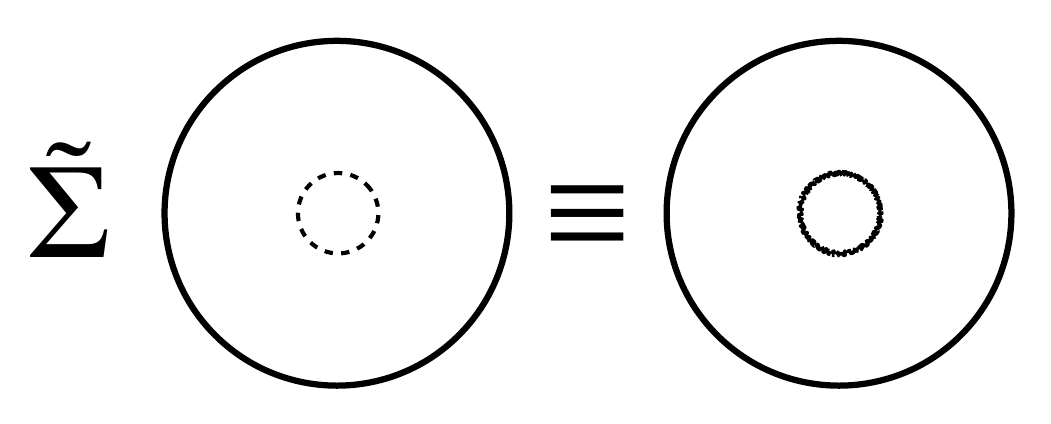}
    \caption{Superposition of black hole microstates. For simplicity, we use the notation $\tilde{\Sigma} \equiv \Sigma_i c_i$. The circle in the center (right) is used to represent the pure state.}
    \label{fig:bh-superpos}
\end{figure}

Consider a pure global state, that is a superposition of microstates having energies chosen as above, schematically shown in \fref{fig:bh-superpos}. In addition, consider some boundary subregion much smaller than the half of the entire Cauchy surface. 
One may expect that the state will not be distinguishable from the thermal canonical ensemble.

Especially, this will be valid for some randomly chosen subspace with dimension given by the exponential of some fraction of the coarse grained entropy of the black hole. This may effectively be assumed.\footnote{Even though, we here refer to observations in \cite{Popescu_2006}, we will later argue that the same conclusion basically arises in holography in a natural way. From the perspective of some low energy observer outside the horizon, the canonical ensemble, i.e. mixed state, arises due to the existence of a large number of superselection sectors associated with the interior degrees of freedom. 

We may say that superselecting integrates out the fine grained degrees of freedom of the full theory. Thus, distant observers having only access to the semiclassical degrees of freedom experience a maximally mixed state, i.e. the thermal black hole.}

From the perspective of the boundary subregion, the states in any such subspace cannot be distinguished from a thermal state.
In other words, the entanglement wedge of that subregion will not contain the black hole. 

Since, to leading order in $G_N$, the canonical and microcanonical ensembles agree with each other, one may describe both ensembles by the same geometry. This will lead to the part describing the black hole interior in \eqref{eq:subspace}, $\mathcal{H}_\text{BH}$.
Operators acting on the black hole would approximately commute with any operator located in the entanglement wedge.\footnote{We will see that such arguments can be understood within the framework of algebraic QFT in AdS/CFT. The sufficiently small dimension of the boundary subspace is guaranteed by the lack of certain operators. Their existence goes back to a large number of superselection sectors under the action of semiclassical observables. 

Importantly, this does not necessarily depend on a small subregion size. The effects are even present when the subregion size is almost equal to the half of the boundary Cauchy surface.}

In addition, there exists a shadow region around the black hole which does not belong to the entanglement wedge as well. Since excluded, one may assume that all interior degrees of freedom are localised in that region. 

On the other hand, according to eigenstate thermalization, the long time average of expectation values of low energy bulk operators within the shadow region should be approximated by the microcanonical, i.e. thermal to leading order in $G_N$, expectation value.

In particular, since all off diagonal terms in the long time average ensemble are highly suppressed, this will even be true for superpositions of eigenstates corresponding to any fraction of the BH entropy. 

The resulting space would not have bulk degrees of freedom outside the horizon, where all included states are in thermal equilibrium. In order to obtain the linearised code space with the exterior part, what remains to be done is including the degrees of freedom outside the horizon.

Given a microstate, we may add all states to the code subspace which can be generated by acting with a set of low energy operators on the state. If we do this for all microstates, it would result in a set of different code subspaces.
Under the action of such operators,
the entire space would decompose into an orthogonal sum of the latter. 

Furthermore, recall that at the canonical level, means when the boundary subregion is sufficiently small, one would not be able to distinguish between microstates.
Any low energy bulk operator acting outside the black hole will only operate on the subspace.

However, operators acting on the black hole may change its state while the exterior state will not be affected.                                                                                                                                                                                                                                                                                                                                                                                                                                                                           This would effectively give rise to the approximate code space \eqref{eq:subspace} factorized into an interior part $\mathcal{
H}_\text{BH}$ and an exterior part $\mathcal{H}_\text{ext}$.

\subsection{Quantum extremal surfaces}
\label{subsec:bdy-mes}

\begin{figure}
  \centering
    \includegraphics[width=0.2\textwidth]{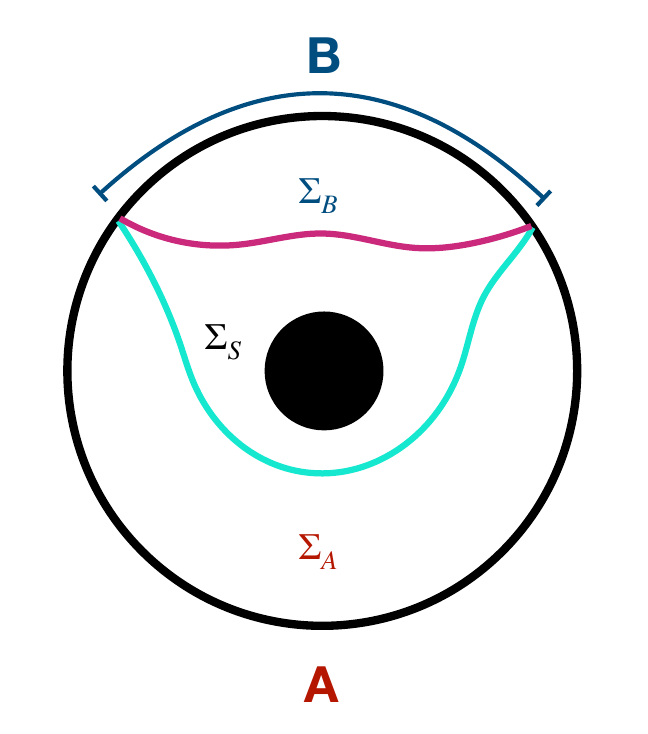}
    \caption{Thermal black hole located at the center. The quantum RT surface for the boundary subregion $A$ may change between two different QESs attached to the boundaries of $A$. The first one is homologous to $A$ with area $\mathcal{A}_2$  (cyan). The second one is homologous to the complement $B$ and has area $\mathcal{A}_1$ (purple).}
    \label{fig:2geodesics}
\end{figure}

Depending on the size of the boundary subregion $A$, its QES may generally change between two quantum RT surfaces which attach to $\partial A$, see \fref{fig:2geodesics}. One of the extremal surfaces is homologous to $A$ and has area $\mathcal{A}_2$. It encloses the bulk region $\Sigma_A$. The other extremal surface, which is homologous to the complementary region $B$ and has area $\mathcal{A}_1$, encloses the bulk region $\Sigma_B$. The shadow region $\Sigma_S$ covers the black hole and lies between the described extremal surfaces.

The QES \cite{Engelhardt:2014gca} minimizes the generalized entropy\footnote{From now on, in order to avoid any confusion, we use the notations $S$ and $S^0$ whenever it will be necessary. The former corresponds to the entanglement entropy for the CFT (quantum gravity) state, whereas the latter is used to denote the entropy computed for semiclassical bulk fields.}
\eqn{
S = \frac{\mathcal{A}}{4 G_N} + S^0,
\label{eq:gee}
}
where $S^0$ is the von Neumann entropy of local quantum fields computed in the fixed background geometry defining the entanglement wedge of the corresponding boundary subregion. 

Consider a pure state in $\mathcal{H}_\text{code} \otimes \mathcal{H}_R$ where $\mathcal{H}_R$ is the space of some purifying external reference system with fixed dimension $d_R$. The reference system is assumed to be sufficiently entangled with the black hole described by $\mathcal{H}_\text{BH}$. Recall that 
\eqn{
\mathcal{H}_\text{ext} \simeq \mathcal{H}_{\Sigma_A} \otimes \mathcal{H}_{\Sigma_S} \otimes \mathcal{H}_{\Sigma_B}
}
corresponds to the exterior part in $\mathcal{H}_\text{code}$.
For being able to decode any low energy bulk operator in region $\Sigma_A \cup \Sigma_S$ from $A$, the entanglement wedge of $A \cup R$ must consist of $\Sigma_A \cup \Sigma_S \cup R$. That means, for decoding the region $\Sigma_A \cup \Sigma_S$ from $A$, one has to satisfy
\eqn{
4 G_N S^0(\Sigma_A \Sigma_S R) + \mathcal{A}_1 < 4 G_N S^0(\Sigma_A R) + \mathcal{A}_2
\label{eq:uneq}
}
for all states in $\mathcal{H}_\text{code} \otimes \mathcal{H}_R$.

In this case, the QES associated with $A$ will be given by $\mathcal{A} = \mathcal{A}_1$.

Assuming that the bulk entropies satisfy $S^0(\Sigma_A R) = S^0(\Sigma_A) = \mathcal{O}(1)$ and $S^0(\Sigma_B) = \mathcal{O}(1)$, to leading order in $G_N$, we may simplify the inequality from above as $4 G_N \log d < \mathcal{A}_2 - \mathcal{A}_1$
where $d = \exp S_\text{bulk} = \exp S^0(R)  = d_R$ corresponds to the dimension of the code space containing the microstates.

However, when $4 G_N \log d > \mathcal{A}_2 - \mathcal{A}_1$, there will be a change in the location of the QES such that $\mathcal{A} = \mathcal{A}_2$. This happens for sufficiently mixed microstates, or, equivalently, highly entangled with the purifying system $R$. Consequently, a single operator reconstruction from $A$ will not be possible. Only if the dimension $d$ of the subspace encoding the microstates is bounded as
\eqn{
\log d < \alpha S_\text{BH},
}
where 
\eqn{
\alpha = \frac{ \mathcal{A}_2 - \mathcal{A}_1 }{ \mathcal{A}_\text{hor}},
\label{eq:alpha}
} 
then any bulk operator in region $\Sigma_S$, as well as acting on the black hole, can be reconstructed 
on $A$ \cite{Hayden:2018khn}.

\subsection{Perfect distinguishability}
\label{subsec:holevo}

We have seen that state dependence of any reconstruction is related to the size of the corresponding boundary subregion. In particular, when the difference between the two minimal surfaces approaches the area of the horizon, a single reconstruction can be realized on region $A$ that works for all microstates.

Of course, if so, it should be possible to distinguish between microstates via performing a finite set of measurements on $A$. In other words, one should be able to distinguish a pure density matrix from the thermal density matrix.
This is an important problem. The indistinguishability of microstates under the action of bulk operators reconstructable on some boundary subregion is a central question in the information puzzle.

A good measure for state distinguishability can be derived from the Holevo bound \cite{nielsen2002quantum}. This particular measure has recently been used to study the distinguishability of BTZ microstates under measurements on a boundary subregion \cite{Bao:2017guc}. 
Recall that in AdS/CFT the thermal ensemble on the boundary,
\eqn{
\rho_\text{therm} = \sum_i p_i \rho_i,
\label{eq:thermal-BH}
}
where the coefficients $p_i$ are the same as in \eqref{eq:pi-Z},
is dual to the eternal black hole with inverse temperature $\beta$. The density matrices $\rho_i = | i \rangle \langle i |$ associated with the microstates satisfy the orthogonality condition. In the high temperature limit, $\beta < 2 \pi$, the dual bulk geometry is determined by the AdS Schwarzschild solution \cite{hawking1983thermodynamics}. 
We again assume that the microstate energies are located in a small energy band around the mass of the thermal black hole. To leading order in $G_N$, the same black hole solution can be used for the expectation values of bulk operators.

Consider the following average of relative entropies
\eqn{
\sum_i c_i S(\rho_{i,A} || \rho_{\text{therm},A} ).
\label{eq:average-relE}
}
The reduced density matrices are obtained by tracing out the complement of $A$. In principal, we may consider any choice for the probability coefficients $c_i$. However, the problem can substantially be simplified if we choose them as in \eqref{eq:thermal-BH}.

The expression \eqref{eq:average-relE} then becomes identical to the Holevo information $\chi$ which is bounded as 
\eqn{
0\leq \chi(A) = S_A- \sum_i p_i S_{i,A} \leq S_\text{Sh},
\label{eq:average-relE2}
}
where $S_A \equiv S(\rho_{\text{therm},A})$ and $S_{i,A} \equiv S(\rho_{i,A})$.
The upper value $S_\text{Sh}$ is the classical Shannon entropy and saturates since the reduced density matrices on $A$ for any pair of microstates have orthogonal support. The lower bound is nothing but the concavity condition. State distinguishability via measurements on the boundary subregion will therefore be captured by Holevo information.

If the subregion size $\ell_A$ approaches the size of the entire boundary, $\ell_\text{bdy}$, one finds perfect distinguishability, $S_\text{Sh} = S_\text{BH}$. 
However, more interestingly, it turns out that the states even become perfectly distinguishable when $\ell_A$ reaches a critical value $\ell_\text{crit} < \ell_\text{bdy}$ \cite{Bao:2017guc}. This behavior agrees with computations based on QESs. Indeed, one has \cite{Hayden:2018khn}
\eqn{
\chi = \alpha S_\text{BH}.
\label{eq:master}
}
Note that the Holevo information is derived by using the result\footnote{The entropy derivation is based on the assumption of modular invariance in the CFT. If we additionally take into account that the arguments in \sref{subsec:largecode} rely on the eigenstate thermalization hypothesis, the equality \eqref{eq:master} then suggests that there should be a connection between modular invariance and  eigenstate thermalization.
 
A connection of this kind has recently been found, showing that the modular invariance of two point functions leads to nontrivial constraints on the behavior of three point functions, i.e. matrix elements, in generic two dimensional CFTs \cite{Hikida:2018khg}. Under general conditions, the latter fulfill the conditions for eigenstate thermalization at very high energies $E \gg c$. 

Interestingly, in the particular case of holographic CFTs with large spectral gaps yielding \eqref{eq:S_iA} this even turns out to be true for relatively low energies $E \geq c/12$. Such findings agree with the expectation from above.

On the other hand, it is argued that boundary reconstructions for large microstate subspaces would require approximate QEC. This aligns with findings indicating that systems satisfying eigenstate thermalization should contain approximate error correcting codes in their spectra \cite{Brandao:2017irx}.
}
\eqn{
S_{i,A} = S_A(\mathrm{min} \lccc \ell_A,\ell_\text{bdy} - \ell_A \rccc)
\label{eq:S_iA}
}
computed in sparse CFTs at large central charge \cite{Asplund:2014coa}. At leading order, excited heavy states in the latter are dual to BTZ microstates with semiclassical thermodynamics \cite{Hartman:2014oaa}. 

On the other hand, the dimensionless parameter in \eqref{eq:alpha} can explicitly be evaluated by computing geodesic lengths in pure AdS. In fact, the bulk computation is similar to the connection between channels in the conformal block expansion and geodesics in the defect geometry arising in the CFT derivation.
In other words, the transition between the two expansion channels in the identity block approximation corresponds to the jump between the two QESs. 

In order to have the necessary degrees of freedom encoded in the entanglement wedge of $A$, the pure state entropy should be determined by the geodesic that winds around the defect.

The fact that microstates become perfectly distinguishable when $\ell_A$ reaches a critical value is a nontrivial result. One might expect that the states will only be perfectly distinguishable if $\ell_A = \ell_\text{bdy}$. 
On the other hand, as usual in QEC, the entanglement wedge explicitly depends on the code space \cite{Harlow:2016vwg}. 

The observations are therefore consistent from the latter point of view. In the present case, the code space containing the microstates is exponentially large. Perfect distinguishability occurs when the degrees of freedom, encoded between the two QESs satisfying $\alpha = 1$, belong to the entanglement wedge of $A$.

In the following section, our main goal will be figuring out the mechanism that gives rise to perfect distinguishability. Any such attempt should be consistent with semiclassical computations based on QESs. 

\section{Superselection sectors}
\label{sec:alphas-intertwiners}

In \sref{subsec:sss}, we start with a general introduction to the notion of superselection rules. 
In \sref{subseq:alg-reg}, we move over to a more rigorous formulation within the framework of algebraic QFT. 
In \sref{subsec:mes-abs} and, particularly, in \sref{subsec:holo-mutinfo}, we review relevant observations related to these concepts in the context of holographic entanglement entropy and mutual information.

\subsection{Selection rules}
\label{subsec:sss}

A selection rule constraints certain transitions in a physical system. This may happen due to the existence of conserved quantities. Opposed to this, a superselection rule can be seen as a more rigorous statement which puts restrictions on any possible measurement \cite{wick1952}. 

More precisely, a superselection rule not only forbids certain dynamical mode transitions generated by the Hamiltonian of the system, but also inhibits more general ones. Consider, for instance, two distinct states $|1 \rangle$ and $|2 \rangle$ of a given quantum system. We say both states are separated by a selection rule if 
\eqn{
\langle 1 | H | 2 \rangle = 0,
}
where $H$ denotes the effective Hamiltonian of the system. Now, suppose that $O$ is taken to be the set of all physically realizable, i.e. of simple complexity, observables. Then, the states $|1 \rangle$ and $|2 \rangle$ are separated by a superselection rule if 
\eqn{
\langle 1 | O| 2 \rangle = 0.
}
There is no way to measure the relative phase difference between both states. In other words, it is not possible to verify or prepare any coherent superposition of the states. A superselection rule therefore puts restrictions on the set of observables $O$ compared to the set of all bounded self adjoint operators acting on the Hilbert space $\mathcal{H}$. Consider, for instance, the following self adjoint operator 
\eqn{
V = | 1 \rangle \langle 2 | + | 2 \rangle \langle 1 |
}
for which obviously $\langle 1 | V | 2 \rangle \neq 0$.
From this perspective, the set of observables will be strictly smaller than set of self adjoint operators. 

Any superselection rule can be characterized by a Hilbert space decomposition in form of a direct sum of orthogonal subspaces,
\eqn{
\mathcal{H} =  \bigoplus_i \mathcal{H}_i.
\label{eq:prodHilbert}
}
Under the action of the observables $O$, states in a particular subspace can only be transformed into states living in the same subspace. 
Such subspaces are referred to as the superselection sectors under the action of $O$.

From the perspective of $O$, there is no way to coherently superimpose the states. To see this, let us introduce the following global pure state
\eqn{
| \Psi \rangle = \sum_{i=1}^M \sqrt{c_i}\ | i \rangle,\qquad M = \exp S(E),
\label{eq:superstate}
}
where $S(E)$ is taken to be the average entropy in the microcanonical ensemble. All eigenstates $| i \rangle \equiv | E_i \rangle$ are assumed to be normalized. 
Under the action of any observable $O$, the expectation value, i.e. ensemble average, of the associated measurement becomes
\eqn{
\langle \Psi | O | \Psi \rangle = \mathrm{Tr}[\rho O],\quad \rho = \sum_i c_i \rho_i.
}
The related density matrix is a weighted sum of all pure density matrices $\rho_i$. 
Hence, when the construction $| \Psi \rangle$ is treated as some quantum state, under the action of $O$, the density matrix $\rho$ defines a maximally mixed state which clearly deviates from
\eqn{
\rho_\Psi = | \Psi \rangle \langle \Psi |.
\label{eq:rho_Psi}
}
Hence, $\mathrm{Tr} \left[ \rho^2 \right] \neq (\mathrm{Tr}[\rho])^2$, and
if $c_i = p_i$ we get the usual thermal mixed state. Accordingly, coherent superpositions between states in subspaces $\mathcal{H}_i$ cannot exist. An observer only having access to the set of observables $O$, will not be able to measure the phase differences between the states $|i \rangle$. 

The concept of superselection sectors is well studied within the framework of algebraic QFT.
In general, one may say that superselection sectors in a theory come into existence when operators contained in it are insufficient to create any finite energy state from the vacuum. 
A representation of an algebra is obtained when the operators of that algebra act on a particular state. It may happen that a theory possesses various disjoint representations. 

For instance, consider some neutral algebra $\mathcal{N}$ of local operators which does not contain any charged operator. In addition, suppose there exist differently charged states in the theory under consideration. Under such conditions, acting with $\mathcal{N}$ will not have any impact on the charged states. 

More precisely, one cannot transform between the charges by performing such operations. The full Hilbert space therefore decomposes into orthogonal subspaces which are labeled by the charges. We say that these subspaces are the superselection sectors under the action of $\mathcal{N}$.

The situation changes if charged operators are added to the algebra. By doing so, the algebra $\mathcal{N}$ can be enlarged to a sufficiently large algebra allowing to reduce the number of superselection sectors. Ideally, one may enlarge it to some complete algebra $\mathcal{F}$ that does not suffer from the existence of superselection sectors at all. 

It should be noted that in certain situations one may principally study the model belonging to the subalgebra $\mathcal{N}$ by restricting to the Hilbert space of neutral states. These are created by acting with operators of the neutral subalgebra on the vacuum sector of the theory. This procedure is  basically sufficient to fully reconstruct the superselection structure as well as the complete algebra $\mathcal{F}$ of the full theory.

The existence of superselection sectors has crucial imprints on the relations between algebras and regions in the theory described by $\mathcal{N}$. There can, for instance, appear some violation of a certain type of duality in the sense that all operators commuting with the algebra of operators in some region are properly included in the algebra of the complement of that region.

\begin{figure}
  \centering
    \includegraphics[width=0.22\textwidth]{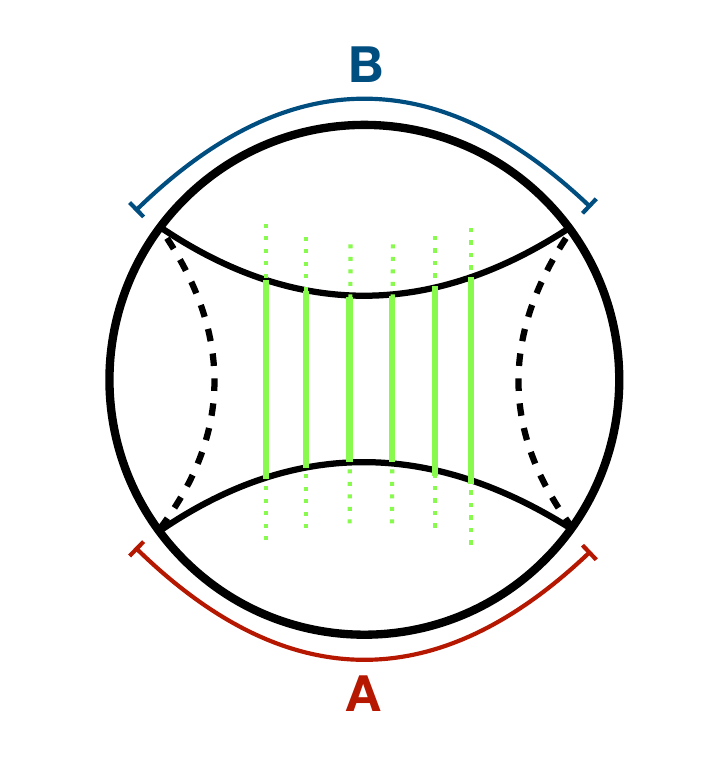}
    \caption{Holographic bulk intertwiners (green, solid lines) are supported on the bulk region which is enclosed by minimal codimension two surfaces.}
    \label{fig:int-global}
\end{figure}
 
Another property that can be violated in the presence of superselection sectors is the notion of additivity. In certain topologically nontrivial cases, for instance, it can happen that operators in some region can be generated by operators in smaller regions inside the bigger region.
More interestingly, superselection sectors may also affect the entanglement structure of the vacuum sector.

In the following, we first describe basic expectations for algebras corresponding to spacetime regions which should be fulfilled in a generic local QFT. Afterwards, we discuss some details related to the comments above.

\subsection{Regions and algebras}
\label{subseq:alg-reg}

Let $A^\prime$ be the set of spacetime points which are spatially separated from the subset $A$. The region $A$ is called causally complete when $(A^\prime)^\prime = A$.

Given causal evolution, any operator located in the causal domain of a certain region belongs to the algebra of that region. Different regions sharing the same causal domain have therefore the same algebra. 
For such type of algebras, the operators inside $A$ should also be included in any other region $B$ when $A \subseteq B$. Such operator algebras satisfy
\eqn{
\mathcal{N} [A] \subseteq \mathcal{N} [B].
\label{eq:prop1}
} 
Due to causality, it may also happen that spatially separated local operators commute such that
\eqn{
\mathcal{N} [A^\prime] \subseteq \mathcal{N}^\prime[A].
\label{eq:prop2}
}
The right hand side above corresponds to the commutant algebra which contains the set of operators commuting with all operators of $\mathcal{N}[A]$. 

A minimal requirement in QFT is that in a specific Hilbert space
representation the operator algebras should be von Neumann algebras, i.e. the double commutant of the algebra is the algebra itself. Such algebras are closed under linear combinations and product operations. All finite dimensional algebras, for instance, are of such type. Demanding this property, one may ask for an enhancement of \eqref{eq:prop2}, such that commutant algebras for complementary causal regions $A$ and $A^\prime$ in the vacuum representation obey the following duality property $\mathcal{N} [A^\prime] = \mathcal{N}^\prime[A]$. Of course, the latter can alternatively be formulated as
\eqn{
\mathcal{N}[A] = \mathcal{N}^\prime[A^\prime],
\label{eq:duality2}
}
assuming the algebra is von Neumann.
However, by doing so, one may be in tension with certain other properties. 

For instance, one may expect that an algebra of operators belonging to some region $A$ can be generated by algebras associated with smaller region included in the former. Such a local generation would violate the mentioned property of additivity. Let us note that, according to the previous assumptions, given two causally complete regions $A$ and $B$, the smallest causally complete region including both regions is $A \vee B = (A \cup B)^{\prime\prime}$. 

Similarly, for two algebras $\mathcal{N}[A]$ and $\mathcal{N}[B]$, the smallest von Neumann algebra containing both algebras is the generated algebra $\mathcal{N}[A] \vee \mathcal{N}[B]$.
The additivity property is then simply written as
\eqn{
\mathcal{N}[A] \vee \mathcal{N}[B] = \mathcal{N}[A \vee B].
\label{eq:additiv}
}
The right hand side is the additive algebra of both regions, where we note that $A \vee B = A \cup B$.
It should be emphasized that the relation above is even expected when the regions $A$ and $B$ are placed on a common Cauchy surface. 

In general, properties like isotonia \eqref{eq:prop1} and causality \eqref{eq:prop2} can be viewed as elementary axioms and should apply in a generic local QFT. 
However, properties such as duality \eqref{eq:duality2} as well as additivity \eqref{eq:additiv} hold only for models with a sufficiently large algebra. It turns out that they basically fail when the theory under consideration has superselection sectors. 

Tensions in respecting both properties, namely, duality \eqref{eq:duality2} and additivity \eqref{eq:additiv}, have explicitly been discussed in the case of theories with global symmetries for which charged operators can be constructed locally in ball shaped regions. In the algebraic approach to QFT, charged superselection sectors of this type are known as Doplicher--Haag--Roberts (DHR) sectors
\cite{doplicher1969fields,*doplicher1971local,*doplicher1974local}, see also \cite{Haag:1992hx}.\footnote{These are sectors in which a charged state representation on the algebra of operators outside a ball region cannot be distinguished from the vacuum representation. In this specific case, the corresponding charge has to be global and should not be associated with a gauge symmetry. 
}  

Consider the case where two ball regions $A$ and $B$ are disconnected. When superselection sectors of the previous type are present, it turns out that the additive algebra is properly included in the commutant algebra of the complementary region, i.e.
\eqn{
\mathcal{N}[A] \vee \mathcal{N}[B] \subset \mathcal{N}^{\prime}[(AB)^{\prime}].
\label{eq:add-dual}
}
Assuming additivity \eqref{eq:additiv}, such an inclusion is a clear manifestation of the mentioned violation of the duality property \eqref{eq:duality2} and thus illustrates that the theory belonging to $\mathcal{N}$ has superselection sectors. 
This goes back to the existence of certain operators, called intertwiners, which are not contained in the additive algebra, but are included in the commutant algebra of the complementary region.

For the commutant algebras of both sides in \eqref{eq:add-dual}, there is a similar loss of duality. The resulting reversed inclusion arises due to the existence of so called twist operators. These are contained in the commutant of the additive algebra, but do not belong to the double commutant algebra of the complementary region.

In topologically nontrivial setups, the existence of such operators is independent of the regularization. This is also the case for certain algebra choices in lattice formulations. Instead, being a macroscopic consequence of the underlying symmetry group, such operators are physical attributes of the continuum theory. 

It is particularly important emphasizing that in such cases both intertwiners and twists can be neutral operators. There is no need for charged operators so that the described properties are already valid for the vacuum sector of the theory. If the regions are sufficiently large, these neutral operators for sure would belong to the additive algebra.

In this regard, the theory associated with $\mathcal{N}$ can be seen as an uncharged neutral sector of a theory described by the complete field algebra $\mathcal{F}$. In more generality, when the latter without superselection sectors is equipped with a global compact symmetry group $G$, the subalgebra $\mathcal{N}$ corresponds to the orbifold under $G$. In other words, it is formed by the operators invariant under the actions of elements of $G$. Under the action of $\mathcal{N}$, the vacuum Hilbert space $\mathcal{H}$ of the theory $\mathcal{F}$ would then decompose into a direct sum over orthogonal sectors
\eqn{
\mathcal{H}_\mathcal{F} = \bigoplus_{i,j} \mathcal{H}_{i,j},
}
where $j$ labels the different irreducible representations under $G$ and $i$ is an index that counts the dimension of the representation $j$. Note that one of the sectors is the vacuum sector of the neutral subalgebra $\mathcal{N}$. 
The composition in the number of representations simply means that the elements of $\mathcal{N}$ do not run between the different base elements of the representation $j$.

\subsection{Measuring the absence}
\label{subsec:mes-abs}

The observations described above are particularly manifest in the entanglement properties of the vacuum sector in QFT. It is important to note that an information theoretic quantity such as entanglement entropy is technically infinite. From the holographic perspective this is reflected in the infinite AdS bulk volume dependence as dictated by the RT prescription \cite{Ryu:2006bv,Ryu:2006ef}. 

Hence, this quantity would not be well suited for making comparisons between two theories described by differently sized algebras, at least in the continuum limit. For instance, it is more appropriate to consider mutual information or related quantities such as relative entropy. For such quantities infinities can be eliminated and thus do not plague the final answer. 

The mutual information between two regions $A$ and $B$ is defined as
\eqn{
I(A,B) = S(A) + S(B) - S(AB).
\label{eq:mutualinfo}
}
We would like to introduce the following conditional expectation property \cite{ohya2004quantum,petz2007quantum}
\eqn{
S_\mathcal{F} (\omega || \phi \circ E) = S_\mathcal{F}(\omega || \omega_\mathcal{N} \circ E) + S_\mathcal{N} (\omega_\mathcal{N} || \phi ),
\label{eq:cep}
}
expressed in terms of relative entropies between certain states.
It is a purely positive quantity. 
The state $\omega_\mathcal{N}$ is the restriction to the neutral subalgebra $\mathcal{N}$ of some state $\omega$ in $\mathcal{F}$ and $\phi$ denotes some other given state in $\mathcal{N}$. The map $E$ is an appropriately chosen conditional expectation between $\mathcal{F}$ and $\mathcal{N}$. See \cite{Faulkner:2020hzi} for a recent discussion on conditional expectations in the context of (approximate) QEC in holography.\footnote{The conditional expectation --- here considered only for one region; details for the case of two ball regions formulated in terms of tensor product algebras are explicitly discussed in \cite{Casini:2019kex} --- defined to be $E: \mathcal{F} \rightarrow \mathcal{N}$, maps positive elements, e.g. self adjoint operators with positive spectrum, to positive elements. Consider $b,b^\prime \in \mathcal{N}$ and $f \in \mathcal{F}$, then 
\eqns{
E(1) = 1,\quad E(b f b^\prime ) = b E(f) b^\prime.
}
Particularly, note that $\mathcal{N}$ is left invariant by $E$. 

Let $\omega$ be some state in $\mathcal{F}$. Computing expectation values in $\mathcal{N}$ with regard to the latter results in the state $\omega_\mathcal{N}$. Suppose, $\phi$ corresponds to some state in $\mathcal{N}$. By using $E$ one can simply construct a state $\phi \circ E$ in $\mathcal{F}$. An invariant state in $\mathcal{F}$ would be given by $\omega \circ E$. By using $E$ one can verify \eqref{eq:cep}. 

Analogous ideas also apply to CFTs where, particularly, in two dimensions, these closely resemble the constructions developed in the case of DHR sectors. More precisely, a primary state which is a local operator, basically defines a sector of the theory. Each of such families correspond to an irreducible representation of the Virasoro algebra. 
Applying a generic operator constructed in terms of the energy momentum tensor leaves them invariant. 

One may therefore properly define a conditional expectation mapping from the complete field theory algebra to the algebra of the energy momentum tensor which is close in this particular case. The subspace invariant under that conditional expectation is spanned by states which are generated by all conformal transformations acting on the vacuum state. It has been argued that such ideas, at least approximately, might also apply in higher dimensions \cite{Casini:2019kex}.
} 

In particular, considering the vacuum states in both algebras, i.e. $\omega$ and $\phi \rightarrow \omega_\mathcal{N}$, makes
equation \eqref{eq:cep} trivial. Of course, both states are invariant under the group $G$ acting unitarily on $\mathcal{F}$. Therefore, $\omega$ and $\phi \circ E$ are identical in $\mathcal{F}$, i.e. $S_\mathcal{N}(\omega_\mathcal{N} || \phi) = 0$.
By definition, $\mathcal{F}$ includes charge creating operators in all irreducible representations, symmetry group operations as well as all elements in $\mathcal{N}$. 

Applying \eqref{eq:cep} to the case of two disconnected ball regions $A$ and $B$, the difference in mutual information between them becomes \cite{Casini:2019kex}
\eqnsplit{
I_\mathcal{F}(A,B) &- I_\mathcal{N}(A,B) 
= S_{\mathcal{F}} (\omega_{AB} || \phi_{AB} \circ E_{AB})\\
&= S_{\mathcal{F}} (\omega_{AB} || \omega_{AB} \circ E_{AB}).
\label{eq:Idiff}
}
The state $\omega_{AB}$ is taken as the vacuum in the algebra $\mathcal{F}[AB] = \mathcal{F}_A \otimes \mathcal{F}_B$, whereas $\phi_{AB}$ is the corresponding one in $\mathcal{N}[AB]$. We have the trivial relation $\omega_{AB} \vert_{\mathcal{N}_A \otimes \mathcal{N}_B} = \phi_{AB}$.
In particular, \eqref{eq:Idiff} allows to compute the difference 
from the neutral algebra in the vacuum sector. 
The theory which belongs to $\mathcal{N}$ can or cannot contain intertwiners in its algebra for the union $AB$. The subalgebra $\mathcal{N}[AB]$ without the intertwiners is additive, so it can be used to compute the mutual information $I_\mathcal{N}$. Here, we should refer to the proper inclusion in \eqref{eq:add-dual}.

On the other hand, the field algebra $\mathcal{F}[AB]$ contains the intertwiners and, therefore, does not satisfy the property of additivity. Hence, the difference \eqref{eq:Idiff} serves as an order parameter for the existence of superselection sectors in the theory that belongs to $\mathcal{N}$.

Assuming a finite set of intertwiners and bringing the regions $A$ and $B$ close to each other, there will still exist vacuum correlations between the regions. However, these are expected to be of the same size for both algebras. The leading divergent terms will therefore cancel. 
The only contributions originate from the intertwiners which are responsible for the difference. The entropy of that region may be seen as the entropy of intertwiner fields.

Note that the relative entropies in \eqref{eq:Idiff} do not correspond to any mutual information. They measure the difference between two states on the algebra $\mathcal{F}[AB]$ which 
contains the intertwiners that belong to the global $\mathcal{N}$, but are absent in the algebra $\mathcal{N}[AB]$ additively formed in the union $AB$. The vacuum state $\omega_{AB}$ differs from $\omega_{AB} \circ E_{AB}$ in the sense that for the latter the intertwiners are projected to the neutral algebras on each region. This is realized by the conditional expectation $E_{AB}$ which simply eliminates all the correlations due to intertwiner contributions. 

On the other hand, one can show that $E(\mathcal{F}[AB]) = \mathcal{N}^\prime[(AB)^\prime]$. Using this relation and the fact that the distinguishability of two invariant states under the conditional expectation is of same level in both algebras, the order parameter can even be computed entirely in $\mathcal{N}$, such that
\eqn{
I_\mathcal{F} - I_\mathcal{N}
=  S_{\mathcal{N}^\prime[(AB)^\prime]} (\omega_{AB} || \omega_{AB} \circ E_{AB}).
}
This is a clear manifestation of the fact that the latter is already intrinsic to $\mathcal{N}$ end can entirely be formulated within the subalgebra. This is even true for the mutual information $I_\mathcal{F}$ \cite{Casini:2019kex}.

\subsection{Holographic mutual information}
\label{subsec:holo-mutinfo}

Before we move on to the BTZ black hole, let us first consider the AdS vacuum.

In general, any QFT with its dual quantum gravity should be seen as a complete theory.
However, the low energy theory describing semiclassical bulk physics is expected to have a large number of superselection sectors.
It is important to emphasize that the quantum RT prescription for computing the generalized entropy should not lead to any violation indicating the existence of superselection sectors.

\begin{figure}
  \centering
    \includegraphics[width=0.21\textwidth]{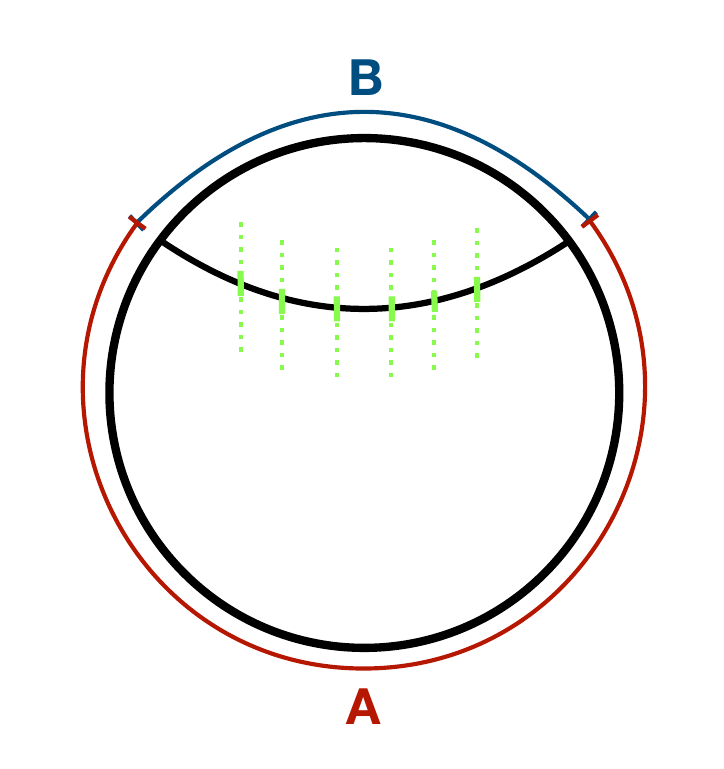}
    \caption{The boundary subregions are complementary to each other. Their entanglement entropy is given by the area of the RT surface (black, solid line). The latter measures the intertwiner correlations (green, dotted lines) ranging across the two subsystems.}
    \label{fig:int-global2}
\end{figure}

Of course, the final result is expected to yield the entropy one would directly obtain from the full CFT computation.
However, restricting any observable to the semiclassical degrees of freedom in the bulk changes the story. It is worth mentioning that switching to the bulk description for this purpose turns out to be very helpful. On the boundary, a strict distinction between low and high energy sectors might be more difficult to achieve.\footnote{Following \cite{Banks:1998dd,*Bena:1999jv,*Hamilton:2005ju,*Hamilton:2006az,*Hamilton:2007wj,*Heemskerk:2012mn,*Kabat:2015swa}, we here assume that there exist boundary operators associated with the (low energy) bulk operator contained in algebras that belong to certain bulk regions. The complete algebra of the boundary CFT (and its dual quantum gravity) should, of course, contain all possible operators.}

Consider two disjoint boundary subregions $A$ and $B$, where $AB$ corresponds to their union. Using the quantum RT prescription \cite{Faulkner:2013ana} and subregion-subregion duality \cite{Czech:2012bh,Dong:2016eik}, one can identify the corresponding bulk regions of their entanglement wedges which we denote by $\Sigma_A$, $\Sigma_B$ and $\Sigma_{AB}$, respectively. In addition, let $\mathcal{N}[\Sigma]$ be the algebra describing the low energy quantum fields in some bulk region $\Sigma$. Furthermore, let $\mathcal{F}[A]$ be the complete algebra of quantum fields in $A$. 

By construction, we have $\mathcal{N}[\Sigma_A] \subset \mathcal{F}[A]$. The algebra generated by the bulk semiclassical fields in $\Sigma_A$ and $\Sigma_B$ is $\mathcal{N}[\Sigma_A] \vee \mathcal{N}[\Sigma_A]$, which is properly included in the algebra of the bulk region $\Sigma_{AB}$, i.e. $\mathcal{N}[\Sigma_{AB}]$.

Accordingly, taking into account the property of commutativity, it can be shown that
\eqn{
\mathcal{N}[\Sigma_A] \vee \mathcal{N}[\Sigma_B] \subset \mathcal{N}^\prime[\Sigma_{AB}^\prime].
}
This is a clear manifestation of duality violation similar as in \eqref{eq:add-dual}, and indicates the existence of superselection sectors under the action of $\mathcal{N}$ describing semiclassical bulk physics. There exist intertwiner operators that belong to $\mathcal{N}[\Sigma_{AB}]$, but not to the additive algebra $\mathcal{N}[\Sigma_A \vee \Sigma_B]$. 

According to the basic properties in holography, the intertwiners, which are contained in $\mathcal{F}$ and induce correlations between the two regions $A$ and $B$ in the boundary field theory, should have bulk duals. These holographic intertwiners are supported on the bulk region $\Sigma_{AB} - \Sigma_A - \Sigma_B$ bounded by minimal codimension two surfaces, see \fref{fig:int-global}. 

Such bulk operators do not belong to the algebra additively formed in the boundary subregions. However, they commute with the algebra associated with the complement of their union, $\mathcal{N}[(AB)^\prime]$.

The bounded region can be seen as a measure for the amount of intertwiners. Importantly, when $B$ is the complement of $A$, the intertwiner correlations are entirely measured by the area of the RT surface, see \fref{fig:int-global2}. As discussed in \cite{Casini:2019kex}, this may eventually lead to the bit thread picture proposed in \cite{Freedman:2016zud}.
We comment on this connection later in the context of the BTZ black hole.

Coming back to the present example, 
we can use the quantum RT formula to compute the mutual information between $A$ and $B$ in $\mathcal{F}$. The same quantity can also be computed in $\mathcal{N}$. In the latter case, the mutual information only depends on the entropies associated with the local quantum fields. The holographic computation then gives \cite{Casini:2019kex}
\eqnsplit{
I_\mathcal{F}(A,B) &- I_\mathcal{N}(A,B) =
\frac{\mathcal{A}(\Sigma_A) + \mathcal{A}(\Sigma_B) -\mathcal{A}(\Sigma_{AB})  }{4 G_N}\\
&+ S^0(\Sigma_A \cup \Sigma_B) - S^0(\Sigma_{AB}).
\label{eq:IF-IN-AdS}
}
The dominant area contributions account for the microscopic correlations of the complete theory containing the intertwiners. The subleading terms can be seen as a measure for the amount of the latter supported on the bulk region bounded by minimal surfaces.

The discussion above particularly shows the powerfulness of the RT prescription. There is no need to work out the leading order correlations relying on the UV degrees of freedom. Indeed, from the point of view of the boundary QFT, this might be a more difficult task. Instead, all the correlations are taken into account by computing minimal codimension two surfaces in the bulk. This nicely illustrates how the RT surface knows about the microscopic degrees of freedom of the underlying quantum gravity.

\section{Superselections and black holes}
\label{sec:sss-bhs}

In this section, we address some of the main observations in the present paper. 
In \sref{subsec:ss-therm}, starting from the previous ideas, we make the connection to black holes and discuss the role of superselection sectors in the thermalization process. 
The subsequent part, \sref{subsec:bh-int}, deals with operator reconstructions in the presence of large microstate subspaces. We argue that intertwining operators play a substantial role in the latter case. 
In \sref{subsec:bit-threads}, we comment on a possible connection to the bit thread picture. 
In \sref{subsec:qnetworks}, we elaborate on some similarities to certain discrete formulations.

\subsection{Superselections and thermalization}
\label{subsec:ss-therm}

Let us now turn to the BTZ black hole. As before, let $O$ be the set of observables that belong to the subalgebra $\mathcal{N}$ describing semiclassical bulk physics. 
We use the same notation for the corresponding operator set in the boundary CFT. Again, all eigenstate energies are fixed in a sufficiently narrow energy band.

In the basis of BTZ microstates, the operators $O$ undergo eigenstate thermalization \cite{Hartman:2014oaa,Hikida:2018khg}. This is expected to be a universal feature of states describing thermodynamically stable black hole geometries. 

Stated differently, all matrix elements satisfy the following relation
\eqn{
\langle i | O | j \rangle = f_O(E_i) \delta_{ij} + e^{-S(E)/2} g_O(E_i,E_j) R_{ij},
\label{eq:eth-rel}
}
where $S(E)$ is the average coarse grained entropy and $R_{ij}$ is a random matrix having zero mean and unit variance. The functions $f,g$ are smooth and depend on the microstate energies. 

To leading order, the average entropy in the microcanonical ensemble, i.e. thermal, is proportional to the central charge of the CFT and is given by the Cardy formula\footnote{The BH entropy is a universal prediction to leading order and serves as an IR constraint on the UV completion of quantum gravity.
As opposed to standard effective field theory, the nontrivial coupling of different energy scales in quantum gravity
basically goes back to the presence of diffeomorphism invariance.

The BH entropy has successfully been derived from a particular string theory setup \cite{Strominger:1996sh}. 
Being independent of the underlying microscopic details, a similar universal behavior is reflected in the precise agreement between the Cardy formula \eqref{eq:cardy} and the BH entropy in AdS$_3$ \cite{Strominger:1997eq}, where the former computes the entropy in a unitary modular invariant two dimensional CFT \cite{Cardy:1986ie}. 

It generally applies when the central charge is fixed and the black hole energies are considered to be very large. Of course, this is very different for the BH formula which should hold in the semiclassical regime, where the central charge of the CFT is large and the energies are of the same order. Indeed, in holographic setups the formula applies beyond the Cardy limit. 

This is what exactly applies to the case of two dimensional large central charge CFTs having a sparse light spectrum \cite{Hartman:2014oaa}. 
}
\eqn{
S(E) \simeq \sqrt{\frac{2 c \pi^2 E}{3}} + \mathcal{O}(1).
\label{eq:cardy}
}
In the semiclassical limit we thus have
\eqn{
\langle i | O | j \rangle \simeq 0\quad \forall\ i \neq j.
\label{eq:eth:simple}
}

Then, similar to what we have seen in the preceding discussion, one would not be able to measure the relative phase differences
between the microstates under the action of $\mathcal{N}$. 

From the perspective of an observer restricted to the semiclassical degrees of freedom, the entire code space would decompose into a large number of orthogonal subspaces, i.e.
\eqn{
\mathcal{H}_\text{code} \simeq \bigoplus_i \mathcal{H}_{\text{code},i}.
\label{eq:sum-orth}
} 
In other words, the operators $O$ would not run between the different bases of the irreducible representations generated by charged local operators in the CFT.

\begin{figure}
  \centering
    \includegraphics[width=0.42\textwidth]{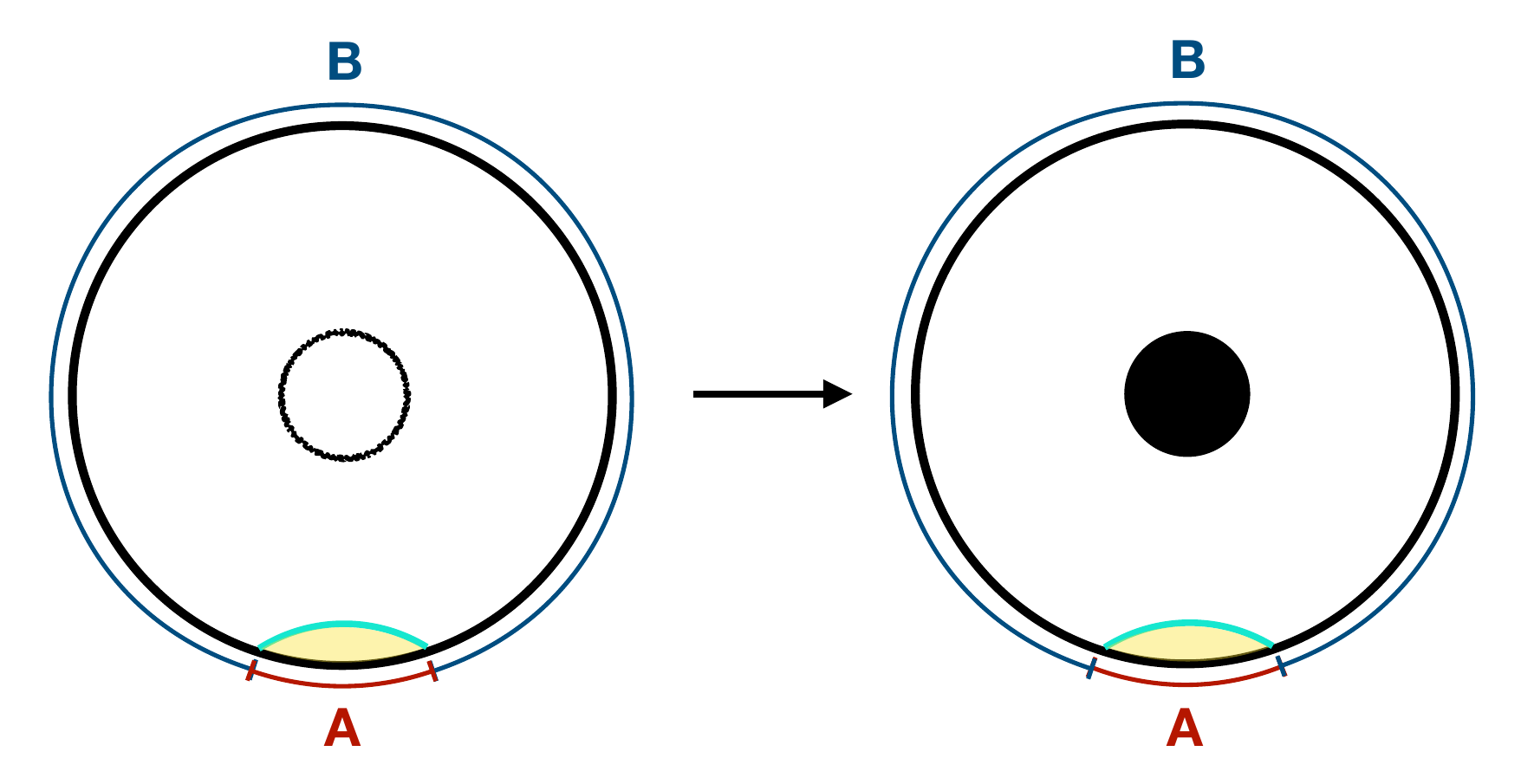}
    \caption{Under the action of the algebra $\mathcal{N}[A]$, the global pure state (left) is effectively seen as the thermal mixed state (right). The entanglement wedge of subregion $A$ corresponds to the colored bulk region.}
    \label{fig:bh-mixed}
\end{figure}

Consider a low energy bulk operator located in the entanglement wedge of subregion $A$ for which we assume $2\ell_A \ll \ell_\text{bdy}$, see \fref{fig:bh-mixed}. 
Recall that in the presence of superselection sectors, the density matrix appearing in the expectation value for such observables evaluated in the basis of \eqref{eq:superstate} turns out to be maximally mixed, see \sref{subsec:sss}. 

In fact, from the perspective of $A$, chosen to be sufficiently small, such an effective thermalization applies in the present case. The charged sectors are invisible under the action of the algebra associated with $A$, so that the entire state looks thermal.\footnote{Even though it seems to be so from the perspective of a global observer, there is no inconsistency in considering a superposition of different microstates, i.e. bulk quantum states, described by \eqref{eq:superstate}, i.e. $\rho_\Psi$. 

Whenever a large number of superselection sectors is present which, for instance, happens under the action of the subalgebra $\mathcal{N}$, the state $\rho_\Psi$ will be seen maximally mixed. Hence, we may write 
\eqns{
\rho_\Psi \circ E = \rho_\text{therm}.
}
The restriction would be realized by an appropriately chosen conditional expectation $E$ that integrates out the contributions relying on the intertwiners, see \ref{subsec:mes-abs}.

We assume that the total energy of the entire system is fixed.
\label{fnote:11}
}

We may say that superselecting integrates out the quantum gravitational degrees of freedom of the complete theory dual to the CFT. 
Distant local observers encounter a thermal black hole. 
Indeed, restrictions on quantum operations, as in the case of superselection rules, can lead to an entanglement structure in pure bipartite systems which resembles that of mixed states, cf. \cite{Bartlett_2006}.

In the following, we revisit certain aspects of the underlying structure of the Hilbert space. 

Let us begin by noticing that whenever $A$ is sufficiently small, i.e. described by a small subalgebra, we would effectively end up with the code space \eqref{eq:subspace} factorized into an interior part and an exterior part which is justified to leading order. 

However, we should still be aware of the fact that the superselection sectors are associated with different microstates, and thus correspond to distinct fine grained internal degrees of freedom.
The energies are contained in a sufficiently narrow energy band. To leading order, one may use the same geometry for all states, namely, the one associated with the mass of the thermal black hole, see \sref{sec:alphas}. 

However, in a strict sense, the black hole states differ from each other.
This would rather lead to a subspace decomposition of the form (see also appendix \ref{subsec:sdecomp})
\eqn{
\mathcal{H}_\text{code} 
\simeq \bigoplus_i \left( \mathcal{H}_{\text{BH},i} \otimes \mathcal{H}_{\text{ext},i}  \right).
\label{eq:Hcode-sum}
}

If the microstates are sufficiently entangled with some external purifying system, we can specify an exterior region surrounding the black hole bounded by QESs, see \fref{fig:2geodesics}.
Any such region exists for all microstates.

Importantly, starting from the global state \eqref{eq:superstate}, these regions will not belong to the entanglement wedge of $A$ as long as the latter is sufficiently small, since the state will look thermal under the action of the corresponding algebra. 

In the bulk, we may therefore localize the fine grained degrees of freedom associated with the different internal states within that bounded region. The structure of the Hilbert space describing the latter can be viewed as being unaffected. This may effectively be assumed, since it is also excluded from the entanglement wedge under the action of the respective algebra.
Then, to leading order, the same QES can be used for all microstates.

Of course, this is an effective picture valid in the case when $A$ is small, leading to the factorized space \eqref{eq:subspace}. 
However, in the light of the arguments raised above, we have to consider a space decomposition into orthogonal subspaces as expressed in \eqref{eq:Hcode-sum}. 

We finalize this part by commenting on certain similarities to the ideas reviewed in \sref{subseq:alg-reg}. 

In a CFT, the complete algebra consists of a set of primary operators and their descendants. In particular, the superselection sectors described above are associated with the primaries generating the microstates. Since any primary is a charged local operator, this is indeed reminiscent of the situation with DHR sectors. 

For a given primary, all members of the respective Verma module can be obtained by using the generators of conformal transformations.
More precisely, all descendants associated with a primary $P_\Delta$ can be 
constructed by linear combinations of certain basis operators
of the form $ U_g P_\Delta U_g^{-1}$, where $g$ parameterizes a conformal transformation. 

\begin{figure}[b]
  \centering
    \includegraphics[width=0.42\textwidth]{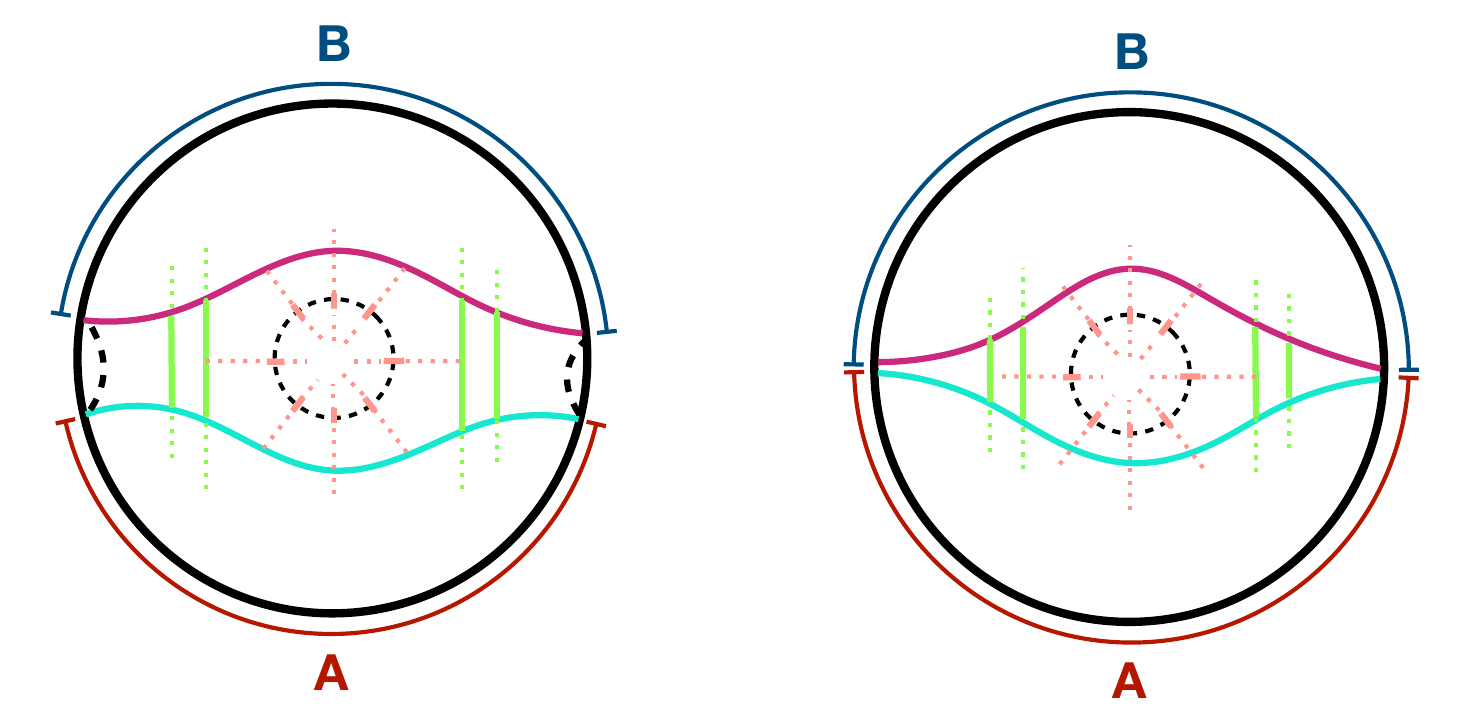}    
        \caption{Holographic bulk intertwiners dual to boundary operators connecting the two disjoint subregions $A$ and $B$ are illustrated as green lines. Black hole intertwiners are shown in form of red bars puncturing the horizon. The associated correlations are indicated by dotted line extensions. It is important to emphasize that the dotted lines extending into the interior do not necessarily represent pure quantum correlations. However, correlations illustrated by the dotted lines in the exterior region will possess a unitary, quantum mechanical description.}
        \label{fig:bh-intertwiners}
\end{figure}

In the particular case of two dimensional CFTs, 
the complete algebra corresponds to the one of the smeared energy momentum tensor. Any operator $F$ of the complete field theory
can be written as a sum over smeared primaries, i.e.
\eqn{
F = \sum_\Delta T_\Delta P_\Delta,
}
where $T_\Delta$ is a generic operator entirely constructed from the energy momentum tensor.

Therefore, under the action of the observable algebra $\mathcal{N}$, which consists of the operators $T_\Delta$, it is not possible to generate transitions between the  charged sectors associated with the interior degrees of freedom.

\subsection{Black hole intertwiners}
\label{subsec:bh-int}

We have discussed that the microstate degrees of freedom effectively localize under the action of $\mathcal{N}$.
The localization happens in a region that surrounds the black hole and is bounded by QESs.

As we have seen in \sref{subsec:holo-mutinfo}, there exist superselection sectors suggesting that certain intertwining operators would be operative on the bounded bulk region. In the following, we wish to elaborate on this aspect in more detail.

We begin with the case of a single microstate and consider an appropriate order parameter, namely, the mutual information between two disjoint boundary subregions. The complementary regions are assumed to be sufficiently small, centered around two points dividing the asymptotic boundary into two equally sized segments. This separation results in two disjoint subregions $A$ and $B$, see \fref{fig:bh-intertwiners} (left). 

The difference in mutual information with respect to $\mathcal{F}$ and $\mathcal{N}$, defined as above, takes the form
\eqn{
I_\mathcal{F}(A,B) - I_\mathcal{N}(A,B) = \frac{2 \mathcal{A}(\Sigma_{A,B})  - \mathcal{A}(\Sigma_{AB})}{4 G_N}  + \ldots\ ,
}
where the subleading entropy is associated with the quantum fields supported on the region bounded by the minimal surfaces. It may be viewed as a measure for the amount of holographic bulk intertwiners. These would be dual to intertwining operators connecting $A$ and $B$ on the boundary. Pictorially, they are represented in form of vertical lines running between the RT surfaces associated with the two disjoint subregions and are basically of the same type as shown in \fref{fig:int-global}. In the CFT, they do not belong to the algebra additively formed in the subregions, but they commute with the algebra $\mathcal{N}$ for the complement of their union.

In addition, motivated by previous arguments, there should also exist operators which intertwine certain endomorphisms associated with the disjoint representations in the CFT. Due to reasons which will become clear below, such black hole intertwiners, as we call them, will be depicted as bulk objects puncturing the horizon. The associated quantum correlations would then extend across the exterior and the interior region as depicted in \fref{fig:bh-intertwiners}. Recall that such correlations are of the kind discussed in \sref{subsec:intOps}. We will come back to this point further below.

Generally, it turns out that not having access to such correlations results in not being able to generate transitions between the internal sectors.
Whenever a boundary subregion is sufficiently small, any bulk operator in the bounded region as well as acting on the black hole cannot be reconstructed from it.

On the other hand, if the corresponding RT surface becomes the upper one shown in \fref{fig:bh-intertwiners} (left), the entire bounded region will belong to the entanglement wedge of $A$. We end up with the situation, where 
both boundary entropies, i.e. $S_{A,B}$, coincide with each other.
Of course, this is expected for a pure state. There will be no superselection sectors present.
All the intertwiner correlations are measured by the area of the minimal surface, see \fref{fig:int-global2}. Bulk operators located in the bounded region can be reconstructed on $A$. In the bipartite case, the difference in mutual information would then be
\eqn{
I_\mathcal{F}(A,B) - I_\mathcal{N}(A,B) \simeq \frac{\mathcal{A}(\Sigma_A)}{2 G_N},
}
where, again, the area term serves as an order parameter for the failure of duality in the theory described by the subalgebra $\mathcal{N}$.

Having considered the disjoint setup, suppose that we now gradually shrink the complementary region of the two intervals. By doing so, the boundaries of $A$ and $B$ shall move to locations slightly below and above, respectively, the points dividing the AdS boundary into two parts, see right panel in \fref{fig:bh-intertwiners}. This decreases the entropy of quantum fields located in the bounded shadow region, i.e. the amount of holographic bulk intertwiners.

Recall that on the boundary, their counterparts are taken to be operators which connect the two minimally disjoint subregions.
In the bulk, intertwiner contributions are measured by the area of minimal surfaces.
It is clear that, when $A$ and $B$ are sufficiently large, but not complementary to each other, we almost end up with the two QESs discussed in \sref{sec:alphas}. 

The relevant correlations relying on the black hole intertwiners should then be supported on the bounded bulk region. Under the action of the algebra belonging to the minimally disjoint subregions $A$ and $B$, there would exist superselection sectors.

These would describe semiclassical bulk physics. Intuitively, this might be expected. Namely, physics in the far distant bulk region should be described by some local QFT defined on a fixed background. However, a proper description in the vicinity of the horizon should require certain entities which would not be captured by an algebra that only describes local physics.
Under the action of the latter, the global state will look maximally mixed. This is basically reflected in the entropy
\eqn{
S \left( \rho_{i,\{A,B\}} \right) \simeq S \left( \rho_{\text{therm},\{A,B\}} \right).
\label{eq:S_iA-small}
}
In what follows, we would like to elaborate more on the black hole intertwiners leading to correlations supported across the near horizon region.

Let us note that given a pure microstate, it is known how an interior operator can be reconstructed in terms CFT operators \cite{Papadodimas:2012aq,*Papadodimas:2013wnh,*Papadodimas:2013jku}. The additional objects required for such reconstructions are the state dependent mirror operators. Recall that in the standard perturbative approach to bulk operator reconstructions there are no such entities necessary \cite{Banks:1998dd,*Bena:1999jv,*Hamilton:2005ju,*Hamilton:2006az,*Hamilton:2007wj,*Heemskerk:2012mn,*Kabat:2015swa}. This is a major difference which is expected to be an intrinsic feature of reconstructing local operators in quantum gravity. However, note that this appears to be so if we restrict to the case of a single black hole microstate.

\begin{figure}[h!]
  \centering
    \includegraphics[width=0.24\textwidth]{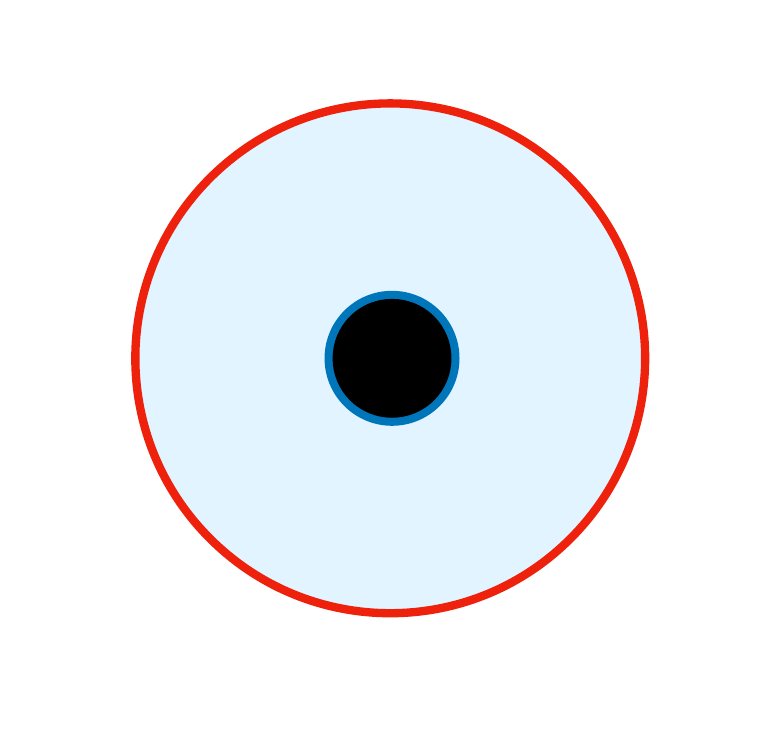}
        \caption{Consider the state \eqref{eq:superstate} which is well defined from the CFT point of view. Under the restriction to $\mathcal{N}$, the holographic bulk dual is the thermal AdS black hole. For a local distant observer the microscopic interior degrees of freedom seem to be encoded on the smooth horizon (blue circle) which cannot be accessed from the boundary. However, the entire information outside (light blue region) and inside would be reconstructable under the action of the complete algebra of the CFT. 
We may therefore expect that there exist certain correlations which rely on operators that are not contained in the subalgebra $\mathcal{N}$. As in the case of bulk intertwiners, the former should be supported across an extended bulk region outside as well as inside the horizon. 
This may be motivated by reminding that for the entire boundary, under the restriction to $\mathcal{N}$, the RT surface becomes the black hole horizon.
Because of entanglement wedge reconstruction, see \sref{subsec:statedep}, any operator located in the exterior bulk region may be reconstructable from the restricted boundary state. This will be possible while no internal degrees of freedom may be accessed. 
Then, according to the expectation above, the (quantum) microscopic correlations responsible for storing the black hole information would not reside in the interior, but extend across the exterior near horizon region, the black hole atmosphere. 
They are thought to be intrinsic quantum properties of the emergent bulk spacetime and would not be described by the subalgebra $\mathcal{N}$.
Indeed, this is consistent with the way how AdS/CFT is functioning. The bulk spacetime dual to the boundary field theory is covering the entire space including the region outside the black hole horizon. Hence, the way how the holographic duality operates already suggests that information would be transferred via hidden nonlocal (from the semiclassical perspective) correlations having support on the exterior bulk region.
}
\label{fig:bh-bdys}
\end{figure}

As we already exemplified in \sref{subsec:intOps}, the black hole intertwiners sketched in \fref{fig:bh-intertwiners} may be related to certain mirror operators that turn out be required for interior bulk operator reconstructions from the boundary in the case of a given particular microstate. This is especially supported by observations suggesting a close connection between nonlocal objects like wormhole threading Wilson lines and mirror operators in the two sided eternal black hole \cite{Guica:2015zpf}. 
In fact, for a single microstate, an analogous setup may be realized by constructing the associated TFD like state as in \cite{Papadodimas:2012aq}.

Interestingly, by examining the concept of superselection sectors and intertwiners in the case of the two sided eternal black hole, an identification of that kind has been proposed in
\cite{Casini:2019kex}. It has been argued that intertwiners formed by charged operators on both sides
would be dual to the factorizable (namely, by high dimension CFT operators) wormhole threading gauge invariant operators in the bulk \cite{Harlow:2015lma,Guica:2015zpf}. 
Here, particular evidence is provided by comparing the mutual information between the left and right sides of the TFD, again computed for $\mathcal{F}$ and $\mathcal{N}$. As in the one sided case, the nonzero difference
\eqn{
I_\mathcal{F}(L,R) - I_\mathcal{N}(L,R) \neq 0
\label{eq:F_N-diff-TFD}
}
is determined by area contributions which indicates the lack of intertwiner correlations under the action of $\mathcal{N}$.
The associated correlations would then be supported on an extended bulk region outside as well as inside the bifurcation surface. This is precisely what we also expect happening at the horizon (wormhole throat) for the one sided black hole.

On the other hand, inspired by the bulk threading Wilson line in the $U(1)$ case, it has been shown that in the gravitational case, there exist similar nonlocal objects extending through the bulk \cite{Harlow:2018tqv}. 

We view such findings as further evidence for the existence of the described black hole intertwiners.
We would like to emphasize that in the present work, the reasoning for the existence of such entities purely relies on algebraic arguments. It is highly interesting to explore how these are explicitly constituted in the CFT.

Let us now consider a large number of superimposed microstates described by \eqref{eq:superstate}. Again, since under the action of a subalgebra associated with some sufficiently small subregion all charged sectors can be viewed as localized within the bounded shadow region, these sectors are expected to play a similar role as DHR sectors. Recall that in the latter case, charges can be localized in ball like regions and are not distinguishable from the vacuum in the complementary region. 

Similarly, in the present situation, the charged state representations cannot be accessed under the action of the algebra outside the bounded region. 
In other words, we cannot distinguish the microstates 
from the boundary subregions. The subsystems have superselection sectors. 
For a single microstate, this is somewhat reflected in the entanglement entropy \eqref{eq:S_iA-small}. 

Consider the expectation value of some bulk operator $\tilde O$ localized in the interior region. Given some microstate, it can be constructed in terms of CFT operators \cite{Papadodimas:2012aq,Papadodimas:2015xma}. Its expectation value in the basis of the state can formally be expressed as
\eqn{
\langle i | \tilde O | i \rangle 
= \langle 0 | \varrho_i(\tilde O) | 0 \rangle,
}
where the transformation $\varrho_i(\tilde O)$ is written in terms of the primary operator,
\eqn{
\varrho_i(\tilde O)  = \psi_{i}^\dag \tilde O \psi_{i},
}
with $|i \rangle = \psi_{i} | 0 \rangle$. 

We should bear in mind that by construction the operator $\tilde O$ itself depends on unmeasured details of the microstate.
Thus, in a strict sense, the introduced transformation should be refined by taking this state dependence (which arises due to the fact that we explicitly restrict ourselves to a single state) into account. For brevity, we do not work out the details.

\begin{figure}[h!]
  \centering
    \includegraphics[width=0.44\textwidth]{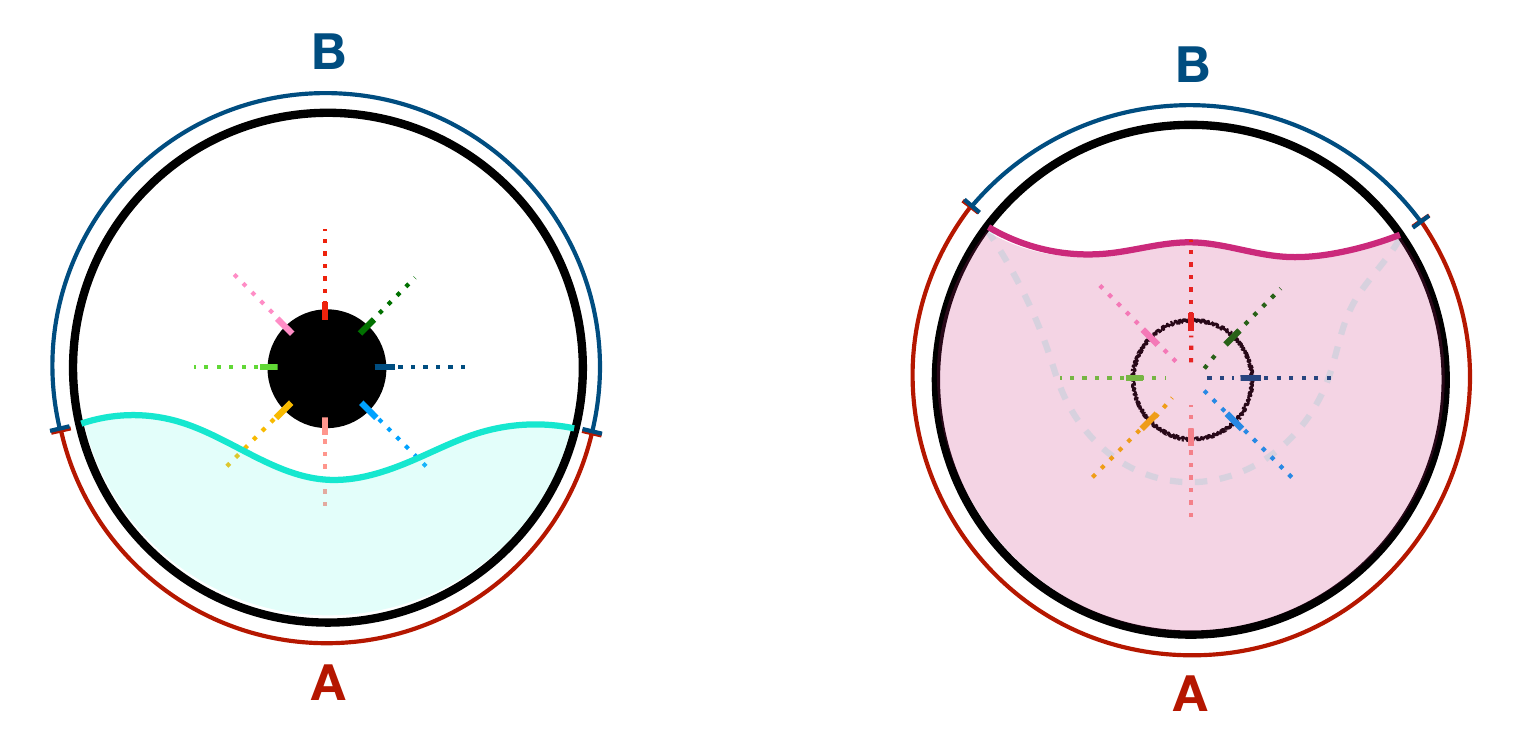}
        \caption{For a sufficiently small subregion $A$, the global state under the action of the associated algebra would look maximally mixed. This is pictorially illustrated using the bulk representation (left). The relevant correlations relying on the black hole intertwiners (colored punctures with dotted line extensions) cannot be accessed from $A$. 
By increasing the subregion size, these start to become accessible.        
At a certain level, all relevant correlations would be encoded. Under the action of the algebra associated with the enlarged subregion $A$, the perceived global state would then maximally depart from the thermal state (right).}
        \label{fig:bh-intertwiners-many}
\end{figure}

So, it is possible to express the charged representation as a composition of the vacuum representation in terms of an appropriately constructed operator which would belong to the full algebra.

Under the action of any operator $O$, the fine grained sector can be seen as localized in the bounded shadow region. An appropriate transformation should not affect the elements of the algebra associated with a given boundary subregion assumed to be sufficiently small, i.e.
\eqn{
S_{i,A} \simeq S_A^\text{vac} = \frac{c}{3} \log \left(  \frac{\ell_A}{\epsilon_\text{UV}} \right).
}
Relying on the simplified ansatz from above, we may then ask for the operator intertwining two different transformations, i.e.
\eqn{
\mathrm{I}_{ij} \varrho_i(\tilde O) = \varrho_j(\tilde O) \mathrm{I}_{ij},
}
which turns out to be a unitary operator of the form
\eqn{
\mathrm{I}_{ij} = \psi_j^\dag \psi_i.
}
It may be seen as the generator of transitions between the two charged representations. As mentioned, the details of the intertwiners should be worked out in more detail. For instance, in doing so one may separate the fine grained part and the coarse grained part for each microstate. From this, it becomes clear that the disjoint coarse grained and fine grained sectors associated with the microstates would be linked with each other. The black hole intertwiners are thought to be operators allowing that kind of transitions.

It is worth mentioning that once such operators would be invocable, it may be possible to access  the off diagonal terms in the matrix ensemble. These contributions would then lead to unitarization in black hole evaporation, cf. e.g. \cite{Penington:2019kki}. Of course, if this becomes so, there would be no superselection sectors under the action of the respective algebra. We discuss these aspects in more detail in \sref{subsec:pagec} as well as possible connections to certain wormhole topologies in \sref{sec:wormholes}.

Coming back to the present setup, since we consider a large number of microstates, there would exist black hole intertwiners associated with the different states. As sketched in \fref{fig:bh-intertwiners}, these would puncture the horizon giving rise to correlations extending across the near horizon region that can be identified as the black hole atmosphere. We briefly comment on this in \fref{fig:bh-bdys}. 

Having said this, starting with a sufficiently small subregion $A$, the global state under the action of the associated algebra in the bulk would look maximally mixed. 
The relevant correlations relying on the intertwiners lie outside the entanglement wedge of $A$, see left panel in \fref{fig:bh-intertwiners-many}. What is observed from the perspective of $A$ is nothing but the thermal mixed state. In such a case, the fine grained entropy for the subsystem is
\eqnsplit{
S(\rho_{\text{therm},A}) &= - \mathrm{Tr} \left[ \rho_{\text{therm},A} \log \rho_{\text{therm},A} \right].
\label{eq:SthermA}
}
In the holographic entropy computation, we would need to impose the usual homology constraint on the RT surface, since the resulting state would be ignorant of the internal degrees of freedom.

\begin{figure}[b]
  \centering
    \includegraphics[width=0.2\textwidth]{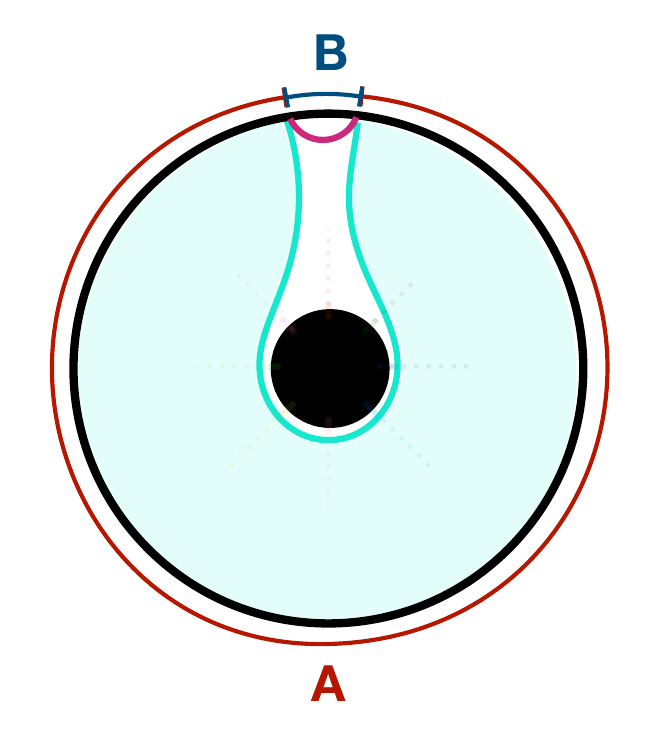}
        \caption{By restricting the global state to the subalgebra $\mathcal{N}$, it (taken to be the boundary state) cannot access the interior degrees of freedom. In this case, the ensemble state is dual to the thermal black hole. The boundary becomes ignorant of nonlocal correlations (faded out) relying on black hole intertwiners. This necessitates imposing the usual homology constraint on the minimal surface as dictated by the RT prescription.}
        \label{fig:bh-intertwiners-cut}
\end{figure}

Note that under the restriction to $\mathcal{N}$, this is of course true even when $\ell_A \rightarrow \ell_\text{bdy}$ as depicted in \fref{fig:bh-intertwiners-cut}. 
The nonlocal correlations are induced by operators which are not contained in $\mathcal{N}$.
Under the action of the latter, the entropies for the two complementary subregions $A$ and $B$ will not coincide and therefore indicate the existence of superselection sectors.

Nevertheless, by gradually increasing $\ell_A$, one would start to encode the intertwiner correlations in the entanglement wedge of $A$. At a certain level, all relevant correlations would be encoded. Under the action of the associated algebra, where the latter enlarges gradually, the state will start to depart from the thermal state \eqref{eq:thermal-BH}, thus, leading to the fine grained entropy
\eqn{
S(\rho_{\Psi,A}) = - \mathrm{Tr} \left[ \rho_{\Psi,A} \log \rho_{\Psi,A} \right].
\label{eq:SpsiA}
}
We portray this situation in the right panel of \fref{fig:bh-intertwiners-many}.

We should emphasize that this observation aligns with findings in the context of microstate distinguishability, see \sref{subsec:holevo}. 

Namely, recall that as soon as $\ell_A = \ell_\text{crit}$, it becomes possible to perfectly distinguish microstates under a finite set of measurements on $A$.
According to the present discussion, this simply happens because all the necessary correlations associated with the intertwiners are encoded in the entanglement wedge of $A$, such that any operator located within the shadow region as well as acting on the black hole can be reconstructed from $A$. 

This may explain why perfect distinguishability already sets in when $\ell_A = \ell_\text{crit}$, and not only if $\ell_A = \ell_\text{bdy}$. The Holevo information saturates in both cases,
\eqn{
\chi(\ell_\text{crit}) = \chi(\ell_\text{bdy}) = S_\text{BH}.
\label{eq:chi2}
}
In particular, it happens to be so for any finite subregion $A$ as long as $\ell_A \geq \ell_\text{crit}$.

It is important to note that \eqref{eq:SpsiA} satisfies the linearity property, means, it may be approximated as 
\eqn{
S(\rho_{\Psi,A}) \approx \overline{ S (\rho_A) },
\label{eq:Slin}
}
cf. e.g. \cite{Almheiri:2016blp}. The right hand side of \eqref{eq:Slin} is defined to be the ensemble of subregion entropies associated with each microstate contributing to the global density matrix \eqref{eq:rho_Psi}, that is
\eqn{
\overline{ S (\rho_A) } \equiv \sum_i c_i S(\rho_{i,A}).
\label{eq:Sensemble}
}
Adopting the approximation from \eqref{eq:Slin}, the Holevo information can be expressed (for $\ell_A \geq \ell_\text{bdy}/2$) as
\eqn{
\chi = S(\rho_{\text{therm},A}) - S(\rho_{\Psi,A}).
\label{eq:newChi}
}
Starting with $\chi = 0$, one gradually reduces the number of invisible internal sectors by increasing the size of $A$. As soon as $\ell_A = \ell_\text{crit}$, the difference becomes maximal, namely $S_\text{BH}$, see \eqref{eq:chi2}. 

Putting the latter relations together, the fine grained entropy associated with $A$
may then be expressed as
\eqn{
S(\rho_{\Psi,A})
\simeq
\begin{cases}
S(\rho_{\text{therm},A}) & \text{for}\ \ell_A < \frac{\ell_\text{bdy}}{2}\\\\
\overline{ S (\rho_A) } & \text{for}\ \ell_A > \frac{\ell_\text{bdy}}{2}
\end{cases}.
\label{eq:S_A-piece}
}
So long $2 \ell_A \leq \ell_\text{bdy}$, the entropy coincides with the thermal result \eqref{eq:SthermA}. For $2\ell_A \geq \ell_\text{bdy}$, the increasing deviation from the thermal answer becomes clear when we rewrite the ensemble as 
\eqn{
\overline{ S (\rho_A) }  = S(\rho_{\text{therm},A})  - \alpha S_\text{BH},
\label{eq:SPsi-alt}
}
where we have used \eqref{eq:newChi}. The maximum is reached when $\alpha = 1$. For instance, having $2 \ell_A \simeq \ell_\text{bdy}$, one would not be able to clearly distinguish the global state \eqref{eq:superstate} from the thermal mixed state \eqref{eq:thermal-BH}.

\subsection{Bit threads}
\label{subsec:bit-threads}

An alternative interpretation in the context of microstate distinguishability
has been proposed in \citep{Bao:2017guc}. It suggests a connection between Holevo information and the bit thread formulation of the RT proposal \cite{Freedman:2016zud}. 

According to the latter, the minimal surface can be related to a maximization procedure, where one maximizes over vector fields with a fixed maximal density per unit area. The vector field represents a flux of bit threads which cross the bulk region and connect the two entangled subregions on the boundary. 
In fact, such a maximization procedure is equivalent to minimizing the minimal surface area traversed with a maximal density of bit threads.   

\begin{figure}
  \centering
    \includegraphics[width=0.22\textwidth]{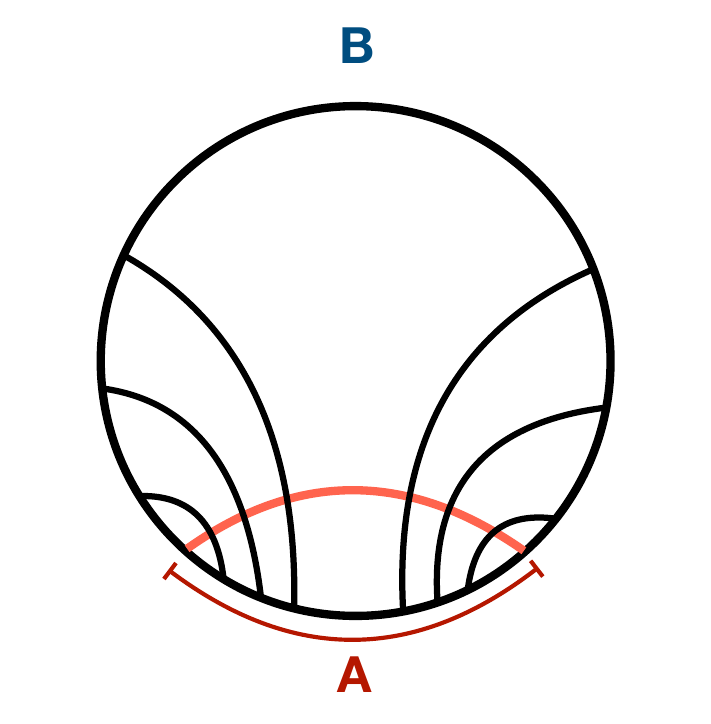}
    \caption{Bit threads (black lines) traverse the minimal surface (red line) with maximal density. They connect $A$ to its complementary region $B$. The number of bit threads reproduces the entanglement entropy obtained via the RT formula.}
    \label{fig:bthreads}
\end{figure}

Following this prescription, it has been shown that the number of bit threads may reproduce the entanglement entropy obtained via the RT formula, see figure \ref{fig:bthreads}. 

Now, consider the case where $2 \ell_A > \ell_\text{bdy}$. 
Due to the homology constraint in the holographic computation of entanglement entropy, the entropies for the complementary subregions $A$ and $B$ satisfy
\eqn{
S_{A} - S_B = \chi.
}
This entropy difference, which, of course, does not apply to the pure state, can be understood in terms of certain excess bit threads attaching to the horizon.

\begin{figure}
  \centering
    \includegraphics[width=0.2\textwidth]{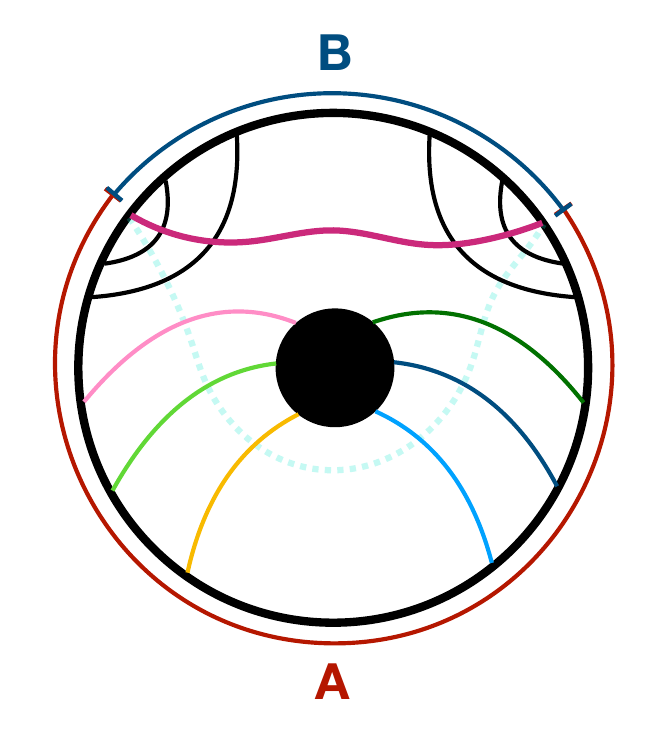}
    \caption{Excess bit threads (colored thin lines) connecting the black hole surface to the boundary subregion $A$.}
    \label{fig:excess}
\end{figure}

Namely, consider the minimal surfaces in the bulk associated with $A$ and $B$ one would get if the global state was maximally mixed, see \fref{fig:excess}. Since the number of bit threads for $A$ would be larger, the excess bit threads will start attaching to the horizon.

If $\ell_A = \ell_\text{crit}$, the horizon would get entirely saturated by Planck size separated bit threads connecting to the boundary subregion $A$. 

Recall that in the latter case, one is able to distinguish microstates on $A$.
Of course, we may write
\eqn{
S_{A} - S_{B} = \alpha S_\text{BH}
}
as well, see also relation \eqref{eq:SPsi-alt}.

In conclusion, by increasing $\ell_A$ there appear more and more bit threads connecting the black hole surface to the boundary subregion $A$.

Referring to what we have discussed above, such excess bit threads may then be seen as the black hole intertwiners from \sref{subsec:bh-int}.
In the AdS vacuum case, a possible connection between holographic bulk intertwiners and bit threads has been proposed in \cite{Casini:2019kex}.

\subsection{Quantum networks}
\label{subsec:qnetworks}

Before we continue with discussing the previous ideas in the context of Hawking evaporation and the information paradox, let us note that intertwining operators also arise in other frameworks describing quantum aspects of black holes. 

A particular setup is based on a spin network construction which can be considered as a graph that consists of a collection of vertices connected by links representing the relation between them. A quantum black hole may be identified as the part of the spin network that is not visible at infinity. It is argued that the black hole seen by a distant observer appears to be a collection of operators that intertwine the links of the network which puncture the horizon and connect the exterior and interior graphs constituting the entire network \cite{Krasnov:2009pd}. 

The dimension of the space associated with such intertwiners, which is related to the Hilbert space of $SU(2)$ Chern--Simons theory on a sphere with punctures, scales with the area of the horizon. It is stated that such degrees of freedom are necessary to make the exterior measurable and, thus, to define a finite black hole entropy as prescribed by the BH formula. In this way, states corresponding to different horizon shapes may be distinguished from the exterior region.\footnote{In this context, based on earlier findings \cite {ashtekar1998quantum,*ashtekar2000quantum}, it has been proposed that states of $SU(2)$ Chern--Simons theory would describe black hole microstates \cite{Engle:2009vc}, see also \cite{smolin1995linking}.}

Similarly, in the present case, the coarse grained entropy of the black hole can be seen as a measure for the amount of operators intertwining the disjoint charged representations. These are associated with microstates and correspond to different bulk quantum states, i.e. horizon shapes.

We have seen that such black hole intertwiners play an important role in boundary reconstructions. Namely, any bulk operator acting on the black hole described by an exponentially large code space can be reconstructed from the boundary subregion, if correlations relying on these intertwiners are encoded in its entanglement wedge. 

It should not be surprising that the BH entropy may be related to the number of certain operators. There is no tension with the usual idea of microstate counting. The number of the relevant intertwiners should be of the same order as the number of microstates. 

On the other hand, it is argued that the coarse grained entropy might be seen as a measure for gravity edge modes operative at the horizon, see e.g. \cite{Takayanagi:2019tvn}.
This may suggest interesting connections between such seemingly different perspectives.

It would be tempting to study these aspects in discrete models of AdS/CFT. A particular realization is provided by tensor networks which are expected to represent a fixed time slice of the bulk AdS space \cite{Swingle:2009bg}. 

Indeed, by counting the number of certain links in the tensor network, it is possible to reproduce the fine grained entropy computed via the RT prescription. From the point of view of spin networks, a connection to the tensor network understanding of AdS/CFT, for instance, has been discussed in \cite{Livine:2017fgq}. 

Referring to the discussion above, we may think of superimposing different  tensor networks describing the disjoint bulk 
spacetimes associated with the pure states. The nodes connecting the 
open graphs, i.e. interior and exterior, along the horizon would then be dressed with 
intertwiners which intertwine the different links associated with the networks. 

It would be interesting to further explore potential connections.

\section{Black hole evaporation}
\label{sec:bh-evap}

In this section, we discuss black hole evaporation in light of the insights obtained in \sref{subsec:bh-int}. 
Relying on a simplified model described in \sref{subsec:1sidedBH}, we argue in \sref{subsec:pagec} that the information transfer from the black hole to its radiation happens via black hole intertwiners.
By computing the von Neumann entropy associated with the radiation system, we derive the Page curve. We show that after the Page time, the fine grained entropy is given by an ensemble of entropies associated with each microstate contributing to the thermal ensemble. This would resolve the state puzzle reviewed in \sref{subsec:holo-BHs}. 
In \sref{subsec:nonloc-smooth}, we discuss certain other related aspects.

\subsection{One sided black hole}
\label{subsec:1sidedBH}

Consider a one sided evaporating black hole symmetrically formed from collapsing matter. 
In order to realize the evaporation process, absorbing boundary conditions may be imposed \cite{Rocha:2008fe,VanRaamsdonk:2013sza}. Reflections at the boundary would otherwise mean that the black hole can reach equilibrium with its radiation such that any change in the black hole size would not be possible. 

In the absorbing case, however, the outgoing modes can be extracted into some auxiliary reservoir $\mathcal{H}_\text{rad}$, simulating a unitary evaporation process. The radiation will not return to the black hole described by some Hilbert space $\mathcal{H}_\text{BH}$. The outgoing modes contained in a spacelike surface anchored at a fixed boundary time will not belong to the spacelike surface at a later time. In other words, the bulk entropy will explicitly depend on the boundary time. 

To tackle the problem at the semiclassical level, the idea has been studying QESs minimizing the generalized entropy. In contrast to the classical RT surface, an initial QES does not remain unchanged at a later boundary time, because of the additional bulk entropy term.
In order to study the information flow during evaporation, the relevant task then becomes the determination of the entanglement wedges for $\mathcal{H}_\text{BH}$ and $\mathcal{H}_\text{rad}$.

It follows that until the Page time almost no information is transferred to the outside via radiation. After the Page time, however, a large part of the interior becomes encoded in the entanglement wedge of the radiation \cite{Penington:2019npb,Almheiri:2019psf}. In particular, after throwing a tiny portion of matter into the black hole, assumed to be happening before the Page time, the fed in information becomes encoded in the (entanglement wedge of the) radiation after one scrambling time \cite{Sekino:2008he}. 

Since the late time radiation is entangled with the interior which, on the other hand, is encoded in the entanglement wedge of the early radiation, this leads to the expected Page transition \cite{Page:1993wv,Page:2013dx}.\footnote{To leading order,
the infalling time (in terms of Eddington--Finkelstein coordinates) of the corresponding QES is one scrambling time before the instance of extracting the Hawking radiation into the auxiliary reservoir \cite{Penington:2019npb}. This is in agreement with the Hayden--Preskill decoding criterion based on toy models simulating the information transfer from evaporating black holes \cite{Hayden:2007cs}.
} 

Before the Page time, the interior should be described by $\mathcal{H}_\text{BH}$. Assume that the microstates are highly entangled with an external purifying system which effectively gives rise to the maximally mixed state, see \sref{subsec:largecode}. In order to reconstruct interior operators from $\mathcal{H}_\text{BH}$, the interior must be encoded in its entanglement wedge. 
It turns out that this is the case if
\eqn{
S_\text{code} < S_\text{BH} - S_\text{rad}.
\label{eq:cond-bPage}
}

Similarly, because of unitarity, after the Page time, the interior should be encoded in the entanglement wedge of $\mathcal{H}_\text{rad}$. 

If that is so, one may perform a state independent reconstruction for the entire code space which is possible if
\eqn{
S_\text{code} < S_\text{rad} - S_\text{BH}.
\label{eq:cond-aPage}
}

The conditions \eqref{eq:cond-bPage} and \eqref{eq:cond-aPage} can be derived by applying the QES prescription to the maximally mixed state, see e.g. \cite{Penington:2019npb}. Following the discussion in \sref{subsec:bdy-mes}, we would simply need to identify the subregions as
\eqn{
\mathcal{H}_{\{A,B\}} \leftrightarrow \mathcal{H}_{\{\text{rad},\text{BH}\}}.
\label{eq:identif}
}

\subsection{Page curve}
\label{subsec:pagec}

Since the evaporation process can be modeled by the construction in \sref{subsec:bdy-mes}, we may deduce the findings summarized above in the context of microstate distinguishability.

Recall that if $\ell_A = \ell_\text{crit}$, it is possible to perfectly distinguish microstates on subregion $A$. This appears to be so due to the fact that the corresponding intertwiner correlations are encoded in the entanglement wedge. An operator acting on the black hole can be reconstructed on $A$. Since these correlations talk to the interior, the relevant degrees of freedom become encoded in $A$.
 
By making the identification in \eqref{eq:identif}, we may then conclude that the information associated with the internal states is transferred to the radiation via black hole intertwiners. Correlations relying on the latter are supported in the atmosphere which may be seen as the near horizon region where the unitarization process for the radiation is operating.

Then, analogous to what we have discussed in \sref{subsec:bh-int}, before Page time, the fine grained entropy of the radiation would be given by
\eqn{
S(\rho_{\text{therm},\text{rad}}) = - \mathrm{Tr} \left[ \rho_{\text{therm},\text{rad}} \log \rho_{\text{therm},\text{rad}} \right],
\label{eq:Srad-beforePage}
}
i.e. the von Neumann entropy of the reduced density matrix derived from the thermal mixed state \eqref{eq:thermal-BH}.

\begin{figure}[h!]
  \centering
    \includegraphics[width=0.38\textwidth]{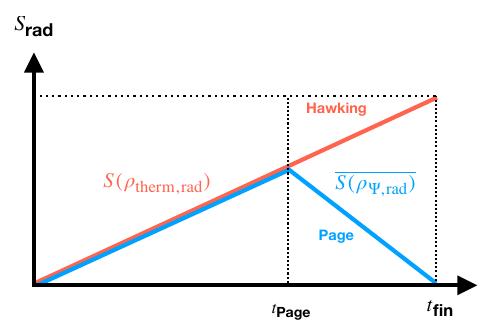}
    \caption{Entropy curve deduced from \eqref{eq:Srad-tot}. Before Page time, the fine grained entropy (blue) coincides with the semiclassical result (orange). The state looks thermal and its von Neumann entropy would increase even after Page time. However, around the Page time, the state of the radiation undergoes a change, since the interior degrees of freedom start to become accessible due to the intertwiner correlations. Such an information encoding gradually purifies the state of the radiation. Consequently, its von Neumann entropy decreases after the Page time. This leads to the expected evolution consistent with unitarity.}
    \label{fig:pagecurve}
\end{figure}

After the Page time, the radiation subsystem starts to access the interior degrees of freedom. By gradually increasing its size, the relevant correlations become encoded such that the global state seen from the perspective of the radiation starts to depart from the thermal mixed state. 
In other words, the number of initially inaccessible sectors gets reduced.

The black hole evaporation process may therefore be viewed as an evolution of the algebraic structure of the Hilbert space.
The state after the Page time would be described by the reduced density matrix derived from \eqref{eq:superstate}, hence
\eqn{
S(\rho_{\Psi,\text{rad}}) = - \mathrm{Tr} \left[ \rho_{\Psi,\text{rad}} \log \rho_{\Psi,\text{rad}} \right],
\label{eq:Srad-afterPage}
}
which leads to the expected Page transition.

So, as in \eqref{eq:S_A-piece}, we get the following fine grained entropy by combining the results \eqref{eq:Srad-beforePage} and \eqref{eq:Srad-afterPage} 
\eqn{
S(\rho_\text{rad}) \simeq 
\begin{cases}
S(\rho_{\text{therm},\text{rad}}) & \text{for}\ t < t_\text{Page}\\\\
\overline{S(\rho_{\Psi,\text{rad}}  )  } & \text{for}\ t > t_\text{Page}
\end{cases}.
\label{eq:Srad-tot}
}
The reduced density matrix before the Page time is obtained under the restriction to the subalgebra $\mathcal{N}$, means, it coincides with the state predicted semiclassically as in Hawking's computations. 

The entropy curve is sketched in \fref{fig:pagecurve}. We should point out that for deriving the Page transition the standard von Neumann entropy is computed. 
This agrees with findings obtained by applying the quantum RT prescription to the thermal state. 
However, recall that in the latter case one runs into an apparent state paradox \cite{Bousso:2019ykv}. The transition can be derived despite the fact that the underlying state is maximally mixed. Of course, this seems to be contradictory, since, precisely as in Hawking's original computation, the latter would suggest information loss. 

In order to resolve the state puzzle, it has been proposed that the generalized entropy takes a different value from what one would usually get via the von Neumann entropy formula \cite{Bousso:2020kmy}. Before the Page time, the entropy would be given by the semiclassical result. After the Page time, it should be given by an ensemble of entropies, each corresponding to a different quantum mechanical theory, see \eqref{eq:Srad-ensemble}. 

How this equality would work out in detail has not been clear so far. The computations above demonstrate how an entropy evolution of such kind may be realized. It suggests that the entropies in the ensemble should be associated with microstates contributing to the thermal ensemble \eqref{eq:thermal-BH}.

\subsection{Nonlocality and smoothness}
\label{subsec:nonloc-smooth}

We have seen that the proposed mechanism resulting in the Page transition also provides a resolution of the state puzzle apparent in semiclassical computations of the generalized entropy.

One might ask whether such a resolution leads to a problem regarding the smoothness of the horizon for an old black hole. The global state is taken to be the superposition of a large number of bulk geometries. However, due to the intrinsic nonlocal nature of the intertwiner correlations, as seen from the semiclassical perspective, any observer falling into the black hole would encounter a smooth horizon. There will be no drama for them. 

Locally, it would not be possible to access the superselection sectors associated with the black hole degrees of freedom. In other words, there is no way to do any local measurement in order to distinguish the different horizon shapes. 

In fact, this may be understood as the basic mechanism which gives rise to the thermal mixed state described by a single classical background, see \sref{subsec:ss-therm}. The presence of a smooth horizon may be ascribed to the effective thermalization under the restriction to the subalgebra $\mathcal{N}$. 

Having identified this property, we may argue that black hole information is topologically protected, i.e. against local perturbations. According to the principle of balanced holography, which states that the information is stored in nonlocal correlations between hidden microscopic degrees of freedom in the interior and exterior, the information storage would be realized similar to the nonlocal protection of topological qubits \cite{Verlinde:2013uja,*Verlinde:2013vja}. 

Indeed, the described transfer mechanism relying on intertwiner correlations would exactly operate in this manner. The virtual local QFT modes in the black hole atmosphere would be captured by the subalgebra $\mathcal{N}$. During the evaporation process, the hidden information encoded in the intertwiner correlations, which, to the contrary, cannot be described by $\mathcal{N}$, would be transferred to the radiation originating from the near horizon region.
We come back to this in \sref{subsec:evapTFD}.

\begin{figure}[b]
  \centering
    \includegraphics[width=0.18\textwidth]{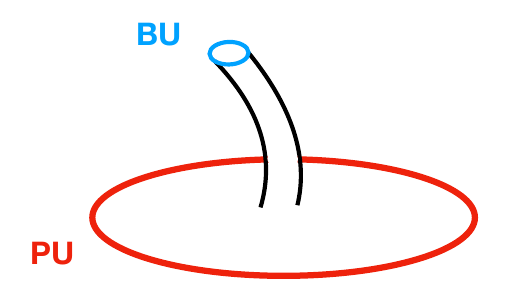}
    \caption{Baby universe (BU) emitted from an initial parent universe (PU). The presence of intertwining operators may be indicating that both sectors are connected via a wormhole.}
    \label{fig:pu-bu}
\end{figure}

Computations of the generalized entropy also suggest that there is no firewall forming at the horizon. For the maximally mixed state the horizon is smooth, even after the Page time when the entanglement wedge of the radiation extends to the interior. Since we precisely arrive at the same conclusion in \sref{subsec:pagec}, without invoking the quantum RT prescription, we may consider this as further evidence supporting the no drama scenario.

Our findings suggest that we necessarily need to take into account the contribution of certain nonlocal correlations extending across the black hole atmosphere. This turns out to be crucial for unitary evaporation and, thus, for resolving the information puzzle. 
Hence, we find that a unitarization mechanism requiring a particular role of the near horizon region would be realized in AdS/CFT.

Let us note that earlier discussions have pointed out the relevance of stringy effects in resolving the information puzzle. Understanding possible connections to the present ideas would be very interesting, particularly, with regard to the failure of the nice slice (i.e. bulk Cauchy surfaces where the low energy degrees of freedom may be decoupled from the high energy one) argument in string theory put forward in \cite{Lowe:1995ac}.

On the other hand, it has been proposed that certain nonlocal observables, so called precursors \cite{Polchinski:1999yd,Freivogel:2002ex}, i.e. infinite sums of local observables, may access the interior degrees of freedom \cite{Hubeny:2002dg}. Such operators might be seen as some type of large Wilson lines \cite{Susskind:1999ey}, although certain questions regarding this interpretation have been raised in \cite{Giddings:2001pt}. 

As detailed out in \sref{subsec:intOps}, extended objects of such kind, which may be expressed in terms of high dimension CFT operators, appear to be crucial for the resolution of the factorization problem in AdS/CFT \cite{Harlow:2015lma,Guica:2015zpf}. Indeed, the black hole intertwiners, thought to be composite as well,
may be viewed as operators giving rise to similar entities. These may be seen as gravitational counterparts that are intrinsic quantum properties of the emergent bulk spacetime. 

In this regard, it may be interesting to further explore and generalize the idea of reconstructing bulk operators from gravitational Wilson lines as, for instance, considered in \cite{Fitzpatrick:2016mtp,Anand:2017dav}.

\section{Wormholes and replicas}
\label{sec:wormholes}

In \sref{subsec:wh-ints}, we elaborate on possible connections between certain wormhole topologies appearing in gravity replica computations of the generalized entropy and the ideas presented in this article, as particularly addressed in \sref{sec:bh-evap}. 
In \sref{subsec:evapTFD}, we discuss how the ideas discussed in this paper might be reflected in the case of an evaporating two sided black hole.

\subsection{Intertwiners and wormholes}
\label{subsec:wh-ints}

Let us begin by reminding that according to what we have seen in \sref{subsec:bh-int}, once the intertwiner correlations become accessible to the boundary subregion, means, if the initial algebra associated with the region gets enlarged, perfect distinguishability between microstates can be achieved. This appears to be happening precisely in the same way as described in \sref{subsec:holevo}. The same follows from applying the quantum RT prescription to the highly mixed state, see \sref{subsec:bdy-mes}.

Considering an evaporating black hole described by a simplified model consistent with the Hayden--Preskill decoding criterion, we have argued that black hole intertwiners may be responsible for the information transfer during the evaporation process. 

Computations of the generalized entropy indicate that the entanglement wedge of the radiation extends to the interior after the Page time. According to the arguments in \sref{subsec:pagec}, this would happen basically due to the fact that quantum correlations across the atmosphere become encoded in the radiation. 

Of course, under the action of any semiclassical observable, one would be ignorant of these nonlocal correlations, since they rely on operators which are not contained in $\mathcal{N}$.
The existence of such operators depends on the age of the black hole, because the number of possible interior states decreases with time.

The described mechanism implies that the density matrix associated with the radiation undergoes a change around the halfway point, thus, leading to the Page transition depicted in \fref{fig:pagecurve}.

The picture of intertwiner correlations threading through wormholes is particularly supported by constructing the TFD like state for a single BTZ microstate. 
Similar to reconstructions of gauge invariant operators inside the two sided eternal black hole, such correlations may be seen as gravitational wormhole threading line operators. Since, in principle, we may interpret the BTZ black hole as a wormhole instead of a one sided black hole, this would indeed resemble the situation described above.
Such a similitude might already point towards a relation between wormholes and black hole intertwiners.
Even more, it might conceivably provide an explanation for the appearance of replica wormholes in gravitational entropy computations \cite{Almheiri:2019qdq,Penington:2019kki}. 

\begin{figure}
  \centering
    \includegraphics[width=0.11\textwidth]{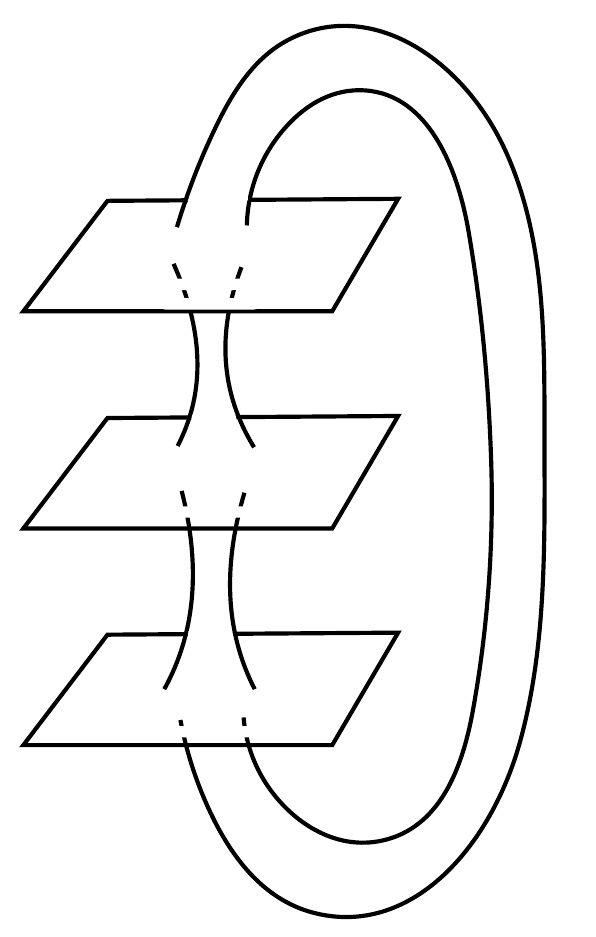}
    \caption{Wormhole connections between nearest neighbour replica manifolds. For simplicity, we only show the case $n=3$ and do not distinguish between bra and ket wormholes.}
    \label{fig:replicas-nearest}
\end{figure}

Let us for a moment return to the discussion in \sref{subsec:bh-int}. 
There, the operators intertwining two different transformations would imply a composition of distinct near horizon geometries associated with the microstates. 
Due to considerations pointed out above, it is suggestive that different sectors will be connected by wormholes. 

Indeed, to some extent, this picture is realized in replica computations in the presence of topology changing baby universe (BU) emissions \cite{Marolf:2020xie,Giddings:2020yes}. 
Namely, from the perspective of an observable parent universe (PU), additional BUs may be viewed as superselection sectors, see \fref{fig:pu-bu}.
Wormhole connections between these and the observable PU would allow retrieving the information. Interestingly, under certain assumptions, which basically align with the replica topologies suggested in \cite{Almheiri:2019qdq,Penington:2019kki}, the entropy of the state describing the PU results in an ensemble average. 

Note that an ensemble of entropies has been obtained in \sref{subsec:pagec}. 
Taking this similarity serious, it might imply that one should take into account wormhole connections that do not necessarily follow from the usual quantum mechanical rules for computing the entropy combined with the rules of wormhole calculus. The latter would only implicate wormhole connections between neighboring bra and ket replica manifolds \cite{Giddings:2020yes}, see sketch in \fref{fig:replicas-nearest}.

Let us be more precise. 

Assume the existence of topology change, i.e. a given initial universe can split by emitting a disconnected BU. In addition, suppose that BUs can be emitted and absorbed by the PU (e.g. asymptotically AdS), but cannot interact with each other resulting in other universes. For simplicity, consider a single type of BU. 

Within this framework, one may study transitions between an initial PU plus a certain number of BUs and a final PU plus some number of BUs. The amplitudes will be summations over different geometries including wormholes. These can be reproduced by defining an appropriate Hamiltonian. It is possible to derive a simple relation between BU states and coupling constants. Consider the following state
\eqn{
|\alpha \rangle = N e^{- \frac{1}{2} (a^\dag - \alpha)^2} | 0 \rangle,
\label{eq:alpha-state}
}
where $N$ denotes some appropriately chosen normalization constant. The so called $\alpha$ state \eqref{eq:alpha-state} diagonalizes the sum of the creation and annihilation operator, i.e. $a^\dag + a$, acting on the BU Fock space with eigenvalue $\alpha$, where $|0\rangle$ denotes its vacuum. It is the eigenvalue of the Hamiltonian
\eqn{
H_\alpha = H + \alpha \int d^3x\ B(x)
}
expressed in terms of the coupling constant $\alpha$ and an operator $B$ describing the effect of the BU emission on the PU fields captured by the Hamiltonian $H$.

\begin{figure}
  \centering
    \includegraphics[width=0.2\textwidth]{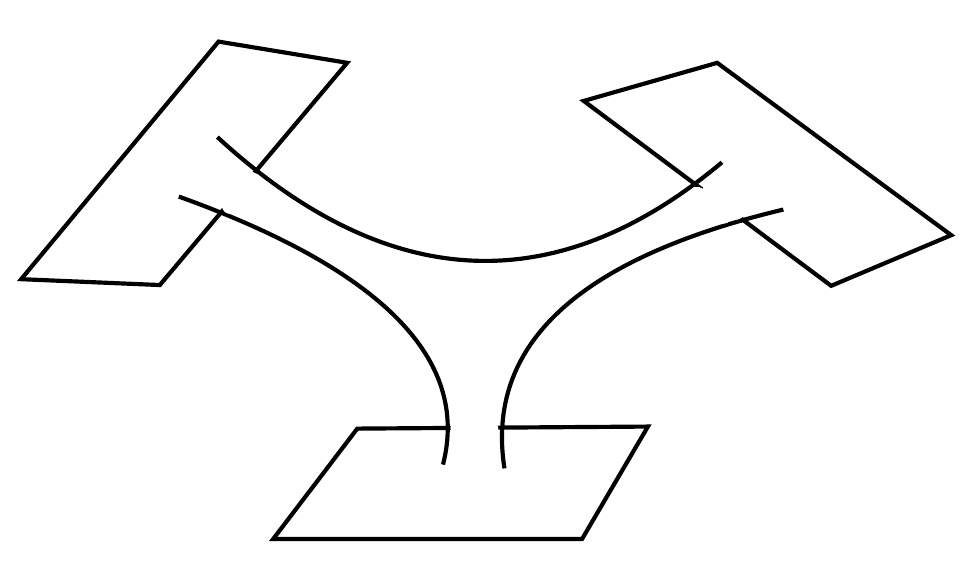}
    \caption{All to all wormhole connections between replica manifolds, again shown for the case $n=3$.}
    \label{fig:replicas-all2all}
\end{figure}

Note that, formally, the operator $B$ might be seen as some support function defining a charged superselection sector localized in a certain region. The representation induced by a charged state on the algebra outside that region would not be distinguishable from the PU state representation. 
This is similar to the effective localization in the bulk in the presence of superselection sectors as discussed before.

Due to simplifying reasons, consider Renyi entropies given in terms of $\mathrm{Tr}[\rho^n]$. Using the state \eqref{eq:alpha-state}, an initial uncorrelated product state of BUs and the PU evolved for some time $t$ can be written as
\eqn{
| \Psi \rangle = \int d\alpha\ \psi(\alpha) U_\alpha | \psi_i \rangle | \alpha \rangle,
\label{eq:B-P-state}
}
where $U_\alpha = \exp\left( -i H_\alpha t \right)$.

Computing the trace associated with the pure state \eqref{eq:B-P-state} living in the full Hilbert space, the standard rules suggest to only consider wormhole connections between neighboring bra and ket replicas which yields the expected purity \cite{Giddings:2020yes}.

On the other hand, if one assumes a sum over all possible wormhole connections as in \cite{Almheiri:2019qdq,Penington:2019kki}, see \fref{fig:replicas-all2all}, one arrives at the following reduced density matrix for the PU
\eqn{
\rho_{\text{PU}} = \mathrm{Tr}_{\text{BU}} \left[ \rho_{\Psi} \right] = \int d\alpha\ |\psi(\alpha)|^2 \rho_{i,\alpha}
\label{eq:rhoPU}
}
with $\rho_{i,\alpha} = U_\alpha \rho_i U_\alpha^\dag$, where $\rho_i =  | \psi_i \rangle \langle \psi_i |$ denotes an initial parent density matrix.
Of course, \eqref{eq:rhoPU} is mixed, thus
$\mathrm{Tr} \left[  \rho_{\text{PU}}^n \right] \neq \left( \mathrm{Tr} \left[  \rho_{\text{PU}} \right] \right)^n$.

However, the interesting point here is the fact that if one assumes all possible wormhole connections between bra and ket replicas as in the case of \eqref{eq:rhoPU}, it follows that
\eqn{
\mathrm{Tr} \left[  \rho_{\text{PU}}^n \right] = \int d\alpha\ |\psi(\alpha)|^2 \mathrm{Tr} \left[  \rho_{i,\alpha}^n \right].
\label{eq:Tr-rhoPu}
}
The expression \eqref{eq:Tr-rhoPu} looks like an ensemble of entropies 
weighted by some probability function. This is quite similar to the findings in \sref{subsec:pagec}, where the radiation entropy after the Page time is expressed as an ensemble of entropies associated with the microstates contributing to the thermal ensemble. 

In particular, viewing the emitted BU as a superselection sector from the perspective of the observable PU, the connecting wormhole might be intertwining the different internal degrees of freedom with the radiation after the Page time.

This might result in a possible interpretation of black hole internal states as BUs emitted from the observable universe.
We would like to emphasize that such a picture might already arise from the fact that the distinguishability measure in \sref{subsec:holevo} can be expressed as the difference between the entropy of the radiation before and after the Page time, see \eqref{eq:S_A-piece}.

Indeed, a similar understanding has been developed in \cite{Balasubramanian:2007qv}, arguing that the thermal density matrix of a single CFT might be viewed as an ensemble of universes, each associated with a microstate, instead of a single eternal black hole. Such findings align with our conclusions in \sref{subsec:ss-therm}, where it is stated that the classical black hole background will be valid so long the observer is restricted to the subalgebra $\mathcal{N}$.

The question we should ask in this context is how such all to all wormhole connections between replica manifolds may be understood in light of the arguments brought above. 

Let us consider the setup in \cite{Almheiri:2019qdq}, where after the Page time it is argued that such (Euclidean) wormholes are responsible for the Page transition. 
Adopting to our findings, we would arrive at the following situation.

At the beginning, means, immediately after the black hole has formed and no radiation has been released yet, the encoding of the information about the interior states would still extend to the atmosphere via hidden (from the perspective of a semiclassical observer) nonlocal correlations relying on black hole intertwiners. Recall that the latter do not belong to $\mathcal{N}$ describing semiclassical bulk physics. These quantum correlations are thought to be intrinsic quantum properties of the emergent bulk spacetime. 

However, once the black hole starts to release its radiation across the near horizon exterior region, it will be correlated with the internal degrees of freedom. Since initially this correlation is minimal, the global state would still look thermal from the perspective of the early Hawking radiation.

At the Page time, the correlation with the interior modes becomes maximal. Since the size of the black hole is almost reduced by a factor of two, there are still hidden intertwiner correlations (also reduced in number) encoded in the atmosphere nonlocally extending to the interior. Due to maximal correlation between the interior degrees of freedom and the radiation, this would be so with the surrounding quantum atmosphere as well. Late time radiation released in that near horizon region would therefore not increase the entropy curve, thus, resulting in the expected Page transition.

Now, according to the picture above, if we replicate the entire manifold for computing the fine grained entropy of the radiation after the Page time, the disjoint near horizon regions defined on all replica sheets should be identified with each other. 

This might require the mentioned all to all wormhole connections running between the replicas and connecting the interior as well as the outer near horizon regions. The radiation system associated with each replica manifold should encounter the same near horizon exterior region, which is indicated by choosing the same coloring in the gravity sector as depicted in \fref{fig:replicas-all2all} (left).
This has to be the case, since it would be highly quantum correlated with the black hole atmosphere in the post Page time regime.

\begin{figure}
  \centering
    \includegraphics[width=0.48\textwidth]{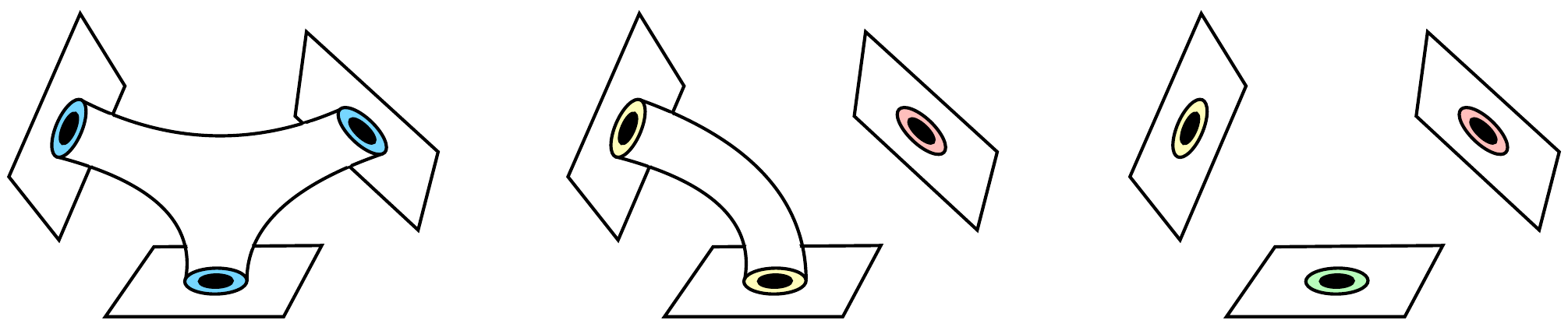}
    \caption{Replica wormholes shown for the case $n=3$ after (left), around (center), and before (right) the Page time. The dark disk including the surrounding annulus corresponds to the region coupled to gravity. 
The black hole atmosphere would be encoded in the latter which is indicated by the coloring.   
    The replica manifolds are prepared in Euclidean time. The outer white region is associated with the external radiation system defined on flat space. For simplicity, we do not show the cyclic gluings along the cuts in the matter region which are not visible here as well.
    The identical coloring indicates that the radiation is maximally quantum correlated with the atmosphere, and therefore with the interior.}
    \label{fig:replicasWHs}
\end{figure}
 
On the other hand, precisely due to the same reasoning, before Page time, this would not need to be so. Namely, from the point of view of the radiation, the global state still looks maximally mixed at that stage. This would mean that by replicating the exterior near horizon region, there is no need to identify them by wormhole connections, see \fref{fig:replicas-all2all} (right) which would correspond to Hawking's semiclassical computation. 

At this stage, almost no information of the interior can be deciphered from the radiation.
Of course, referring to the intertwiner arguments, around the Page time, one would expect an intermediate state where partially connected replica manifolds may contribute to the fine grained entropy, see \fref{fig:replicas-all2all} (center).

We would like to close the present discussion by mentioning that earlier studies based on simplified toy models have argued that in order to predict unitarity in black hole evaporation, one should take into account certain 
effects that, viewed from the semiclassical point of view, somehow nonlocally transfer information from the internal state to the radiation.

By now, we hope to have demonstrated how such a unitarization mechanism may explicitly be realized in AdS/CFT. The relevant correlations for the transfer process turn out to be extended across the black hole atmosphere. Importantly, this picture, where we only compute the standard von Neumann entropy, leads to the same results deduced semiclassically by employing the QES prescription. 

In fact, the possible connection between wormholes and intertwiners suggests that topology change may be related to such near horizon quantum gravitational effects. 

Our findings indicate that contributing wormholes should be comparable to the size of the black hole which is in line with the suggestions relying on the computations in \cite{Almheiri:2019qdq,Penington:2019kki}.

\subsection{Evaporating two sided black hole}
\label{subsec:evapTFD}

It will be interesting to figure out how the ideas discussed in the previous sections might apply to an evaporating two sided black hole described by \eqref{eq:TFD}. 

This process has recently been studied by embedding the latter inside an ambient space \cite{Verlinde:2020upt}, arguing that after the Page time, the density matrix of the black hole is given by a mixed state, the so called thermomixed double (TMD)
\eqn{
\rho_\text{TMD} = \sum_i p_i | i \rangle_L \langle i | \otimes | i \rangle_R \langle i |.
\label{eq:TMD}
}
The geometric representation of the
gravitational saddle point associated with \eqref{eq:TMD} is proposed to be containing the island due to entanglement with the environment. Indeed, it resembles the density matrix for the two sided black hole in JT gravity.

Consider some entangled state of the black hole plus environment which can be written as
\eqn{
| \Psi \rangle = \sum_{i} \sqrt{p_i} |i\rangle_L |i\rangle_R | \phi_{i} \rangle_\varepsilon,
}
namely, in terms of the two CFT states and the state $|\phi_{i}\rangle$ associated with the environment (radiation), $\varepsilon$.

Assuming the CFTs decohere independently, say into $\varepsilon_L$ and $\varepsilon_R$, each side would be in a thermal mixed state, \eqref{eq:thermFromTFD}, after the Page time. The state of both CFTs would be the factorized thermal state $\rho_L \otimes \rho_R$, thus, no mutual information would be shared between both sides, i.e.
\eqn{
I_{\rho_L \otimes \rho_R} (L,R) = 0.
}

Alternatively, each CFT could again be in a thermal mixed state \eqref{eq:thermFromTFD}, but 
share its entanglement in a balanced way, means half of it with the radiation $\varepsilon$ (not decoupled into $\varepsilon_{L,R}$) and the other half with the other CFT. In such a case, the ensemble average of the matrix element associated with the radiation state would satisfy 
$\overline{ \langle \phi_{i} | \phi_{j} \rangle  } = \delta_{i j}$,
and the density matrix for both CFTs would be described by \eqref{eq:TMD}. Here, we should in particular note that
\eqn{
I_\mathrm{TFD} (L,R) - I_\mathrm{TMD} (L,R) = S_\text{BH}.
\label{eq:I_TFD_TMD}
}
It has been argued that such a balanced decoherence may be realized due to topological protection of black hole information \cite{Verlinde:2013uja,*Verlinde:2013vja}. In fact, as we have discussed in \sref{subsec:nonloc-smooth}, the information transfer via black hole intertwiners may be seen as an explicit realization of such a protection.

It is worth emphasizing that according to the findings in \sref{subsec:pagec}, there would exist superselection sectors from the perspective of the state describing the black hole after the Page time, i.e. $\mathcal{H}_\text{BH}$. This simply follows from the arguments in \sref{subsec:bh-int}, and is indeed reflected in the expression \eqref{eq:I_TFD_TMD}. The presence of superselection sectors would precisely lead to such a difference \cite{Casini:2019kex}, see discussion around \eqref{eq:F_N-diff-TFD}. We may therefore write
\eqn{
\rho_\mathrm{TFD}  \circ E =  \rho_\mathrm{TMD},
}
where the TMD would be generated by some conditional expectation $E$ from the TFD.

\section{Final remarks}
\label{sec:fin}

We would like to conclude this paper with a few final remarks.\\

\textbf{\small Unitarity and superselections ---}
We have seen that eigenstate thermalization, being a characteristic feature of quantum gravity, is intimately connected to the existence of a large number of invisible sectors encountered at the semiclassical level. The presence of these superselection sectors turns out to be important in the context of operator reconstructions. 

More precisely, their existence proves to be crucial for understanding unitary evaporation of black holes and has therefore direct implications for the microscopic resolution of the black hole information paradox.

The importance of superselection sectors in the context of black hole unitarity has also been emphasized in earlier studies. For instance, it has been argued that coherent particle scattering from (extremal) black holes would require that the associated ground states have superselection sectors \cite{Giddings:1992kn}. This would particularly be needed for having a finite black hole entropy as predicted by the BH formula.\\

\textbf{\small Topological protection ---}
The proposed transfer mechanism relying on black hole intertwiners may be seen as an explicit realization of topological protection of black hole information \cite{Verlinde:2013uja,*Verlinde:2013vja}.

Due to the very same reasoning, it also suggests a certain type of quantum secret sharing scheme in the Hawking radiation, see e.g. \cite{Balasubramanian:2020hfs}. At late times, reconstructing operators located in the black hole interior would be impossible by only accessing a certain subsystem of the radiation. 

According to what we have seen, in such a case, one would miss a substantial part of the intertwiner correlations. However, radiation released across the quantum atmosphere may only be maximally correlated with the internal degrees of freedom, as long as the entire system is taken into account.\\

\textbf{\small Nontrivial corrections ---}
On may ask how intertwiners would affect subleading corrections to the entanglement entropy of a particular subsystem. It has been found that there occur nontrivial corrections in bipartitions where two different RT surfaces may compete with each other resulting in a phase transition \cite{Marolf:2020vsi,Dong:2020iod}.

Understanding potential connections would be interesting. In fact, contributions from black hole intertwiners would indeed go beyond usual bulk quantum field effects described by $\mathcal{N}$. 

On the other hand, nontrivial corrections of such kind also arise in gravity computations of the generalized entropy, which go back to replica manifolds partially connected via wormholes \cite{Penington:2019kki}. 

It is worth emphasizing that according to the arguments raised in \sref{subsec:wh-ints}, we may expect contributions of such type around the Page time due to partial encoding of the intertwiner correlations, cf. \fref{fig:replicasWHs}. 

In fact, having a nontrivial center for the algebra associated with a given subsystem, as it has been the case in the present considerations, may lead to a substantial entropy decrease around the transition point \cite{Bianchi:2019stn}.\\

\textbf{\small Nonperturbative effects ---}
In the presence of gauge symmetries, the underlying Hilbert space decomposes into orthogonal subsectors. This is surely so for gravity due to diffeomorphism invariance. 

The space decomposition in \eqref{eq:Hcode-sum} may be derived from certain discrete formulations relying on intertwiners that are attached to vertices in the corresponding quantum network. The same role will be played by the proposed black hole intertwiners behaving as certain type of edge operators. 

As we have discussed, the latter may give rise to correlations which are nonlocally extended (from the perspective of semiclassical observers) across the black hole exterior and the interior, thus resulting in the notion of a quantum atmosphere outside the horizon. 

The proposed mechanism for information transfer to the radiation relies on the existence of the latter. This is what turns out to be the case in AdS/CFT. 

For the two sided eternal black hole, such correlations would range between the two boundaries which are threading through the wormhole in the bulk. These may be viewed as gravitational Wilson lines whose existence would indicate the connectedness through the bulk. 

Considering the two sided case, where the bulk is described by JT gravity, there precisely exist such nonlocal objects which turn out to be factorizable in the boundary theory \cite{Harlow:2018tqv}. 

For the two sided eternal black hole similar operators appear in the presence of $U(1)$ gauge symmetry \cite{Harlow:2015lma,Guica:2015zpf}, again factorizable in terms of high dimension CFT operators. Interestingly, such Wilson lines may be directly related to the mirror operators in the Papadodimas--Raju proposal for reconstructing interior operators \cite{Guica:2015zpf}.

Relying on the findings in this paper and the observations described above, this may explain why there is no information as well as firewall problem in the two sided eternal black hole. 

The entanglement wedge of one side always includes the region behind the horizon. This would be so due to the presence of gauge invariant operators connecting both sides of the bifurcation surface that can eventually be composed in the boundary field theory in terms of high dimension operators. This may indicate the emergence of such gravity operators. 

Since one may construct a TFD like state for a single microstate, such nonlocal correlations would also be present for the pure one sided black hole. Recall that the thermal mixed state can be obtained by integrating out one side of the TFD. We may say that doing so would cut through the intertwiner correlations.

In the one sided case, after the Page time, the exterior will be connected to the interior via black hole intertwiners through the wormhole. Accordingly, the quantum atmosphere of the old black hole would be maximally correlated with the radiation.

It is particularly worth mentioning that such nonlocal correlations ranging across the atmosphere to the interior would indicate the underlying nonperturbative effects. 

Of course, this is expected for resolving the black hole information paradox. In the CFT, the exact conformal blocks (i.e. the universal building blocks for correlation functions) going beyond the usual leading order contribution in the large central charge limit can be constructed in terms of gravitational Wilson lines \cite{Fitzpatrick:2016mtp}. 

Computations of correlators in the semiclassical limit, however, already indicate information loss in the boundary field theory \cite{Fitzpatrick:2016ive}. This is resolved by nonperturbative effects at finite central charge which may precisely be realized by black hole intertwiners leading to nonlocal correlations across the horizon.\\

\textbf{\small Ensemble of entropies ---}
The ensemble \eqref{eq:Srad-tot} follows from purely algebraic arguments without resorting to any averaging over different quantum mechanical theories. The CFT state seen under certain operator restrictions has superselection sectors. Transitions between the disjoint representations are not allowed at the semiclassical level.

One might ask whether or not the described mechanism for information transfer explicitly discussed for the simplified model in \sref{sec:bh-evap} may apply to real black hole evaporation. Of course, the BTZ black hole is stable and does not evaporate. However, as we have mentioned, by imposing absorbing boundary conditions, this may be realized. 

In the simplified setup, apparently this condition has not been imposed. The idea was resorting to entanglement wedge reconstruction. Namely, depending on which entanglement wedge contains the static black hole, the relevant degrees of freedom may be attributed to one of the two subsystems. In fact, after making the identification \eqref{eq:identif}, the conditions \eqref{eq:cond-bPage} and \eqref{eq:cond-aPage} can be derived by using the quantum RT prescription, which is consistent with unitary black hole evaporation.

We shall say that the same conclusions also apply when absorbing boundary conditions are imposed. For simplicity, let us consider a pure BTZ state. Take a second state, but assume it to be describing some external bath. Then, we let the boundaries of 
the black hole state be absorbing, where we extract the outgoing radiation modes into the bath. 

According to semiclassical computations of the fine grained radiation entropy, the quantum RT surface for the latter would initially be empty. However, due to the minimization condition, after the Page time the minimal surface would be nonempty and become the horizon of the shrinking black hole with the interior belonging to the entanglement wedge of the radiation.

Of course, this is not the whole story. We need to go beyond the situation describing a single microstate.

We can do so, and consider instead a superposition of microstates as in \eqref{eq:superstate}. The transfer mechanism discussed in \sref{subsec:pagec} would then also apply to the absorbing boundary case above. After the Page time, the quantum atmosphere surrounding the old black hole, but as well as the interior, would belong to the radiation. 

This is expected to be so, since the latter becomes maximally correlated with the atmosphere. At that stage, immediately after the Page transition, there still exist intertwiners associated with the microstates whose correlations range from the atmosphere into the interior. This may explain why, semiclassically, the entanglement wedge of the radiation extends to the interior of the black hole (i.e. island) which is obtained by applying the quantum RT prescription. 

Importantly, the mechanism above would particularly imply that such a purifying bulk region will not only be restricted to the interior, but extends to the exterior near horizon region as well. This expectation which does not rely on the QES prescription may provide an explanation for the existence of such a near horizon region appearing in computations of the generalized entropy. 

\section{Acknowledgements}

I would like to thank Jibril Ben Achour, Juan Maldacena and Tadashi Takayanagi for valuable comments and correspondence.
I thank Yuya Kusuki, Kotaro Tamaoka, Tomonori Ugajin and Zixia Wei for related discussions and Tadashi Takayanagi for the kind support.
I gratefully acknowledge the support from the Japan Society for the Promotion of Science (JSPS) and the Alexander von Humboldt (AvH) foundation. 
I am supported by a JSPS Grant-in-Aid for Scientific Research under the Grant No.~{19F19813}.

\section{Appendix}

\subsection{General canonical principle}
\label{subsec:equi}

In the present part, we recap the central arguments underlying the general canonical principle mentioned in \sref{subsec:statedep}.

Consider a large, isolated quantum system $S$ described by its Hilbert space $\mathcal{H}_S$. Let $S$ be decomposed into two parts $A$ and $B$, with the assumption that
\eqn{
d_A \ll d_B,
}
where $d_{A,B} = |\mathcal{H}_{A,B}|$. 

Furthermore, suppose that the state of the system $S$ satisfies some global constraint $C$, which can be realized by restricting the states of $A$ and $B$ to some subspace $\mathcal{H}_C$ with dimension $d_C$, i.e.
\eqn{
\mathcal{H}_S = \mathcal{H}_A \otimes \mathcal{H}_B \supseteq \mathcal{H}_C.
}
Next, define the following maximally mixed state
\eqn{
\mathcal{E}_C = \frac{\mathbf{1}_C}{d_C},
}
where $\mathbf{1}_C$ denotes the identity projection on $\mathcal{H}_C$. Since each pure state in $\mathcal{E}_C$ is equally probable, $\mathcal{E}_C$ is called the equiprobable state of $S$ corresponding to the global constraint $C$. Let further be $\Omega_A$ the canonical state of $A$ under the restriction $C$. It corresponds to the quantum state of $A$ when $S$ is in $\mathcal{E}_C$ and is therefore given by tracing out $B$ in the equiprobable state of $S$, i.e.
\eqn{
\Omega_A = \mathrm{Tr}_B (\mathcal{E}_C).
}
Under these assumptions, given a sufficiently small subsystem $A$ of some large system $S$, it turns out that the distance between the quantum and canonical states of $A$ is such that \cite{Popescu_2006} 
\eqn{
\rho_A = \mathrm{Tr}_B (\rho_S) \approx \Omega_A,
}
where $\rho_S$ is taken to be a pure state of $S$.
In other words, from the perspective of the sufficiently small subsystem $A$, any pure state of the entire quantum system $S$ cannot be distinguished from the mixed state $\mathcal{E}_C$.

\subsection{Hilbert space decomposition}
\label{subsec:sdecomp}

In general, a given Hilbert space $\mathcal{H}$ may decompose into orthogonal subspaces if a certain subalgebra has a nontrivial center, which results in a direct sum of tensor products \cite{Casini:2013rba}.

Let us consider the factorized Hilbert space
\eqn{
\mathcal{H} = \mathcal{H}_\text{coarse} \otimes \mathcal{H}_\text{fine}.
}
Suppose that the two subsystems are noninteracting, however, they are quantum correlated. 

The system's total Hamiltonian shall be given by
\eqn{
H = H_\text{coarse} + H_\text{fine}.
}
The energy eigenvalues are $E_{\text{coarse}_i} + E_{\text{fine}_j}$.

The intersection of the corresponding subalgebras, i.e. the center of the one associated with the coarse grained sector, becomes nontrivial on the energy eigenspace $\mathcal{H}(E) \subset \mathcal{H}$, namely $H_\text{coarse}$. 

In such a case, the Hilbert space acquires the following structure
\eqn{
\mathcal{H}(E) = \bigoplus_{i,j} \left( \mathcal{H}_\text{coarse} \left( E_{\text{coarse}_i}) \otimes \mathcal{H}_\text{fine} (E_{\text{fine}_j}  \right) \right),
}
where $E_{\text{fine}_j} = E - E_{\text{coarse}_i}$. 

As discussed in the main text, a decomposition of such type applies to a large subspace of microstates, having an energy eigenspace that corresponds to the code space, where
\eqn{
\mathcal{H}_{\text{fine},i} , \mathcal{H}_{\text{coarse},i} \rightarrow \mathcal{H}_{\text{BH},i} , \mathcal{H}_{\text{ext},i}
}
would lead to the space decomposition in \eqref{eq:Hcode-sum}.

Under the action of a sufficiently small subalgebra one effectively obtains the factorized Hilbert space 
\eqn{
\mathcal{H} = \mathcal{H}_\text{BH} \otimes \mathcal{H}_\text{ext},
}
see footnote \ref{fnote:11}.

Accordingly, in order to describe the superposition of a large number of microstates by a single bulk geometry, the existence of certain (intertwining) operators in the full algebra turns out to be indispensable. 

Even though there is, strictly speaking, a nontrivial center for the algebra associated with a small boundary subregion, one effectively arrives at a direct product of a coarse grained and a fine grained sector. 

On the other hand, given a single microstate, one may also consider a factorization of the full boundary Hilbert space,
\eqn{
\mathcal{H} = \mathcal{H}_A \otimes \mathcal{H}_{\bar A},
} 
where $A$ and $\bar A$ denote the complementary subregions. In general, for any superposition of different microstates, which would give rise to
\eqn{
\mathcal{H} = \bigoplus_i \left( \mathcal{H}_{A_i} \otimes \mathcal{H}_{\bar A_i}   \right),
}
there would be different RT surfaces dividing the bulk space into the regions $\Sigma_{A_i}$ and $\Sigma_{\bar A_i}$. 

Any subalgebra would then have a nontrivial center with elements corresponding to area operators associated with the minimal surfaces. 

As already commented on in footnote \ref{footnote:cpsace}, opposed to standard subsystem QEC, this situation may be addressed within the more general framework of operator algebra QEC \cite{beny2007generalization}. 

In the case of exponentially large code spaces, for instance, an approximate version of the latter has been discussed in \cite{Hayden:2018khn}.

\bibliography{biblio}

\begin{thebibliography}{135}%
\makeatletter
\providecommand \@ifxundefined [1]{%
 \@ifx{#1\undefined}
}%
\providecommand \@ifnum [1]{%
 \ifnum #1\expandafter \@firstoftwo
 \else \expandafter \@secondoftwo
 \fi
}%
\providecommand \@ifx [1]{%
 \ifx #1\expandafter \@firstoftwo
 \else \expandafter \@secondoftwo
 \fi
}%
\providecommand \natexlab [1]{#1}%
\providecommand \enquote  [1]{``#1''}%
\providecommand \bibnamefont  [1]{#1}%
\providecommand \bibfnamefont [1]{#1}%
\providecommand \citenamefont [1]{#1}%
\providecommand \href@noop [0]{\@secondoftwo}%
\providecommand \href [0]{\begingroup \@sanitize@url \@href}%
\providecommand \@href[1]{\@@startlink{#1}\@@href}%
\providecommand \@@href[1]{\endgroup#1\@@endlink}%
\providecommand \@sanitize@url [0]{\catcode `\\12\catcode `\$12\catcode
  `\&12\catcode `\#12\catcode `\^12\catcode `\_12\catcode `\%12\relax}%
\providecommand \@@startlink[1]{}%
\providecommand \@@endlink[0]{}%
\providecommand \url  [0]{\begingroup\@sanitize@url \@url }%
\providecommand \@url [1]{\endgroup\@href {#1}{\urlprefix }}%
\providecommand \urlprefix  [0]{URL }%
\providecommand \Eprint [0]{\href }%
\providecommand \doibase [0]{http://dx.doi.org/}%
\providecommand \selectlanguage [0]{\@gobble}%
\providecommand \bibinfo  [0]{\@secondoftwo}%
\providecommand \bibfield  [0]{\@secondoftwo}%
\providecommand \translation [1]{[#1]}%
\providecommand \BibitemOpen [0]{}%
\providecommand \bibitemStop [0]{}%
\providecommand \bibitemNoStop [0]{.\EOS\space}%
\providecommand \EOS [0]{\spacefactor3000\relax}%
\providecommand \BibitemShut  [1]{\csname bibitem#1\endcsname}%
\let\auto@bib@innerbib\@empty
\bibitem [{\citenamefont {Maldacena}(1999)}]{Maldacena:1997re}%
  \BibitemOpen
  \bibfield  {author} {\bibinfo {author} {\bibfnamefont {J.~M.}\ \bibnamefont
  {Maldacena}},\ }\href {\doibase 10.1023/A:1026654312961,
  10.4310/ATMP.1998.v2.n2.a1} {\bibfield  {journal} {\bibinfo  {journal} {Int.
  J. Theor. Phys.}\ }\textbf {\bibinfo {volume} {38}},\ \bibinfo {pages} {1113}
  (\bibinfo {year} {1999})},\ \bibinfo {note} {[Adv. Theor. Math.
  Phys.2,231(1998)]},\ \Eprint {http://arxiv.org/abs/hep-th/9711200}
  {arXiv:hep-th/9711200 [hep-th]} \BibitemShut {NoStop}%
\bibitem [{\citenamefont {Gubser}\ \emph {et~al.}(1998)\citenamefont {Gubser},
  \citenamefont {Klebanov},\ and\ \citenamefont {Polyakov}}]{Gubser:1998bc}%
  \BibitemOpen
  \bibfield  {author} {\bibinfo {author} {\bibfnamefont {S.~S.}\ \bibnamefont
  {Gubser}}, \bibinfo {author} {\bibfnamefont {I.~R.}\ \bibnamefont
  {Klebanov}}, \ and\ \bibinfo {author} {\bibfnamefont {A.~M.}\ \bibnamefont
  {Polyakov}},\ }\href {\doibase 10.1016/S0370-2693(98)00377-3} {\bibfield
  {journal} {\bibinfo  {journal} {Phys. Lett.}\ }\textbf {\bibinfo {volume}
  {B428}},\ \bibinfo {pages} {105} (\bibinfo {year} {1998})},\ \Eprint
  {http://arxiv.org/abs/hep-th/9802109} {arXiv:hep-th/9802109 [hep-th]}
  \BibitemShut {NoStop}%
\bibitem [{\citenamefont {Witten}(1998)}]{Witten:1998qj}%
  \BibitemOpen
  \bibfield  {author} {\bibinfo {author} {\bibfnamefont {E.}~\bibnamefont
  {Witten}},\ }\href {\doibase 10.4310/ATMP.1998.v2.n2.a2} {\bibfield
  {journal} {\bibinfo  {journal} {Adv. Theor. Math. Phys.}\ }\textbf {\bibinfo
  {volume} {2}},\ \bibinfo {pages} {253} (\bibinfo {year} {1998})},\ \Eprint
  {http://arxiv.org/abs/hep-th/9802150} {arXiv:hep-th/9802150 [hep-th]}
  \BibitemShut {NoStop}%
\bibitem [{\citenamefont {'t~Hooft}(1993)}]{tHooft:1993dmi}%
  \BibitemOpen
  \bibfield  {author} {\bibinfo {author} {\bibfnamefont {G.}~\bibnamefont
  {'t~Hooft}},\ }\href@noop {} {\bibfield  {journal} {\bibinfo  {journal}
  {Conf. Proc. C}\ }\textbf {\bibinfo {volume} {930308}},\ \bibinfo {pages}
  {284} (\bibinfo {year} {1993})},\ \Eprint
  {http://arxiv.org/abs/gr-qc/9310026} {arXiv:gr-qc/9310026} \BibitemShut
  {NoStop}%
\bibitem [{\citenamefont {Susskind}(1995)}]{Susskind:1994vu}%
  \BibitemOpen
  \bibfield  {author} {\bibinfo {author} {\bibfnamefont {L.}~\bibnamefont
  {Susskind}},\ }\href {\doibase 10.1063/1.531249} {\bibfield  {journal}
  {\bibinfo  {journal} {J. Math. Phys.}\ }\textbf {\bibinfo {volume} {36}},\
  \bibinfo {pages} {6377} (\bibinfo {year} {1995})},\ \Eprint
  {http://arxiv.org/abs/hep-th/9409089} {arXiv:hep-th/9409089} \BibitemShut
  {NoStop}%
\bibitem [{\citenamefont {Ryu}\ and\ \citenamefont
  {Takayanagi}(2006{\natexlab{a}})}]{Ryu:2006bv}%
  \BibitemOpen
  \bibfield  {author} {\bibinfo {author} {\bibfnamefont {S.}~\bibnamefont
  {Ryu}}\ and\ \bibinfo {author} {\bibfnamefont {T.}~\bibnamefont
  {Takayanagi}},\ }\href {\doibase 10.1103/PhysRevLett.96.181602} {\bibfield
  {journal} {\bibinfo  {journal} {Phys. Rev. Lett.}\ }\textbf {\bibinfo
  {volume} {96}},\ \bibinfo {pages} {181602} (\bibinfo {year}
  {2006}{\natexlab{a}})},\ \Eprint {http://arxiv.org/abs/hep-th/0603001}
  {arXiv:hep-th/0603001} \BibitemShut {NoStop}%
\bibitem [{\citenamefont {Ryu}\ and\ \citenamefont
  {Takayanagi}(2006{\natexlab{b}})}]{Ryu:2006ef}%
  \BibitemOpen
  \bibfield  {author} {\bibinfo {author} {\bibfnamefont {S.}~\bibnamefont
  {Ryu}}\ and\ \bibinfo {author} {\bibfnamefont {T.}~\bibnamefont
  {Takayanagi}},\ }\href {\doibase 10.1088/1126-6708/2006/08/045} {\bibfield
  {journal} {\bibinfo  {journal} {JHEP}\ }\textbf {\bibinfo {volume} {08}},\
  \bibinfo {pages} {045} (\bibinfo {year} {2006}{\natexlab{b}})},\ \Eprint
  {http://arxiv.org/abs/hep-th/0605073} {arXiv:hep-th/0605073 [hep-th]}
  \BibitemShut {NoStop}%
\bibitem [{\citenamefont {Hubeny}\ \emph {et~al.}(2007)\citenamefont {Hubeny},
  \citenamefont {Rangamani},\ and\ \citenamefont {Takayanagi}}]{Hubeny:2007xt}%
  \BibitemOpen
  \bibfield  {author} {\bibinfo {author} {\bibfnamefont {V.~E.}\ \bibnamefont
  {Hubeny}}, \bibinfo {author} {\bibfnamefont {M.}~\bibnamefont {Rangamani}}, \
  and\ \bibinfo {author} {\bibfnamefont {T.}~\bibnamefont {Takayanagi}},\
  }\href {\doibase 10.1088/1126-6708/2007/07/062} {\bibfield  {journal}
  {\bibinfo  {journal} {JHEP}\ }\textbf {\bibinfo {volume} {07}},\ \bibinfo
  {pages} {062} (\bibinfo {year} {2007})},\ \Eprint
  {http://arxiv.org/abs/0705.0016} {arXiv:0705.0016 [hep-th]} \BibitemShut
  {NoStop}%
\bibitem [{\citenamefont {Lewkowycz}\ and\ \citenamefont
  {Maldacena}(2013)}]{Lewkowycz:2013nqa}%
  \BibitemOpen
  \bibfield  {author} {\bibinfo {author} {\bibfnamefont {A.}~\bibnamefont
  {Lewkowycz}}\ and\ \bibinfo {author} {\bibfnamefont {J.}~\bibnamefont
  {Maldacena}},\ }\href {\doibase 10.1007/JHEP08(2013)090} {\bibfield
  {journal} {\bibinfo  {journal} {JHEP}\ }\textbf {\bibinfo {volume} {08}},\
  \bibinfo {pages} {090} (\bibinfo {year} {2013})},\ \Eprint
  {http://arxiv.org/abs/1304.4926} {arXiv:1304.4926 [hep-th]} \BibitemShut
  {NoStop}%
\bibitem [{\citenamefont {Faulkner}\ \emph {et~al.}(2013)\citenamefont
  {Faulkner}, \citenamefont {Lewkowycz},\ and\ \citenamefont
  {Maldacena}}]{Faulkner:2013ana}%
  \BibitemOpen
  \bibfield  {author} {\bibinfo {author} {\bibfnamefont {T.}~\bibnamefont
  {Faulkner}}, \bibinfo {author} {\bibfnamefont {A.}~\bibnamefont {Lewkowycz}},
  \ and\ \bibinfo {author} {\bibfnamefont {J.}~\bibnamefont {Maldacena}},\
  }\href {\doibase 10.1007/JHEP11(2013)074} {\bibfield  {journal} {\bibinfo
  {journal} {JHEP}\ }\textbf {\bibinfo {volume} {11}},\ \bibinfo {pages} {074}
  (\bibinfo {year} {2013})},\ \Eprint {http://arxiv.org/abs/1307.2892}
  {arXiv:1307.2892 [hep-th]} \BibitemShut {NoStop}%
\bibitem [{\citenamefont {Engelhardt}\ and\ \citenamefont
  {Wall}(2015)}]{Engelhardt:2014gca}%
  \BibitemOpen
  \bibfield  {author} {\bibinfo {author} {\bibfnamefont {N.}~\bibnamefont
  {Engelhardt}}\ and\ \bibinfo {author} {\bibfnamefont {A.~C.}\ \bibnamefont
  {Wall}},\ }\href {\doibase 10.1007/JHEP01(2015)073} {\bibfield  {journal}
  {\bibinfo  {journal} {JHEP}\ }\textbf {\bibinfo {volume} {01}},\ \bibinfo
  {pages} {073} (\bibinfo {year} {2015})},\ \Eprint
  {http://arxiv.org/abs/1408.3203} {arXiv:1408.3203 [hep-th]} \BibitemShut
  {NoStop}%
\bibitem [{\citenamefont {Almheiri}\ \emph {et~al.}(2015)\citenamefont
  {Almheiri}, \citenamefont {Dong},\ and\ \citenamefont
  {Harlow}}]{Almheiri:2014lwa}%
  \BibitemOpen
  \bibfield  {author} {\bibinfo {author} {\bibfnamefont {A.}~\bibnamefont
  {Almheiri}}, \bibinfo {author} {\bibfnamefont {X.}~\bibnamefont {Dong}}, \
  and\ \bibinfo {author} {\bibfnamefont {D.}~\bibnamefont {Harlow}},\ }\href
  {\doibase 10.1007/JHEP04(2015)163} {\bibfield  {journal} {\bibinfo  {journal}
  {JHEP}\ }\textbf {\bibinfo {volume} {04}},\ \bibinfo {pages} {163} (\bibinfo
  {year} {2015})},\ \Eprint {http://arxiv.org/abs/1411.7041} {arXiv:1411.7041
  [hep-th]} \BibitemShut {NoStop}%
\bibitem [{\citenamefont {Harlow}(2017)}]{Harlow:2016vwg}%
  \BibitemOpen
  \bibfield  {author} {\bibinfo {author} {\bibfnamefont {D.}~\bibnamefont
  {Harlow}},\ }\href {\doibase 10.1007/s00220-017-2904-z} {\bibfield  {journal}
  {\bibinfo  {journal} {Commun. Math. Phys.}\ }\textbf {\bibinfo {volume}
  {354}},\ \bibinfo {pages} {865} (\bibinfo {year} {2017})},\ \Eprint
  {http://arxiv.org/abs/1607.03901} {arXiv:1607.03901 [hep-th]} \BibitemShut
  {NoStop}%
\bibitem [{\citenamefont {Hayden}\ and\ \citenamefont
  {Penington}(2019)}]{Hayden:2018khn}%
  \BibitemOpen
  \bibfield  {author} {\bibinfo {author} {\bibfnamefont {P.}~\bibnamefont
  {Hayden}}\ and\ \bibinfo {author} {\bibfnamefont {G.}~\bibnamefont
  {Penington}},\ }\href {\doibase 10.1007/JHEP12(2019)007} {\bibfield
  {journal} {\bibinfo  {journal} {JHEP}\ }\textbf {\bibinfo {volume} {12}},\
  \bibinfo {pages} {007} (\bibinfo {year} {2019})},\ \Eprint
  {http://arxiv.org/abs/1807.06041} {arXiv:1807.06041 [hep-th]} \BibitemShut
  {NoStop}%
\bibitem [{\citenamefont {Hawking}(1975)}]{hawking1975particle}%
  \BibitemOpen
  \bibfield  {author} {\bibinfo {author} {\bibfnamefont {S.~W.}\ \bibnamefont
  {Hawking}},\ }\href@noop {} {\bibfield  {journal} {\bibinfo  {journal}
  {Communications in mathematical physics}\ }\textbf {\bibinfo {volume} {43}},\
  \bibinfo {pages} {199} (\bibinfo {year} {1975})}\BibitemShut {NoStop}%
\bibitem [{\citenamefont {Hawking}(1976)}]{Hawking:1976ra}%
  \BibitemOpen
  \bibfield  {author} {\bibinfo {author} {\bibfnamefont {S.}~\bibnamefont
  {Hawking}},\ }\href {\doibase 10.1103/PhysRevD.14.2460} {\bibfield  {journal}
  {\bibinfo  {journal} {Phys. Rev. D}\ }\textbf {\bibinfo {volume} {14}},\
  \bibinfo {pages} {2460} (\bibinfo {year} {1976})}\BibitemShut {NoStop}%
\bibitem [{\citenamefont {Bekenstein}(1973)}]{Bekenstein:1973ur}%
  \BibitemOpen
  \bibfield  {author} {\bibinfo {author} {\bibfnamefont {J.~D.}\ \bibnamefont
  {Bekenstein}},\ }\href {\doibase 10.1103/PhysRevD.7.2333} {\bibfield
  {journal} {\bibinfo  {journal} {Phys. Rev.}\ }\textbf {\bibinfo {volume}
  {D7}},\ \bibinfo {pages} {2333} (\bibinfo {year} {1973})}\BibitemShut
  {NoStop}%
\bibitem [{\citenamefont {Page}(1993)}]{Page:1993wv}%
  \BibitemOpen
  \bibfield  {author} {\bibinfo {author} {\bibfnamefont {D.~N.}\ \bibnamefont
  {Page}},\ }\href {\doibase 10.1103/PhysRevLett.71.3743} {\bibfield  {journal}
  {\bibinfo  {journal} {Phys. Rev. Lett.}\ }\textbf {\bibinfo {volume} {71}},\
  \bibinfo {pages} {3743} (\bibinfo {year} {1993})},\ \Eprint
  {http://arxiv.org/abs/hep-th/9306083} {arXiv:hep-th/9306083} \BibitemShut
  {NoStop}%
\bibitem [{\citenamefont {Page}(2013)}]{Page:2013dx}%
  \BibitemOpen
  \bibfield  {author} {\bibinfo {author} {\bibfnamefont {D.~N.}\ \bibnamefont
  {Page}},\ }\href {\doibase 10.1088/1475-7516/2013/09/028} {\bibfield
  {journal} {\bibinfo  {journal} {JCAP}\ }\textbf {\bibinfo {volume} {09}},\
  \bibinfo {pages} {028} (\bibinfo {year} {2013})},\ \Eprint
  {http://arxiv.org/abs/1301.4995} {arXiv:1301.4995 [hep-th]} \BibitemShut
  {NoStop}%
\bibitem [{\citenamefont {Penington}(2019)}]{Penington:2019npb}%
  \BibitemOpen
  \bibfield  {author} {\bibinfo {author} {\bibfnamefont {G.}~\bibnamefont
  {Penington}},\ }\href@noop {} {\  (\bibinfo {year} {2019})},\ \Eprint
  {http://arxiv.org/abs/1905.08255} {arXiv:1905.08255 [hep-th]} \BibitemShut
  {NoStop}%
\bibitem [{\citenamefont {Almheiri}\ \emph {et~al.}(2019)\citenamefont
  {Almheiri}, \citenamefont {Engelhardt}, \citenamefont {Marolf},\ and\
  \citenamefont {Maxfield}}]{Almheiri:2019psf}%
  \BibitemOpen
  \bibfield  {author} {\bibinfo {author} {\bibfnamefont {A.}~\bibnamefont
  {Almheiri}}, \bibinfo {author} {\bibfnamefont {N.}~\bibnamefont
  {Engelhardt}}, \bibinfo {author} {\bibfnamefont {D.}~\bibnamefont {Marolf}},
  \ and\ \bibinfo {author} {\bibfnamefont {H.}~\bibnamefont {Maxfield}},\
  }\href {\doibase 10.1007/JHEP12(2019)063} {\bibfield  {journal} {\bibinfo
  {journal} {JHEP}\ }\textbf {\bibinfo {volume} {12}},\ \bibinfo {pages} {063}
  (\bibinfo {year} {2019})},\ \Eprint {http://arxiv.org/abs/1905.08762}
  {arXiv:1905.08762 [hep-th]} \BibitemShut {NoStop}%
\bibitem [{\citenamefont {Czech}\ \emph {et~al.}(2012)\citenamefont {Czech},
  \citenamefont {Karczmarek}, \citenamefont {Nogueira},\ and\ \citenamefont
  {Van~Raamsdonk}}]{Czech:2012bh}%
  \BibitemOpen
  \bibfield  {author} {\bibinfo {author} {\bibfnamefont {B.}~\bibnamefont
  {Czech}}, \bibinfo {author} {\bibfnamefont {J.~L.}\ \bibnamefont
  {Karczmarek}}, \bibinfo {author} {\bibfnamefont {F.}~\bibnamefont
  {Nogueira}}, \ and\ \bibinfo {author} {\bibfnamefont {M.}~\bibnamefont
  {Van~Raamsdonk}},\ }\href {\doibase 10.1088/0264-9381/29/15/155009}
  {\bibfield  {journal} {\bibinfo  {journal} {Class. Quant. Grav.}\ }\textbf
  {\bibinfo {volume} {29}},\ \bibinfo {pages} {155009} (\bibinfo {year}
  {2012})},\ \Eprint {http://arxiv.org/abs/1204.1330} {arXiv:1204.1330
  [hep-th]} \BibitemShut {NoStop}%
\bibitem [{\citenamefont {Wall}(2014)}]{Wall:2012uf}%
  \BibitemOpen
  \bibfield  {author} {\bibinfo {author} {\bibfnamefont {A.~C.}\ \bibnamefont
  {Wall}},\ }\href {\doibase 10.1088/0264-9381/31/22/225007} {\bibfield
  {journal} {\bibinfo  {journal} {Class. Quant. Grav.}\ }\textbf {\bibinfo
  {volume} {31}},\ \bibinfo {pages} {225007} (\bibinfo {year} {2014})},\
  \Eprint {http://arxiv.org/abs/1211.3494} {arXiv:1211.3494 [hep-th]}
  \BibitemShut {NoStop}%
\bibitem [{\citenamefont {Headrick}\ \emph {et~al.}(2014)\citenamefont
  {Headrick}, \citenamefont {Hubeny}, \citenamefont {Lawrence},\ and\
  \citenamefont {Rangamani}}]{Headrick:2014cta}%
  \BibitemOpen
  \bibfield  {author} {\bibinfo {author} {\bibfnamefont {M.}~\bibnamefont
  {Headrick}}, \bibinfo {author} {\bibfnamefont {V.~E.}\ \bibnamefont
  {Hubeny}}, \bibinfo {author} {\bibfnamefont {A.}~\bibnamefont {Lawrence}}, \
  and\ \bibinfo {author} {\bibfnamefont {M.}~\bibnamefont {Rangamani}},\ }\href
  {\doibase 10.1007/JHEP12(2014)162} {\bibfield  {journal} {\bibinfo  {journal}
  {JHEP}\ }\textbf {\bibinfo {volume} {12}},\ \bibinfo {pages} {162} (\bibinfo
  {year} {2014})},\ \Eprint {http://arxiv.org/abs/1408.6300} {arXiv:1408.6300
  [hep-th]} \BibitemShut {NoStop}%
\bibitem [{\citenamefont {Dong}\ \emph {et~al.}(2016)\citenamefont {Dong},
  \citenamefont {Harlow},\ and\ \citenamefont {Wall}}]{Dong:2016eik}%
  \BibitemOpen
  \bibfield  {author} {\bibinfo {author} {\bibfnamefont {X.}~\bibnamefont
  {Dong}}, \bibinfo {author} {\bibfnamefont {D.}~\bibnamefont {Harlow}}, \ and\
  \bibinfo {author} {\bibfnamefont {A.~C.}\ \bibnamefont {Wall}},\ }\href
  {\doibase 10.1103/PhysRevLett.117.021601} {\bibfield  {journal} {\bibinfo
  {journal} {Phys. Rev. Lett.}\ }\textbf {\bibinfo {volume} {117}},\ \bibinfo
  {pages} {021601} (\bibinfo {year} {2016})},\ \Eprint
  {http://arxiv.org/abs/1601.05416} {arXiv:1601.05416 [hep-th]} \BibitemShut
  {NoStop}%
\bibitem [{\citenamefont {Jafferis}\ \emph {et~al.}(2016)\citenamefont
  {Jafferis}, \citenamefont {Lewkowycz}, \citenamefont {Maldacena},\ and\
  \citenamefont {Suh}}]{Jafferis:2015del}%
  \BibitemOpen
  \bibfield  {author} {\bibinfo {author} {\bibfnamefont {D.~L.}\ \bibnamefont
  {Jafferis}}, \bibinfo {author} {\bibfnamefont {A.}~\bibnamefont {Lewkowycz}},
  \bibinfo {author} {\bibfnamefont {J.}~\bibnamefont {Maldacena}}, \ and\
  \bibinfo {author} {\bibfnamefont {S.~J.}\ \bibnamefont {Suh}},\ }\href
  {\doibase 10.1007/JHEP06(2016)004} {\bibfield  {journal} {\bibinfo  {journal}
  {JHEP}\ }\textbf {\bibinfo {volume} {06}},\ \bibinfo {pages} {004} (\bibinfo
  {year} {2016})},\ \Eprint {http://arxiv.org/abs/1512.06431} {arXiv:1512.06431
  [hep-th]} \BibitemShut {NoStop}%
\bibitem [{\citenamefont {Almheiri}\ \emph
  {et~al.}(2020{\natexlab{a}})\citenamefont {Almheiri}, \citenamefont
  {Mahajan}, \citenamefont {Maldacena},\ and\ \citenamefont
  {Zhao}}]{Almheiri:2019hni}%
  \BibitemOpen
  \bibfield  {author} {\bibinfo {author} {\bibfnamefont {A.}~\bibnamefont
  {Almheiri}}, \bibinfo {author} {\bibfnamefont {R.}~\bibnamefont {Mahajan}},
  \bibinfo {author} {\bibfnamefont {J.}~\bibnamefont {Maldacena}}, \ and\
  \bibinfo {author} {\bibfnamefont {Y.}~\bibnamefont {Zhao}},\ }\href {\doibase
  10.1007/JHEP03(2020)149} {\bibfield  {journal} {\bibinfo  {journal} {JHEP}\
  }\textbf {\bibinfo {volume} {03}},\ \bibinfo {pages} {149} (\bibinfo {year}
  {2020}{\natexlab{a}})},\ \Eprint {http://arxiv.org/abs/1908.10996}
  {arXiv:1908.10996 [hep-th]} \BibitemShut {NoStop}%
\bibitem [{\citenamefont {Almheiri}\ \emph
  {et~al.}(2020{\natexlab{b}})\citenamefont {Almheiri}, \citenamefont
  {Hartman}, \citenamefont {Maldacena}, \citenamefont {Shaghoulian},\ and\
  \citenamefont {Tajdini}}]{Almheiri:2019qdq}%
  \BibitemOpen
  \bibfield  {author} {\bibinfo {author} {\bibfnamefont {A.}~\bibnamefont
  {Almheiri}}, \bibinfo {author} {\bibfnamefont {T.}~\bibnamefont {Hartman}},
  \bibinfo {author} {\bibfnamefont {J.}~\bibnamefont {Maldacena}}, \bibinfo
  {author} {\bibfnamefont {E.}~\bibnamefont {Shaghoulian}}, \ and\ \bibinfo
  {author} {\bibfnamefont {A.}~\bibnamefont {Tajdini}},\ }\href {\doibase
  10.1007/JHEP05(2020)013} {\bibfield  {journal} {\bibinfo  {journal} {JHEP}\
  }\textbf {\bibinfo {volume} {05}},\ \bibinfo {pages} {013} (\bibinfo {year}
  {2020}{\natexlab{b}})},\ \Eprint {http://arxiv.org/abs/1911.12333}
  {arXiv:1911.12333 [hep-th]} \BibitemShut {NoStop}%
\bibitem [{\citenamefont {Penington}\ \emph {et~al.}(2019)\citenamefont
  {Penington}, \citenamefont {Shenker}, \citenamefont {Stanford},\ and\
  \citenamefont {Yang}}]{Penington:2019kki}%
  \BibitemOpen
  \bibfield  {author} {\bibinfo {author} {\bibfnamefont {G.}~\bibnamefont
  {Penington}}, \bibinfo {author} {\bibfnamefont {S.~H.}\ \bibnamefont
  {Shenker}}, \bibinfo {author} {\bibfnamefont {D.}~\bibnamefont {Stanford}}, \
  and\ \bibinfo {author} {\bibfnamefont {Z.}~\bibnamefont {Yang}},\ }\href@noop
  {} {\  (\bibinfo {year} {2019})},\ \Eprint {http://arxiv.org/abs/1911.11977}
  {arXiv:1911.11977 [hep-th]} \BibitemShut {NoStop}%
\bibitem [{\citenamefont {Anegawa}\ and\ \citenamefont
  {Iizuka}(2020)}]{Anegawa:2020ezn}%
  \BibitemOpen
  \bibfield  {author} {\bibinfo {author} {\bibfnamefont {T.}~\bibnamefont
  {Anegawa}}\ and\ \bibinfo {author} {\bibfnamefont {N.}~\bibnamefont
  {Iizuka}},\ }\href {\doibase 10.1007/JHEP07(2020)036} {\bibfield  {journal}
  {\bibinfo  {journal} {JHEP}\ }\textbf {\bibinfo {volume} {07}},\ \bibinfo
  {pages} {036} (\bibinfo {year} {2020})},\ \Eprint
  {http://arxiv.org/abs/2004.01601} {arXiv:2004.01601 [hep-th]} \BibitemShut
  {NoStop}%
\bibitem [{\citenamefont {Hashimoto}\ \emph {et~al.}(2020)\citenamefont
  {Hashimoto}, \citenamefont {Iizuka},\ and\ \citenamefont
  {Matsuo}}]{Hashimoto:2020cas}%
  \BibitemOpen
  \bibfield  {author} {\bibinfo {author} {\bibfnamefont {K.}~\bibnamefont
  {Hashimoto}}, \bibinfo {author} {\bibfnamefont {N.}~\bibnamefont {Iizuka}}, \
  and\ \bibinfo {author} {\bibfnamefont {Y.}~\bibnamefont {Matsuo}},\ }\href
  {\doibase 10.1007/JHEP06(2020)085} {\bibfield  {journal} {\bibinfo  {journal}
  {JHEP}\ }\textbf {\bibinfo {volume} {06}},\ \bibinfo {pages} {085} (\bibinfo
  {year} {2020})},\ \Eprint {http://arxiv.org/abs/2004.05863} {arXiv:2004.05863
  [hep-th]} \BibitemShut {NoStop}%
\bibitem [{\citenamefont {Hartman}\ \emph {et~al.}(2020)\citenamefont
  {Hartman}, \citenamefont {Shaghoulian},\ and\ \citenamefont
  {Strominger}}]{Hartman:2020swn}%
  \BibitemOpen
  \bibfield  {author} {\bibinfo {author} {\bibfnamefont {T.}~\bibnamefont
  {Hartman}}, \bibinfo {author} {\bibfnamefont {E.}~\bibnamefont
  {Shaghoulian}}, \ and\ \bibinfo {author} {\bibfnamefont {A.}~\bibnamefont
  {Strominger}},\ }\href {\doibase 10.1007/JHEP07(2020)022} {\bibfield
  {journal} {\bibinfo  {journal} {JHEP}\ }\textbf {\bibinfo {volume} {07}},\
  \bibinfo {pages} {022} (\bibinfo {year} {2020})},\ \Eprint
  {http://arxiv.org/abs/2004.13857} {arXiv:2004.13857 [hep-th]} \BibitemShut
  {NoStop}%
\bibitem [{\citenamefont {Almheiri}\ \emph
  {et~al.}(2013{\natexlab{a}})\citenamefont {Almheiri}, \citenamefont {Marolf},
  \citenamefont {Polchinski},\ and\ \citenamefont {Sully}}]{Almheiri:2012rt}%
  \BibitemOpen
  \bibfield  {author} {\bibinfo {author} {\bibfnamefont {A.}~\bibnamefont
  {Almheiri}}, \bibinfo {author} {\bibfnamefont {D.}~\bibnamefont {Marolf}},
  \bibinfo {author} {\bibfnamefont {J.}~\bibnamefont {Polchinski}}, \ and\
  \bibinfo {author} {\bibfnamefont {J.}~\bibnamefont {Sully}},\ }\href
  {\doibase 10.1007/JHEP02(2013)062} {\bibfield  {journal} {\bibinfo  {journal}
  {JHEP}\ }\textbf {\bibinfo {volume} {02}},\ \bibinfo {pages} {062} (\bibinfo
  {year} {2013}{\natexlab{a}})},\ \Eprint {http://arxiv.org/abs/1207.3123}
  {arXiv:1207.3123 [hep-th]} \BibitemShut {NoStop}%
\bibitem [{\citenamefont {Almheiri}\ \emph
  {et~al.}(2013{\natexlab{b}})\citenamefont {Almheiri}, \citenamefont {Marolf},
  \citenamefont {Polchinski}, \citenamefont {Stanford},\ and\ \citenamefont
  {Sully}}]{Almheiri:2013hfa}%
  \BibitemOpen
  \bibfield  {author} {\bibinfo {author} {\bibfnamefont {A.}~\bibnamefont
  {Almheiri}}, \bibinfo {author} {\bibfnamefont {D.}~\bibnamefont {Marolf}},
  \bibinfo {author} {\bibfnamefont {J.}~\bibnamefont {Polchinski}}, \bibinfo
  {author} {\bibfnamefont {D.}~\bibnamefont {Stanford}}, \ and\ \bibinfo
  {author} {\bibfnamefont {J.}~\bibnamefont {Sully}},\ }\href {\doibase
  10.1007/JHEP09(2013)018} {\bibfield  {journal} {\bibinfo  {journal} {JHEP}\
  }\textbf {\bibinfo {volume} {09}},\ \bibinfo {pages} {018} (\bibinfo {year}
  {2013}{\natexlab{b}})},\ \Eprint {http://arxiv.org/abs/1304.6483}
  {arXiv:1304.6483 [hep-th]} \BibitemShut {NoStop}%
\bibitem [{\citenamefont {Susskind}\ \emph {et~al.}(1993)\citenamefont
  {Susskind}, \citenamefont {Thorlacius},\ and\ \citenamefont
  {Uglum}}]{Susskind:1993if}%
  \BibitemOpen
  \bibfield  {author} {\bibinfo {author} {\bibfnamefont {L.}~\bibnamefont
  {Susskind}}, \bibinfo {author} {\bibfnamefont {L.}~\bibnamefont
  {Thorlacius}}, \ and\ \bibinfo {author} {\bibfnamefont {J.}~\bibnamefont
  {Uglum}},\ }\href {\doibase 10.1103/PhysRevD.48.3743} {\bibfield  {journal}
  {\bibinfo  {journal} {Phys. Rev. D}\ }\textbf {\bibinfo {volume} {48}},\
  \bibinfo {pages} {3743} (\bibinfo {year} {1993})},\ \Eprint
  {http://arxiv.org/abs/hep-th/9306069} {arXiv:hep-th/9306069} \BibitemShut
  {NoStop}%
\bibitem [{\citenamefont {Susskind}\ and\ \citenamefont
  {Thorlacius}(1994)}]{Susskind:1993mu}%
  \BibitemOpen
  \bibfield  {author} {\bibinfo {author} {\bibfnamefont {L.}~\bibnamefont
  {Susskind}}\ and\ \bibinfo {author} {\bibfnamefont {L.}~\bibnamefont
  {Thorlacius}},\ }\href {\doibase 10.1103/PhysRevD.49.966} {\bibfield
  {journal} {\bibinfo  {journal} {Phys. Rev. D}\ }\textbf {\bibinfo {volume}
  {49}},\ \bibinfo {pages} {966} (\bibinfo {year} {1994})},\ \Eprint
  {http://arxiv.org/abs/hep-th/9308100} {arXiv:hep-th/9308100} \BibitemShut
  {NoStop}%
\bibitem [{\citenamefont {Lowe}\ \emph {et~al.}(1995)\citenamefont {Lowe},
  \citenamefont {Polchinski}, \citenamefont {Susskind}, \citenamefont
  {Thorlacius},\ and\ \citenamefont {Uglum}}]{Lowe:1995ac}%
  \BibitemOpen
  \bibfield  {author} {\bibinfo {author} {\bibfnamefont {D.~A.}\ \bibnamefont
  {Lowe}}, \bibinfo {author} {\bibfnamefont {J.}~\bibnamefont {Polchinski}},
  \bibinfo {author} {\bibfnamefont {L.}~\bibnamefont {Susskind}}, \bibinfo
  {author} {\bibfnamefont {L.}~\bibnamefont {Thorlacius}}, \ and\ \bibinfo
  {author} {\bibfnamefont {J.}~\bibnamefont {Uglum}},\ }\href {\doibase
  10.1103/PhysRevD.52.6997} {\bibfield  {journal} {\bibinfo  {journal} {Phys.
  Rev. D}\ }\textbf {\bibinfo {volume} {52}},\ \bibinfo {pages} {6997}
  (\bibinfo {year} {1995})},\ \Eprint {http://arxiv.org/abs/hep-th/9506138}
  {arXiv:hep-th/9506138} \BibitemShut {NoStop}%
\bibitem [{\citenamefont {'t~Hooft}(1985)}]{tHooft:1984kcu}%
  \BibitemOpen
  \bibfield  {author} {\bibinfo {author} {\bibfnamefont {G.}~\bibnamefont
  {'t~Hooft}},\ }\href {\doibase 10.1016/0550-3213(85)90418-3} {\bibfield
  {journal} {\bibinfo  {journal} {Nucl. Phys. B}\ }\textbf {\bibinfo {volume}
  {256}},\ \bibinfo {pages} {727} (\bibinfo {year} {1985})}\BibitemShut
  {NoStop}%
\bibitem [{\citenamefont {Lieb}\ and\ \citenamefont
  {Ruskai}(1973)}]{lieb1973proof}%
  \BibitemOpen
  \bibfield  {author} {\bibinfo {author} {\bibfnamefont {E.~H.}\ \bibnamefont
  {Lieb}}\ and\ \bibinfo {author} {\bibfnamefont {M.~B.}\ \bibnamefont
  {Ruskai}},\ }\href@noop {} {\bibfield  {journal} {\bibinfo  {journal} {Les
  rencontres physiciens-math{\'e}maticiens de Strasbourg-RCP25}\ }\textbf
  {\bibinfo {volume} {19}},\ \bibinfo {pages} {36} (\bibinfo {year}
  {1973})}\BibitemShut {NoStop}%
\bibitem [{\citenamefont {Maldacena}(2003)}]{Maldacena:2001kr}%
  \BibitemOpen
  \bibfield  {author} {\bibinfo {author} {\bibfnamefont {J.~M.}\ \bibnamefont
  {Maldacena}},\ }\href {\doibase 10.1088/1126-6708/2003/04/021} {\bibfield
  {journal} {\bibinfo  {journal} {JHEP}\ }\textbf {\bibinfo {volume} {04}},\
  \bibinfo {pages} {021} (\bibinfo {year} {2003})},\ \Eprint
  {http://arxiv.org/abs/hep-th/0106112} {arXiv:hep-th/0106112} \BibitemShut
  {NoStop}%
\bibitem [{\citenamefont {Maldacena}\ and\ \citenamefont
  {Susskind}(2013)}]{Maldacena:2013xja}%
  \BibitemOpen
  \bibfield  {author} {\bibinfo {author} {\bibfnamefont {J.}~\bibnamefont
  {Maldacena}}\ and\ \bibinfo {author} {\bibfnamefont {L.}~\bibnamefont
  {Susskind}},\ }\href {\doibase 10.1002/prop.201300020} {\bibfield  {journal}
  {\bibinfo  {journal} {Fortsch. Phys.}\ }\textbf {\bibinfo {volume} {61}},\
  \bibinfo {pages} {781} (\bibinfo {year} {2013})},\ \Eprint
  {http://arxiv.org/abs/1306.0533} {arXiv:1306.0533 [hep-th]} \BibitemShut
  {NoStop}%
\bibitem [{\citenamefont {Van~Raamsdonk}(2009)}]{VanRaamsdonk:2009ar}%
  \BibitemOpen
  \bibfield  {author} {\bibinfo {author} {\bibfnamefont {M.}~\bibnamefont
  {Van~Raamsdonk}},\ }\href@noop {} {\  (\bibinfo {year} {2009})},\ \Eprint
  {http://arxiv.org/abs/0907.2939} {arXiv:0907.2939 [hep-th]} \BibitemShut
  {NoStop}%
\bibitem [{\citenamefont {Verlinde}(2020)}]{Verlinde:2020upt}%
  \BibitemOpen
  \bibfield  {author} {\bibinfo {author} {\bibfnamefont {H.}~\bibnamefont
  {Verlinde}},\ }\href@noop {} {\  (\bibinfo {year} {2020})},\ \Eprint
  {http://arxiv.org/abs/2003.13117} {arXiv:2003.13117 [hep-th]} \BibitemShut
  {NoStop}%
\bibitem [{\citenamefont {Bousso}\ and\ \citenamefont
  {Tomasevic}(2019)}]{Bousso:2019ykv}%
  \BibitemOpen
  \bibfield  {author} {\bibinfo {author} {\bibfnamefont {R.}~\bibnamefont
  {Bousso}}\ and\ \bibinfo {author} {\bibfnamefont {M.}~\bibnamefont
  {Tomasevic}},\ }\href@noop {} {\  (\bibinfo {year} {2019})},\ \Eprint
  {http://arxiv.org/abs/1911.06305} {arXiv:1911.06305 [hep-th]} \BibitemShut
  {NoStop}%
\bibitem [{\citenamefont {Bousso}\ and\ \citenamefont
  {Wildenhain}(2020)}]{Bousso:2020kmy}%
  \BibitemOpen
  \bibfield  {author} {\bibinfo {author} {\bibfnamefont {R.}~\bibnamefont
  {Bousso}}\ and\ \bibinfo {author} {\bibfnamefont {E.}~\bibnamefont
  {Wildenhain}},\ }\href@noop {} {\  (\bibinfo {year} {2020})},\ \Eprint
  {http://arxiv.org/abs/2006.16289} {arXiv:2006.16289 [hep-th]} \BibitemShut
  {NoStop}%
\bibitem [{\citenamefont {Afkhami-Jeddi}\ \emph {et~al.}(2020)\citenamefont
  {Afkhami-Jeddi}, \citenamefont {Cohn}, \citenamefont {Hartman},\ and\
  \citenamefont {Tajdini}}]{Afkhami-Jeddi:2020ezh}%
  \BibitemOpen
  \bibfield  {author} {\bibinfo {author} {\bibfnamefont {N.}~\bibnamefont
  {Afkhami-Jeddi}}, \bibinfo {author} {\bibfnamefont {H.}~\bibnamefont {Cohn}},
  \bibinfo {author} {\bibfnamefont {T.}~\bibnamefont {Hartman}}, \ and\
  \bibinfo {author} {\bibfnamefont {A.}~\bibnamefont {Tajdini}},\ }\href@noop
  {} {\  (\bibinfo {year} {2020})},\ \Eprint {http://arxiv.org/abs/2006.04839}
  {arXiv:2006.04839 [hep-th]} \BibitemShut {NoStop}%
\bibitem [{\citenamefont {Maloney}\ and\ \citenamefont
  {Witten}(2020)}]{Maloney:2020nni}%
  \BibitemOpen
  \bibfield  {author} {\bibinfo {author} {\bibfnamefont {A.}~\bibnamefont
  {Maloney}}\ and\ \bibinfo {author} {\bibfnamefont {E.}~\bibnamefont
  {Witten}},\ }\href@noop {} {\  (\bibinfo {year} {2020})},\ \Eprint
  {http://arxiv.org/abs/2006.04855} {arXiv:2006.04855 [hep-th]} \BibitemShut
  {NoStop}%
\bibitem [{\citenamefont {Cotler}\ and\ \citenamefont
  {Jensen}(2020)}]{Cotler:2020ugk}%
  \BibitemOpen
  \bibfield  {author} {\bibinfo {author} {\bibfnamefont {J.}~\bibnamefont
  {Cotler}}\ and\ \bibinfo {author} {\bibfnamefont {K.}~\bibnamefont
  {Jensen}},\ }\href@noop {} {\  (\bibinfo {year} {2020})},\ \Eprint
  {http://arxiv.org/abs/2006.08648} {arXiv:2006.08648 [hep-th]} \BibitemShut
  {NoStop}%
\bibitem [{\citenamefont {Maxfield}\ and\ \citenamefont
  {Turiaci}(2020)}]{Maxfield:2020ale}%
  \BibitemOpen
  \bibfield  {author} {\bibinfo {author} {\bibfnamefont {H.}~\bibnamefont
  {Maxfield}}\ and\ \bibinfo {author} {\bibfnamefont {G.~J.}\ \bibnamefont
  {Turiaci}},\ }\href@noop {} {\  (\bibinfo {year} {2020})},\ \Eprint
  {http://arxiv.org/abs/2006.11317} {arXiv:2006.11317 [hep-th]} \BibitemShut
  {NoStop}%
\bibitem [{\citenamefont {Teitelboim}(1983)}]{Teitelboim:1983ux}%
  \BibitemOpen
  \bibfield  {author} {\bibinfo {author} {\bibfnamefont {C.}~\bibnamefont
  {Teitelboim}},\ }\href {\doibase 10.1016/0370-2693(83)90012-6} {\bibfield
  {journal} {\bibinfo  {journal} {Phys. Lett. B}\ }\textbf {\bibinfo {volume}
  {126}},\ \bibinfo {pages} {41} (\bibinfo {year} {1983})}\BibitemShut
  {NoStop}%
\bibitem [{\citenamefont {Jackiw}(1985)}]{Jackiw:1984je}%
  \BibitemOpen
  \bibfield  {author} {\bibinfo {author} {\bibfnamefont {R.}~\bibnamefont
  {Jackiw}},\ }\href {\doibase 10.1016/0550-3213(85)90448-1} {\bibfield
  {journal} {\bibinfo  {journal} {Nucl. Phys. B}\ }\textbf {\bibinfo {volume}
  {252}},\ \bibinfo {pages} {343} (\bibinfo {year} {1985})}\BibitemShut
  {NoStop}%
\bibitem [{\citenamefont {Saad}\ \emph {et~al.}(2018)\citenamefont {Saad},
  \citenamefont {Shenker},\ and\ \citenamefont {Stanford}}]{Saad:2018bqo}%
  \BibitemOpen
  \bibfield  {author} {\bibinfo {author} {\bibfnamefont {P.}~\bibnamefont
  {Saad}}, \bibinfo {author} {\bibfnamefont {S.~H.}\ \bibnamefont {Shenker}}, \
  and\ \bibinfo {author} {\bibfnamefont {D.}~\bibnamefont {Stanford}},\
  }\href@noop {} {\  (\bibinfo {year} {2018})},\ \Eprint
  {http://arxiv.org/abs/1806.06840} {arXiv:1806.06840 [hep-th]} \BibitemShut
  {NoStop}%
\bibitem [{\citenamefont {Saad}\ \emph {et~al.}(2019)\citenamefont {Saad},
  \citenamefont {Shenker},\ and\ \citenamefont {Stanford}}]{Saad:2019lba}%
  \BibitemOpen
  \bibfield  {author} {\bibinfo {author} {\bibfnamefont {P.}~\bibnamefont
  {Saad}}, \bibinfo {author} {\bibfnamefont {S.~H.}\ \bibnamefont {Shenker}}, \
  and\ \bibinfo {author} {\bibfnamefont {D.}~\bibnamefont {Stanford}},\
  }\href@noop {} {\  (\bibinfo {year} {2019})},\ \Eprint
  {http://arxiv.org/abs/1903.11115} {arXiv:1903.11115 [hep-th]} \BibitemShut
  {NoStop}%
\bibitem [{\citenamefont {Stanford}\ and\ \citenamefont
  {Witten}(2019)}]{Stanford:2019vob}%
  \BibitemOpen
  \bibfield  {author} {\bibinfo {author} {\bibfnamefont {D.}~\bibnamefont
  {Stanford}}\ and\ \bibinfo {author} {\bibfnamefont {E.}~\bibnamefont
  {Witten}},\ }\href@noop {} {\  (\bibinfo {year} {2019})},\ \Eprint
  {http://arxiv.org/abs/1907.03363} {arXiv:1907.03363 [hep-th]} \BibitemShut
  {NoStop}%
\bibitem [{\citenamefont {Giddings}(2012)}]{Giddings:2012bm}%
  \BibitemOpen
  \bibfield  {author} {\bibinfo {author} {\bibfnamefont {S.~B.}\ \bibnamefont
  {Giddings}},\ }\href {\doibase 10.1103/PhysRevD.85.124063} {\bibfield
  {journal} {\bibinfo  {journal} {Phys. Rev. D}\ }\textbf {\bibinfo {volume}
  {85}},\ \bibinfo {pages} {124063} (\bibinfo {year} {2012})},\ \Eprint
  {http://arxiv.org/abs/1201.1037} {arXiv:1201.1037 [hep-th]} \BibitemShut
  {NoStop}%
\bibitem [{\citenamefont {Giddings}\ and\ \citenamefont
  {Shi}(2013)}]{Giddings:2012dh}%
  \BibitemOpen
  \bibfield  {author} {\bibinfo {author} {\bibfnamefont {S.~B.}\ \bibnamefont
  {Giddings}}\ and\ \bibinfo {author} {\bibfnamefont {Y.}~\bibnamefont {Shi}},\
  }\href {\doibase 10.1103/PhysRevD.87.064031} {\bibfield  {journal} {\bibinfo
  {journal} {Phys. Rev. D}\ }\textbf {\bibinfo {volume} {87}},\ \bibinfo
  {pages} {064031} (\bibinfo {year} {2013})},\ \Eprint
  {http://arxiv.org/abs/1205.4732} {arXiv:1205.4732 [hep-th]} \BibitemShut
  {NoStop}%
\bibitem [{\citenamefont {Giddings}(2013)}]{Giddings:2012gc}%
  \BibitemOpen
  \bibfield  {author} {\bibinfo {author} {\bibfnamefont {S.~B.}\ \bibnamefont
  {Giddings}},\ }\href {\doibase 10.1103/PhysRevD.88.064023} {\bibfield
  {journal} {\bibinfo  {journal} {Phys. Rev. D}\ }\textbf {\bibinfo {volume}
  {88}},\ \bibinfo {pages} {064023} (\bibinfo {year} {2013})},\ \Eprint
  {http://arxiv.org/abs/1211.7070} {arXiv:1211.7070 [hep-th]} \BibitemShut
  {NoStop}%
\bibitem [{\citenamefont {Page}(1976)}]{Page:1976df}%
  \BibitemOpen
  \bibfield  {author} {\bibinfo {author} {\bibfnamefont {D.~N.}\ \bibnamefont
  {Page}},\ }\href {\doibase 10.1103/PhysRevD.13.198} {\bibfield  {journal}
  {\bibinfo  {journal} {Phys. Rev. D}\ }\textbf {\bibinfo {volume} {13}},\
  \bibinfo {pages} {198} (\bibinfo {year} {1976})}\BibitemShut {NoStop}%
\bibitem [{\citenamefont {Giddings}(2016)}]{Giddings:2015uzr}%
  \BibitemOpen
  \bibfield  {author} {\bibinfo {author} {\bibfnamefont {S.~B.}\ \bibnamefont
  {Giddings}},\ }\href {\doibase 10.1016/j.physletb.2015.12.076} {\bibfield
  {journal} {\bibinfo  {journal} {Phys. Lett. B}\ }\textbf {\bibinfo {volume}
  {754}},\ \bibinfo {pages} {39} (\bibinfo {year} {2016})},\ \Eprint
  {http://arxiv.org/abs/1511.08221} {arXiv:1511.08221 [hep-th]} \BibitemShut
  {NoStop}%
\bibitem [{\citenamefont {Almheiri}\ \emph
  {et~al.}(2020{\natexlab{c}})\citenamefont {Almheiri}, \citenamefont
  {Hartman}, \citenamefont {Maldacena}, \citenamefont {Shaghoulian},\ and\
  \citenamefont {Tajdini}}]{Almheiri:2020cfm}%
  \BibitemOpen
  \bibfield  {author} {\bibinfo {author} {\bibfnamefont {A.}~\bibnamefont
  {Almheiri}}, \bibinfo {author} {\bibfnamefont {T.}~\bibnamefont {Hartman}},
  \bibinfo {author} {\bibfnamefont {J.}~\bibnamefont {Maldacena}}, \bibinfo
  {author} {\bibfnamefont {E.}~\bibnamefont {Shaghoulian}}, \ and\ \bibinfo
  {author} {\bibfnamefont {A.}~\bibnamefont {Tajdini}},\ }\href@noop {} {\
  (\bibinfo {year} {2020}{\natexlab{c}})},\ \Eprint
  {http://arxiv.org/abs/2006.06872} {arXiv:2006.06872 [hep-th]} \BibitemShut
  {NoStop}%
\bibitem [{\citenamefont {Banks}\ \emph {et~al.}(1998)\citenamefont {Banks},
  \citenamefont {Douglas}, \citenamefont {Horowitz},\ and\ \citenamefont
  {Martinec}}]{Banks:1998dd}%
  \BibitemOpen
  \bibfield  {author} {\bibinfo {author} {\bibfnamefont {T.}~\bibnamefont
  {Banks}}, \bibinfo {author} {\bibfnamefont {M.~R.}\ \bibnamefont {Douglas}},
  \bibinfo {author} {\bibfnamefont {G.~T.}\ \bibnamefont {Horowitz}}, \ and\
  \bibinfo {author} {\bibfnamefont {E.~J.}\ \bibnamefont {Martinec}},\
  }\href@noop {} {\  (\bibinfo {year} {1998})},\ \Eprint
  {http://arxiv.org/abs/hep-th/9808016} {arXiv:hep-th/9808016} \BibitemShut
  {NoStop}%
\bibitem [{\citenamefont {Bena}(2000)}]{Bena:1999jv}%
  \BibitemOpen
  \bibfield  {author} {\bibinfo {author} {\bibfnamefont {I.}~\bibnamefont
  {Bena}},\ }\href {\doibase 10.1103/PhysRevD.62.066007} {\bibfield  {journal}
  {\bibinfo  {journal} {Phys. Rev. D}\ }\textbf {\bibinfo {volume} {62}},\
  \bibinfo {pages} {066007} (\bibinfo {year} {2000})},\ \Eprint
  {http://arxiv.org/abs/hep-th/9905186} {arXiv:hep-th/9905186} \BibitemShut
  {NoStop}%
\bibitem [{\citenamefont {Hamilton}\ \emph
  {et~al.}(2006{\natexlab{a}})\citenamefont {Hamilton}, \citenamefont {Kabat},
  \citenamefont {Lifschytz},\ and\ \citenamefont {Lowe}}]{Hamilton:2005ju}%
  \BibitemOpen
  \bibfield  {author} {\bibinfo {author} {\bibfnamefont {A.}~\bibnamefont
  {Hamilton}}, \bibinfo {author} {\bibfnamefont {D.~N.}\ \bibnamefont {Kabat}},
  \bibinfo {author} {\bibfnamefont {G.}~\bibnamefont {Lifschytz}}, \ and\
  \bibinfo {author} {\bibfnamefont {D.~A.}\ \bibnamefont {Lowe}},\ }\href
  {\doibase 10.1103/PhysRevD.73.086003} {\bibfield  {journal} {\bibinfo
  {journal} {Phys. Rev. D}\ }\textbf {\bibinfo {volume} {73}},\ \bibinfo
  {pages} {086003} (\bibinfo {year} {2006}{\natexlab{a}})},\ \Eprint
  {http://arxiv.org/abs/hep-th/0506118} {arXiv:hep-th/0506118} \BibitemShut
  {NoStop}%
\bibitem [{\citenamefont {Hamilton}\ \emph
  {et~al.}(2006{\natexlab{b}})\citenamefont {Hamilton}, \citenamefont {Kabat},
  \citenamefont {Lifschytz},\ and\ \citenamefont {Lowe}}]{Hamilton:2006az}%
  \BibitemOpen
  \bibfield  {author} {\bibinfo {author} {\bibfnamefont {A.}~\bibnamefont
  {Hamilton}}, \bibinfo {author} {\bibfnamefont {D.~N.}\ \bibnamefont {Kabat}},
  \bibinfo {author} {\bibfnamefont {G.}~\bibnamefont {Lifschytz}}, \ and\
  \bibinfo {author} {\bibfnamefont {D.~A.}\ \bibnamefont {Lowe}},\ }\href
  {\doibase 10.1103/PhysRevD.74.066009} {\bibfield  {journal} {\bibinfo
  {journal} {Phys. Rev. D}\ }\textbf {\bibinfo {volume} {74}},\ \bibinfo
  {pages} {066009} (\bibinfo {year} {2006}{\natexlab{b}})},\ \Eprint
  {http://arxiv.org/abs/hep-th/0606141} {arXiv:hep-th/0606141} \BibitemShut
  {NoStop}%
\bibitem [{\citenamefont {Hamilton}\ \emph {et~al.}(2008)\citenamefont
  {Hamilton}, \citenamefont {Kabat}, \citenamefont {Lifschytz},\ and\
  \citenamefont {Lowe}}]{Hamilton:2007wj}%
  \BibitemOpen
  \bibfield  {author} {\bibinfo {author} {\bibfnamefont {A.}~\bibnamefont
  {Hamilton}}, \bibinfo {author} {\bibfnamefont {D.~N.}\ \bibnamefont {Kabat}},
  \bibinfo {author} {\bibfnamefont {G.}~\bibnamefont {Lifschytz}}, \ and\
  \bibinfo {author} {\bibfnamefont {D.~A.}\ \bibnamefont {Lowe}},\ }\href@noop
  {} {\bibfield  {journal} {\bibinfo  {journal} {AMS/IP Stud. Adv. Math.}\
  }\textbf {\bibinfo {volume} {44}},\ \bibinfo {pages} {85} (\bibinfo {year}
  {2008})},\ \Eprint {http://arxiv.org/abs/0710.4334} {arXiv:0710.4334
  [hep-th]} \BibitemShut {NoStop}%
\bibitem [{\citenamefont {Heemskerk}\ \emph {et~al.}(2012)\citenamefont
  {Heemskerk}, \citenamefont {Marolf}, \citenamefont {Polchinski},\ and\
  \citenamefont {Sully}}]{Heemskerk:2012mn}%
  \BibitemOpen
  \bibfield  {author} {\bibinfo {author} {\bibfnamefont {I.}~\bibnamefont
  {Heemskerk}}, \bibinfo {author} {\bibfnamefont {D.}~\bibnamefont {Marolf}},
  \bibinfo {author} {\bibfnamefont {J.}~\bibnamefont {Polchinski}}, \ and\
  \bibinfo {author} {\bibfnamefont {J.}~\bibnamefont {Sully}},\ }\href
  {\doibase 10.1007/JHEP10(2012)165} {\bibfield  {journal} {\bibinfo  {journal}
  {JHEP}\ }\textbf {\bibinfo {volume} {10}},\ \bibinfo {pages} {165} (\bibinfo
  {year} {2012})},\ \Eprint {http://arxiv.org/abs/1201.3664} {arXiv:1201.3664
  [hep-th]} \BibitemShut {NoStop}%
\bibitem [{\citenamefont {Kabat}\ and\ \citenamefont
  {Lifschytz}(2015)}]{Kabat:2015swa}%
  \BibitemOpen
  \bibfield  {author} {\bibinfo {author} {\bibfnamefont {D.}~\bibnamefont
  {Kabat}}\ and\ \bibinfo {author} {\bibfnamefont {G.}~\bibnamefont
  {Lifschytz}},\ }\href {\doibase 10.1007/JHEP09(2015)059} {\bibfield
  {journal} {\bibinfo  {journal} {JHEP}\ }\textbf {\bibinfo {volume} {09}},\
  \bibinfo {pages} {059} (\bibinfo {year} {2015})},\ \Eprint
  {http://arxiv.org/abs/1505.03755} {arXiv:1505.03755 [hep-th]} \BibitemShut
  {NoStop}%
\bibitem [{\citenamefont {Papadodimas}\ and\ \citenamefont
  {Raju}(2013)}]{Papadodimas:2012aq}%
  \BibitemOpen
  \bibfield  {author} {\bibinfo {author} {\bibfnamefont {K.}~\bibnamefont
  {Papadodimas}}\ and\ \bibinfo {author} {\bibfnamefont {S.}~\bibnamefont
  {Raju}},\ }\href {\doibase 10.1007/JHEP10(2013)212} {\bibfield  {journal}
  {\bibinfo  {journal} {JHEP}\ }\textbf {\bibinfo {volume} {10}},\ \bibinfo
  {pages} {212} (\bibinfo {year} {2013})},\ \Eprint
  {http://arxiv.org/abs/1211.6767} {arXiv:1211.6767 [hep-th]} \BibitemShut
  {NoStop}%
\bibitem [{\citenamefont {Papadodimas}\ and\ \citenamefont
  {Raju}(2014{\natexlab{a}})}]{Papadodimas:2013wnh}%
  \BibitemOpen
  \bibfield  {author} {\bibinfo {author} {\bibfnamefont {K.}~\bibnamefont
  {Papadodimas}}\ and\ \bibinfo {author} {\bibfnamefont {S.}~\bibnamefont
  {Raju}},\ }\href {\doibase 10.1103/PhysRevLett.112.051301} {\bibfield
  {journal} {\bibinfo  {journal} {Phys. Rev. Lett.}\ }\textbf {\bibinfo
  {volume} {112}},\ \bibinfo {pages} {051301} (\bibinfo {year}
  {2014}{\natexlab{a}})},\ \Eprint {http://arxiv.org/abs/1310.6334}
  {arXiv:1310.6334 [hep-th]} \BibitemShut {NoStop}%
\bibitem [{\citenamefont {Papadodimas}\ and\ \citenamefont
  {Raju}(2014{\natexlab{b}})}]{Papadodimas:2013jku}%
  \BibitemOpen
  \bibfield  {author} {\bibinfo {author} {\bibfnamefont {K.}~\bibnamefont
  {Papadodimas}}\ and\ \bibinfo {author} {\bibfnamefont {S.}~\bibnamefont
  {Raju}},\ }\href {\doibase 10.1103/PhysRevD.89.086010} {\bibfield  {journal}
  {\bibinfo  {journal} {Phys. Rev. D}\ }\textbf {\bibinfo {volume} {89}},\
  \bibinfo {pages} {086010} (\bibinfo {year} {2014}{\natexlab{b}})},\ \Eprint
  {http://arxiv.org/abs/1310.6335} {arXiv:1310.6335 [hep-th]} \BibitemShut
  {NoStop}%
\bibitem [{\citenamefont {Papadodimas}\ and\ \citenamefont
  {Raju}(2015)}]{Papadodimas:2015xma}%
  \BibitemOpen
  \bibfield  {author} {\bibinfo {author} {\bibfnamefont {K.}~\bibnamefont
  {Papadodimas}}\ and\ \bibinfo {author} {\bibfnamefont {S.}~\bibnamefont
  {Raju}},\ }\href {\doibase 10.1103/PhysRevLett.115.211601} {\bibfield
  {journal} {\bibinfo  {journal} {Phys. Rev. Lett.}\ }\textbf {\bibinfo
  {volume} {115}},\ \bibinfo {pages} {211601} (\bibinfo {year} {2015})},\
  \Eprint {http://arxiv.org/abs/1502.06692} {arXiv:1502.06692 [hep-th]}
  \BibitemShut {NoStop}%
\bibitem [{\citenamefont {Kabat}\ \emph {et~al.}(2012)\citenamefont {Kabat},
  \citenamefont {Lifschytz}, \citenamefont {Roy},\ and\ \citenamefont
  {Sarkar}}]{Kabat:2012hp}%
  \BibitemOpen
  \bibfield  {author} {\bibinfo {author} {\bibfnamefont {D.}~\bibnamefont
  {Kabat}}, \bibinfo {author} {\bibfnamefont {G.}~\bibnamefont {Lifschytz}},
  \bibinfo {author} {\bibfnamefont {S.}~\bibnamefont {Roy}}, \ and\ \bibinfo
  {author} {\bibfnamefont {D.}~\bibnamefont {Sarkar}},\ }\href {\doibase
  10.1103/PhysRevD.86.026004} {\bibfield  {journal} {\bibinfo  {journal} {Phys.
  Rev. D}\ }\textbf {\bibinfo {volume} {86}},\ \bibinfo {pages} {026004}
  (\bibinfo {year} {2012})},\ \Eprint {http://arxiv.org/abs/1204.0126}
  {arXiv:1204.0126 [hep-th]} \BibitemShut {NoStop}%
\bibitem [{\citenamefont {Kabat}\ and\ \citenamefont
  {Lifschytz}(2013)}]{Kabat:2012av}%
  \BibitemOpen
  \bibfield  {author} {\bibinfo {author} {\bibfnamefont {D.}~\bibnamefont
  {Kabat}}\ and\ \bibinfo {author} {\bibfnamefont {G.}~\bibnamefont
  {Lifschytz}},\ }\href {\doibase 10.1103/PhysRevD.87.086004} {\bibfield
  {journal} {\bibinfo  {journal} {Phys. Rev. D}\ }\textbf {\bibinfo {volume}
  {87}},\ \bibinfo {pages} {086004} (\bibinfo {year} {2013})},\ \Eprint
  {http://arxiv.org/abs/1212.3788} {arXiv:1212.3788 [hep-th]} \BibitemShut
  {NoStop}%
\bibitem [{\citenamefont {Harlow}(2016)}]{Harlow:2015lma}%
  \BibitemOpen
  \bibfield  {author} {\bibinfo {author} {\bibfnamefont {D.}~\bibnamefont
  {Harlow}},\ }\href {\doibase 10.1007/JHEP01(2016)122} {\bibfield  {journal}
  {\bibinfo  {journal} {JHEP}\ }\textbf {\bibinfo {volume} {01}},\ \bibinfo
  {pages} {122} (\bibinfo {year} {2016})},\ \Eprint
  {http://arxiv.org/abs/1510.07911} {arXiv:1510.07911 [hep-th]} \BibitemShut
  {NoStop}%
\bibitem [{\citenamefont {Guica}\ and\ \citenamefont
  {Jafferis}(2017)}]{Guica:2015zpf}%
  \BibitemOpen
  \bibfield  {author} {\bibinfo {author} {\bibfnamefont {M.}~\bibnamefont
  {Guica}}\ and\ \bibinfo {author} {\bibfnamefont {D.~L.}\ \bibnamefont
  {Jafferis}},\ }\href {\doibase 10.21468/SciPostPhys.3.2.016} {\bibfield
  {journal} {\bibinfo  {journal} {SciPost Phys.}\ }\textbf {\bibinfo {volume}
  {3}},\ \bibinfo {pages} {016} (\bibinfo {year} {2017})},\ \Eprint
  {http://arxiv.org/abs/1511.05627} {arXiv:1511.05627 [hep-th]} \BibitemShut
  {NoStop}%
\bibitem [{\citenamefont {Harlow}\ and\ \citenamefont
  {Jafferis}(2020)}]{Harlow:2018tqv}%
  \BibitemOpen
  \bibfield  {author} {\bibinfo {author} {\bibfnamefont {D.}~\bibnamefont
  {Harlow}}\ and\ \bibinfo {author} {\bibfnamefont {D.}~\bibnamefont
  {Jafferis}},\ }\href {\doibase 10.1007/JHEP02(2020)177} {\bibfield  {journal}
  {\bibinfo  {journal} {JHEP}\ }\textbf {\bibinfo {volume} {02}},\ \bibinfo
  {pages} {177} (\bibinfo {year} {2020})},\ \Eprint
  {http://arxiv.org/abs/1804.01081} {arXiv:1804.01081 [hep-th]} \BibitemShut
  {NoStop}%
\bibitem [{\citenamefont {Nomura}(2020{\natexlab{a}})}]{Nomura:2019qps}%
  \BibitemOpen
  \bibfield  {author} {\bibinfo {author} {\bibfnamefont {Y.}~\bibnamefont
  {Nomura}},\ }\href {\doibase 10.1103/PhysRevD.101.066024} {\bibfield
  {journal} {\bibinfo  {journal} {Phys. Rev. D}\ }\textbf {\bibinfo {volume}
  {101}},\ \bibinfo {pages} {066024} (\bibinfo {year} {2020}{\natexlab{a}})},\
  \Eprint {http://arxiv.org/abs/1908.05728} {arXiv:1908.05728 [hep-th]}
  \BibitemShut {NoStop}%
\bibitem [{\citenamefont {Nomura}(2020{\natexlab{b}})}]{Nomura:2019dlz}%
  \BibitemOpen
  \bibfield  {author} {\bibinfo {author} {\bibfnamefont {Y.}~\bibnamefont
  {Nomura}},\ }\href {\doibase 10.1103/PhysRevD.102.026001} {\bibfield
  {journal} {\bibinfo  {journal} {Phys. Rev. D}\ }\textbf {\bibinfo {volume}
  {102}},\ \bibinfo {pages} {026001} (\bibinfo {year} {2020}{\natexlab{b}})},\
  \Eprint {http://arxiv.org/abs/1911.13120} {arXiv:1911.13120 [hep-th]}
  \BibitemShut {NoStop}%
\bibitem [{\citenamefont {Deutsch}(1991)}]{deutsch1991quantum}%
  \BibitemOpen
  \bibfield  {author} {\bibinfo {author} {\bibfnamefont {J.~M.}\ \bibnamefont
  {Deutsch}},\ }\href@noop {} {\bibfield  {journal} {\bibinfo  {journal}
  {Physical Review A}\ }\textbf {\bibinfo {volume} {43}},\ \bibinfo {pages}
  {2046} (\bibinfo {year} {1991})}\BibitemShut {NoStop}%
\bibitem [{\citenamefont {Srednicki}(1994)}]{srednicki1994chaos}%
  \BibitemOpen
  \bibfield  {author} {\bibinfo {author} {\bibfnamefont {M.}~\bibnamefont
  {Srednicki}},\ }\href@noop {} {\bibfield  {journal} {\bibinfo  {journal}
  {Physical Review E}\ }\textbf {\bibinfo {volume} {50}},\ \bibinfo {pages}
  {888} (\bibinfo {year} {1994})}\BibitemShut {NoStop}%
\bibitem [{\citenamefont {Popescu}\ \emph {et~al.}(2006)\citenamefont
  {Popescu}, \citenamefont {Short},\ and\ \citenamefont
  {Winter}}]{Popescu_2006}%
  \BibitemOpen
  \bibfield  {author} {\bibinfo {author} {\bibfnamefont {S.}~\bibnamefont
  {Popescu}}, \bibinfo {author} {\bibfnamefont {A.~J.}\ \bibnamefont {Short}},
  \ and\ \bibinfo {author} {\bibfnamefont {A.}~\bibnamefont {Winter}},\ }\href
  {\doibase 10.1038/nphys444} {\bibfield  {journal} {\bibinfo  {journal}
  {Nature Physics}\ }\textbf {\bibinfo {volume} {2}},\ \bibinfo {pages}
  {754–758} (\bibinfo {year} {2006})}\BibitemShut {NoStop}%
\bibitem [{\citenamefont {Nielsen}\ and\ \citenamefont
  {Chuang}(2002)}]{nielsen2002quantum}%
  \BibitemOpen
  \bibfield  {author} {\bibinfo {author} {\bibfnamefont {M.~A.}\ \bibnamefont
  {Nielsen}}\ and\ \bibinfo {author} {\bibfnamefont {I.}~\bibnamefont
  {Chuang}},\ }\href@noop {} {\enquote {\bibinfo {title} {Quantum computation
  and quantum information},}\ } (\bibinfo {year} {2002})\BibitemShut {NoStop}%
\bibitem [{\citenamefont {Bao}\ and\ \citenamefont
  {Ooguri}(2017)}]{Bao:2017guc}%
  \BibitemOpen
  \bibfield  {author} {\bibinfo {author} {\bibfnamefont {N.}~\bibnamefont
  {Bao}}\ and\ \bibinfo {author} {\bibfnamefont {H.}~\bibnamefont {Ooguri}},\
  }\href {\doibase 10.1103/PhysRevD.96.066017} {\bibfield  {journal} {\bibinfo
  {journal} {Phys. Rev.}\ }\textbf {\bibinfo {volume} {D96}},\ \bibinfo {pages}
  {066017} (\bibinfo {year} {2017})},\ \Eprint
  {http://arxiv.org/abs/1705.07943} {arXiv:1705.07943 [hep-th]} \BibitemShut
  {NoStop}%
\bibitem [{\citenamefont {Hawking}\ and\ \citenamefont
  {Page}(1983)}]{hawking1983thermodynamics}%
  \BibitemOpen
  \bibfield  {author} {\bibinfo {author} {\bibfnamefont {S.~W.}\ \bibnamefont
  {Hawking}}\ and\ \bibinfo {author} {\bibfnamefont {D.~N.}\ \bibnamefont
  {Page}},\ }\href@noop {} {\bibfield  {journal} {\bibinfo  {journal}
  {Communications in Mathematical Physics}\ }\textbf {\bibinfo {volume} {87}},\
  \bibinfo {pages} {577} (\bibinfo {year} {1983})}\BibitemShut {NoStop}%
\bibitem [{\citenamefont {Hikida}\ \emph {et~al.}(2018)\citenamefont {Hikida},
  \citenamefont {Kusuki},\ and\ \citenamefont {Takayanagi}}]{Hikida:2018khg}%
  \BibitemOpen
  \bibfield  {author} {\bibinfo {author} {\bibfnamefont {Y.}~\bibnamefont
  {Hikida}}, \bibinfo {author} {\bibfnamefont {Y.}~\bibnamefont {Kusuki}}, \
  and\ \bibinfo {author} {\bibfnamefont {T.}~\bibnamefont {Takayanagi}},\
  }\href {\doibase 10.1103/PhysRevD.98.026003} {\bibfield  {journal} {\bibinfo
  {journal} {Phys. Rev.}\ }\textbf {\bibinfo {volume} {D98}},\ \bibinfo {pages}
  {026003} (\bibinfo {year} {2018})},\ \Eprint
  {http://arxiv.org/abs/1804.09658} {arXiv:1804.09658 [hep-th]} \BibitemShut
  {NoStop}%
\bibitem [{\citenamefont {Brandao}\ \emph {et~al.}(2019)\citenamefont
  {Brandao}, \citenamefont {Crosson}, \citenamefont {Sahinoglu},\ and\
  \citenamefont {Bowen}}]{Brandao:2017irx}%
  \BibitemOpen
  \bibfield  {author} {\bibinfo {author} {\bibfnamefont {F.~G. S.~L.}\
  \bibnamefont {Brandao}}, \bibinfo {author} {\bibfnamefont {E.}~\bibnamefont
  {Crosson}}, \bibinfo {author} {\bibfnamefont {M.~B.}\ \bibnamefont
  {Sahinoglu}}, \ and\ \bibinfo {author} {\bibfnamefont {J.}~\bibnamefont
  {Bowen}},\ }\href {\doibase 10.1103/PhysRevLett.123.110502} {\bibfield
  {journal} {\bibinfo  {journal} {Phys. Rev. Lett.}\ }\textbf {\bibinfo
  {volume} {123}},\ \bibinfo {pages} {110502} (\bibinfo {year} {2019})},\
  \Eprint {http://arxiv.org/abs/1710.04631} {arXiv:1710.04631 [quant-ph]}
  \BibitemShut {NoStop}%
\bibitem [{\citenamefont {Asplund}\ \emph {et~al.}(2015)\citenamefont
  {Asplund}, \citenamefont {Bernamonti}, \citenamefont {Galli},\ and\
  \citenamefont {Hartman}}]{Asplund:2014coa}%
  \BibitemOpen
  \bibfield  {author} {\bibinfo {author} {\bibfnamefont {C.~T.}\ \bibnamefont
  {Asplund}}, \bibinfo {author} {\bibfnamefont {A.}~\bibnamefont {Bernamonti}},
  \bibinfo {author} {\bibfnamefont {F.}~\bibnamefont {Galli}}, \ and\ \bibinfo
  {author} {\bibfnamefont {T.}~\bibnamefont {Hartman}},\ }\href {\doibase
  10.1007/JHEP02(2015)171} {\bibfield  {journal} {\bibinfo  {journal} {JHEP}\
  }\textbf {\bibinfo {volume} {02}},\ \bibinfo {pages} {171} (\bibinfo {year}
  {2015})},\ \Eprint {http://arxiv.org/abs/1410.1392} {arXiv:1410.1392
  [hep-th]} \BibitemShut {NoStop}%
\bibitem [{\citenamefont {Hartman}\ \emph {et~al.}(2014)\citenamefont
  {Hartman}, \citenamefont {Keller},\ and\ \citenamefont
  {Stoica}}]{Hartman:2014oaa}%
  \BibitemOpen
  \bibfield  {author} {\bibinfo {author} {\bibfnamefont {T.}~\bibnamefont
  {Hartman}}, \bibinfo {author} {\bibfnamefont {C.~A.}\ \bibnamefont {Keller}},
  \ and\ \bibinfo {author} {\bibfnamefont {B.}~\bibnamefont {Stoica}},\ }\href
  {\doibase 10.1007/JHEP09(2014)118} {\bibfield  {journal} {\bibinfo  {journal}
  {JHEP}\ }\textbf {\bibinfo {volume} {09}},\ \bibinfo {pages} {118} (\bibinfo
  {year} {2014})},\ \Eprint {http://arxiv.org/abs/1405.5137} {arXiv:1405.5137
  [hep-th]} \BibitemShut {NoStop}%
\bibitem [{\citenamefont {Wick}\ \emph {et~al.}(1952)\citenamefont {Wick},
  \citenamefont {Wightman},\ and\ \citenamefont {Wigner}}]{wick1952}%
  \BibitemOpen
  \bibfield  {author} {\bibinfo {author} {\bibfnamefont {G.~C.}\ \bibnamefont
  {Wick}}, \bibinfo {author} {\bibfnamefont {A.~S.}\ \bibnamefont {Wightman}},
  \ and\ \bibinfo {author} {\bibfnamefont {E.~P.}\ \bibnamefont {Wigner}},\
  }\href {\doibase 10.1103/PhysRev.88.101} {\bibfield  {journal} {\bibinfo
  {journal} {Phys. Rev.}\ }\textbf {\bibinfo {volume} {88}},\ \bibinfo {pages}
  {101} (\bibinfo {year} {1952})}\BibitemShut {NoStop}%
\bibitem [{\citenamefont {Doplicher}\ \emph {et~al.}(1969)\citenamefont
  {Doplicher}, \citenamefont {Haag},\ and\ \citenamefont
  {Roberts}}]{doplicher1969fields}%
  \BibitemOpen
  \bibfield  {author} {\bibinfo {author} {\bibfnamefont {S.}~\bibnamefont
  {Doplicher}}, \bibinfo {author} {\bibfnamefont {R.}~\bibnamefont {Haag}}, \
  and\ \bibinfo {author} {\bibfnamefont {J.~E.}\ \bibnamefont {Roberts}},\
  }\href@noop {} {\bibfield  {journal} {\bibinfo  {journal} {Communications in
  Mathematical Physics}\ }\textbf {\bibinfo {volume} {15}},\ \bibinfo {pages}
  {173} (\bibinfo {year} {1969})}\BibitemShut {NoStop}%
\bibitem [{\citenamefont {Doplicher}\ \emph {et~al.}(1971)\citenamefont
  {Doplicher}, \citenamefont {Haag},\ and\ \citenamefont
  {Roberts}}]{doplicher1971local}%
  \BibitemOpen
  \bibfield  {author} {\bibinfo {author} {\bibfnamefont {S.}~\bibnamefont
  {Doplicher}}, \bibinfo {author} {\bibfnamefont {R.}~\bibnamefont {Haag}}, \
  and\ \bibinfo {author} {\bibfnamefont {J.~E.}\ \bibnamefont {Roberts}},\
  }\href@noop {} {\bibfield  {journal} {\bibinfo  {journal} {Communications in
  Mathematical Physics}\ }\textbf {\bibinfo {volume} {23}},\ \bibinfo {pages}
  {199} (\bibinfo {year} {1971})}\BibitemShut {NoStop}%
\bibitem [{\citenamefont {Doplicher}\ \emph {et~al.}(1974)\citenamefont
  {Doplicher}, \citenamefont {Haag},\ and\ \citenamefont
  {Roberts}}]{doplicher1974local}%
  \BibitemOpen
  \bibfield  {author} {\bibinfo {author} {\bibfnamefont {S.}~\bibnamefont
  {Doplicher}}, \bibinfo {author} {\bibfnamefont {R.}~\bibnamefont {Haag}}, \
  and\ \bibinfo {author} {\bibfnamefont {J.~E.}\ \bibnamefont {Roberts}},\
  }\href@noop {} {\bibfield  {journal} {\bibinfo  {journal} {Communications in
  Mathematical Physics}\ }\textbf {\bibinfo {volume} {35}},\ \bibinfo {pages}
  {49} (\bibinfo {year} {1974})}\BibitemShut {NoStop}%
\bibitem [{\citenamefont {Haag}(1992)}]{Haag:1992hx}%
  \BibitemOpen
  \bibfield  {author} {\bibinfo {author} {\bibfnamefont {R.}~\bibnamefont
  {Haag}},\ }\href@noop {} {\emph {\bibinfo {title} {{Local quantum physics:
  Fields, particles, algebras}}}}\ (\bibinfo {year} {1992})\BibitemShut
  {NoStop}%
\bibitem [{\citenamefont {Ohya}\ and\ \citenamefont
  {Petz}(2004)}]{ohya2004quantum}%
  \BibitemOpen
  \bibfield  {author} {\bibinfo {author} {\bibfnamefont {M.}~\bibnamefont
  {Ohya}}\ and\ \bibinfo {author} {\bibfnamefont {D.}~\bibnamefont {Petz}},\
  }\href@noop {} {\emph {\bibinfo {title} {Quantum entropy and its use}}}\
  (\bibinfo  {publisher} {Springer Science \& Business Media},\ \bibinfo {year}
  {2004})\BibitemShut {NoStop}%
\bibitem [{\citenamefont {Petz}(2007)}]{petz2007quantum}%
  \BibitemOpen
  \bibfield  {author} {\bibinfo {author} {\bibfnamefont {D.}~\bibnamefont
  {Petz}},\ }\href@noop {} {\emph {\bibinfo {title} {Quantum information theory
  and quantum statistics}}}\ (\bibinfo  {publisher} {Springer Science \&
  Business Media},\ \bibinfo {year} {2007})\BibitemShut {NoStop}%
\bibitem [{\citenamefont {Faulkner}(2020)}]{Faulkner:2020hzi}%
  \BibitemOpen
  \bibfield  {author} {\bibinfo {author} {\bibfnamefont {T.}~\bibnamefont
  {Faulkner}},\ }\href@noop {} {\  (\bibinfo {year} {2020})},\ \Eprint
  {http://arxiv.org/abs/2008.04810} {arXiv:2008.04810 [hep-th]} \BibitemShut
  {NoStop}%
\bibitem [{\citenamefont {Casini}\ \emph {et~al.}(2019)\citenamefont {Casini},
  \citenamefont {Huerta}, \citenamefont {Magan},\ and\ \citenamefont
  {Pontello}}]{Casini:2019kex}%
  \BibitemOpen
  \bibfield  {author} {\bibinfo {author} {\bibfnamefont {H.}~\bibnamefont
  {Casini}}, \bibinfo {author} {\bibfnamefont {M.}~\bibnamefont {Huerta}},
  \bibinfo {author} {\bibfnamefont {J.~M.}\ \bibnamefont {Magan}}, \ and\
  \bibinfo {author} {\bibfnamefont {D.}~\bibnamefont {Pontello}},\ }\href@noop
  {} {\  (\bibinfo {year} {2019})},\ \Eprint {http://arxiv.org/abs/1905.10487}
  {arXiv:1905.10487 [hep-th]} \BibitemShut {NoStop}%
\bibitem [{\citenamefont {Freedman}\ and\ \citenamefont
  {Headrick}(2017)}]{Freedman:2016zud}%
  \BibitemOpen
  \bibfield  {author} {\bibinfo {author} {\bibfnamefont {M.}~\bibnamefont
  {Freedman}}\ and\ \bibinfo {author} {\bibfnamefont {M.}~\bibnamefont
  {Headrick}},\ }\href {\doibase 10.1007/s00220-016-2796-3} {\bibfield
  {journal} {\bibinfo  {journal} {Commun. Math. Phys.}\ }\textbf {\bibinfo
  {volume} {352}},\ \bibinfo {pages} {407} (\bibinfo {year} {2017})},\ \Eprint
  {http://arxiv.org/abs/1604.00354} {arXiv:1604.00354 [hep-th]} \BibitemShut
  {NoStop}%
\bibitem [{\citenamefont {Strominger}\ and\ \citenamefont
  {Vafa}(1996)}]{Strominger:1996sh}%
  \BibitemOpen
  \bibfield  {author} {\bibinfo {author} {\bibfnamefont {A.}~\bibnamefont
  {Strominger}}\ and\ \bibinfo {author} {\bibfnamefont {C.}~\bibnamefont
  {Vafa}},\ }\href {\doibase 10.1016/0370-2693(96)00345-0} {\bibfield
  {journal} {\bibinfo  {journal} {Phys. Lett.}\ }\textbf {\bibinfo {volume}
  {B379}},\ \bibinfo {pages} {99} (\bibinfo {year} {1996})},\ \Eprint
  {http://arxiv.org/abs/hep-th/9601029} {arXiv:hep-th/9601029 [hep-th]}
  \BibitemShut {NoStop}%
\bibitem [{\citenamefont {Strominger}(1998)}]{Strominger:1997eq}%
  \BibitemOpen
  \bibfield  {author} {\bibinfo {author} {\bibfnamefont {A.}~\bibnamefont
  {Strominger}},\ }\href {\doibase 10.1088/1126-6708/1998/02/009} {\bibfield
  {journal} {\bibinfo  {journal} {JHEP}\ }\textbf {\bibinfo {volume} {02}},\
  \bibinfo {pages} {009} (\bibinfo {year} {1998})},\ \Eprint
  {http://arxiv.org/abs/hep-th/9712251} {arXiv:hep-th/9712251 [hep-th]}
  \BibitemShut {NoStop}%
\bibitem [{\citenamefont {Cardy}(1986)}]{Cardy:1986ie}%
  \BibitemOpen
  \bibfield  {author} {\bibinfo {author} {\bibfnamefont {J.~L.}\ \bibnamefont
  {Cardy}},\ }\href {\doibase 10.1016/0550-3213(86)90552-3} {\bibfield
  {journal} {\bibinfo  {journal} {Nucl. Phys.}\ }\textbf {\bibinfo {volume}
  {B270}},\ \bibinfo {pages} {186} (\bibinfo {year} {1986})}\BibitemShut
  {NoStop}%
\bibitem [{\citenamefont {Bartlett}\ \emph {et~al.}(2006)\citenamefont
  {Bartlett}, \citenamefont {Doherty}, \citenamefont {Spekkens},\ and\
  \citenamefont {Wiseman}}]{Bartlett_2006}%
  \BibitemOpen
  \bibfield  {author} {\bibinfo {author} {\bibfnamefont {S.~D.}\ \bibnamefont
  {Bartlett}}, \bibinfo {author} {\bibfnamefont {A.~C.}\ \bibnamefont
  {Doherty}}, \bibinfo {author} {\bibfnamefont {R.~W.}\ \bibnamefont
  {Spekkens}}, \ and\ \bibinfo {author} {\bibfnamefont {H.~M.}\ \bibnamefont
  {Wiseman}},\ }\href {\doibase 10.1103/physreva.73.022311} {\bibfield
  {journal} {\bibinfo  {journal} {Physical Review A}\ }\textbf {\bibinfo
  {volume} {73}} (\bibinfo {year} {2006}),\
  10.1103/physreva.73.022311}\BibitemShut {NoStop}%
\bibitem [{\citenamefont {Almheiri}\ \emph {et~al.}(2017)\citenamefont
  {Almheiri}, \citenamefont {Dong},\ and\ \citenamefont
  {Swingle}}]{Almheiri:2016blp}%
  \BibitemOpen
  \bibfield  {author} {\bibinfo {author} {\bibfnamefont {A.}~\bibnamefont
  {Almheiri}}, \bibinfo {author} {\bibfnamefont {X.}~\bibnamefont {Dong}}, \
  and\ \bibinfo {author} {\bibfnamefont {B.}~\bibnamefont {Swingle}},\ }\href
  {\doibase 10.1007/JHEP02(2017)074} {\bibfield  {journal} {\bibinfo  {journal}
  {JHEP}\ }\textbf {\bibinfo {volume} {02}},\ \bibinfo {pages} {074} (\bibinfo
  {year} {2017})},\ \Eprint {http://arxiv.org/abs/1606.04537} {arXiv:1606.04537
  [hep-th]} \BibitemShut {NoStop}%
\bibitem [{\citenamefont {Krasnov}\ and\ \citenamefont
  {Rovelli}(2009)}]{Krasnov:2009pd}%
  \BibitemOpen
  \bibfield  {author} {\bibinfo {author} {\bibfnamefont {K.}~\bibnamefont
  {Krasnov}}\ and\ \bibinfo {author} {\bibfnamefont {C.}~\bibnamefont
  {Rovelli}},\ }\href {\doibase 10.1088/0264-9381/26/24/245009} {\bibfield
  {journal} {\bibinfo  {journal} {Class. Quant. Grav.}\ }\textbf {\bibinfo
  {volume} {26}},\ \bibinfo {pages} {245009} (\bibinfo {year} {2009})},\
  \Eprint {http://arxiv.org/abs/0905.4916} {arXiv:0905.4916 [gr-qc]}
  \BibitemShut {NoStop}%
\bibitem [{\citenamefont {Ashtekar}\ \emph {et~al.}(1998)\citenamefont
  {Ashtekar}, \citenamefont {Baez}, \citenamefont {Corichi},\ and\
  \citenamefont {Krasnov}}]{ashtekar1998quantum}%
  \BibitemOpen
  \bibfield  {author} {\bibinfo {author} {\bibfnamefont {A.}~\bibnamefont
  {Ashtekar}}, \bibinfo {author} {\bibfnamefont {J.}~\bibnamefont {Baez}},
  \bibinfo {author} {\bibfnamefont {A.}~\bibnamefont {Corichi}}, \ and\
  \bibinfo {author} {\bibfnamefont {K.}~\bibnamefont {Krasnov}},\ }\href@noop
  {} {\bibfield  {journal} {\bibinfo  {journal} {Physical Review Letters}\
  }\textbf {\bibinfo {volume} {80}},\ \bibinfo {pages} {904} (\bibinfo {year}
  {1998})}\BibitemShut {NoStop}%
\bibitem [{\citenamefont {Ashtekar}\ \emph {et~al.}(2000)\citenamefont
  {Ashtekar}, \citenamefont {Baez},\ and\ \citenamefont
  {Krasnov}}]{ashtekar2000quantum}%
  \BibitemOpen
  \bibfield  {author} {\bibinfo {author} {\bibfnamefont {A.}~\bibnamefont
  {Ashtekar}}, \bibinfo {author} {\bibfnamefont {J.}~\bibnamefont {Baez}}, \
  and\ \bibinfo {author} {\bibfnamefont {K.}~\bibnamefont {Krasnov}},\
  }\href@noop {} {\bibfield  {journal} {\bibinfo  {journal} {arXiv preprint
  gr-qc/0005126}\ } (\bibinfo {year} {2000})}\BibitemShut {NoStop}%
\bibitem [{\citenamefont {Engle}\ \emph {et~al.}(2010)\citenamefont {Engle},
  \citenamefont {Perez},\ and\ \citenamefont {Noui}}]{Engle:2009vc}%
  \BibitemOpen
  \bibfield  {author} {\bibinfo {author} {\bibfnamefont {J.}~\bibnamefont
  {Engle}}, \bibinfo {author} {\bibfnamefont {A.}~\bibnamefont {Perez}}, \ and\
  \bibinfo {author} {\bibfnamefont {K.}~\bibnamefont {Noui}},\ }\href {\doibase
  10.1103/PhysRevLett.105.031302} {\bibfield  {journal} {\bibinfo  {journal}
  {Phys. Rev. Lett.}\ }\textbf {\bibinfo {volume} {105}},\ \bibinfo {pages}
  {031302} (\bibinfo {year} {2010})},\ \Eprint {http://arxiv.org/abs/0905.3168}
  {arXiv:0905.3168 [gr-qc]} \BibitemShut {NoStop}%
\bibitem [{\citenamefont {Smolin}(1995)}]{smolin1995linking}%
  \BibitemOpen
  \bibfield  {author} {\bibinfo {author} {\bibfnamefont {L.}~\bibnamefont
  {Smolin}},\ }\href@noop {} {\bibfield  {journal} {\bibinfo  {journal}
  {Journal of Mathematical Physics}\ }\textbf {\bibinfo {volume} {36}},\
  \bibinfo {pages} {6417} (\bibinfo {year} {1995})}\BibitemShut {NoStop}%
\bibitem [{\citenamefont {Takayanagi}\ and\ \citenamefont
  {Tamaoka}(2020)}]{Takayanagi:2019tvn}%
  \BibitemOpen
  \bibfield  {author} {\bibinfo {author} {\bibfnamefont {T.}~\bibnamefont
  {Takayanagi}}\ and\ \bibinfo {author} {\bibfnamefont {K.}~\bibnamefont
  {Tamaoka}},\ }\href {\doibase 10.1007/JHEP02(2020)167} {\bibfield  {journal}
  {\bibinfo  {journal} {JHEP}\ }\textbf {\bibinfo {volume} {02}},\ \bibinfo
  {pages} {167} (\bibinfo {year} {2020})},\ \Eprint
  {http://arxiv.org/abs/1912.01636} {arXiv:1912.01636 [hep-th]} \BibitemShut
  {NoStop}%
\bibitem [{\citenamefont {Swingle}(2012)}]{Swingle:2009bg}%
  \BibitemOpen
  \bibfield  {author} {\bibinfo {author} {\bibfnamefont {B.}~\bibnamefont
  {Swingle}},\ }\href {\doibase 10.1103/PhysRevD.86.065007} {\bibfield
  {journal} {\bibinfo  {journal} {Phys. Rev. D}\ }\textbf {\bibinfo {volume}
  {86}},\ \bibinfo {pages} {065007} (\bibinfo {year} {2012})},\ \Eprint
  {http://arxiv.org/abs/0905.1317} {arXiv:0905.1317 [cond-mat.str-el]}
  \BibitemShut {NoStop}%
\bibitem [{\citenamefont {Livine}(2018)}]{Livine:2017fgq}%
  \BibitemOpen
  \bibfield  {author} {\bibinfo {author} {\bibfnamefont {E.~R.}\ \bibnamefont
  {Livine}},\ }\href {\doibase 10.1103/PhysRevD.97.026009} {\bibfield
  {journal} {\bibinfo  {journal} {Phys. Rev. D}\ }\textbf {\bibinfo {volume}
  {97}},\ \bibinfo {pages} {026009} (\bibinfo {year} {2018})},\ \Eprint
  {http://arxiv.org/abs/1709.08511} {arXiv:1709.08511 [gr-qc]} \BibitemShut
  {NoStop}%
\bibitem [{\citenamefont {Rocha}(2008)}]{Rocha:2008fe}%
  \BibitemOpen
  \bibfield  {author} {\bibinfo {author} {\bibfnamefont {J.~V.}\ \bibnamefont
  {Rocha}},\ }\href {\doibase 10.1088/1126-6708/2008/08/075} {\bibfield
  {journal} {\bibinfo  {journal} {JHEP}\ }\textbf {\bibinfo {volume} {08}},\
  \bibinfo {pages} {075} (\bibinfo {year} {2008})},\ \Eprint
  {http://arxiv.org/abs/0804.0055} {arXiv:0804.0055 [hep-th]} \BibitemShut
  {NoStop}%
\bibitem [{\citenamefont {Van~Raamsdonk}(2014)}]{VanRaamsdonk:2013sza}%
  \BibitemOpen
  \bibfield  {author} {\bibinfo {author} {\bibfnamefont {M.}~\bibnamefont
  {Van~Raamsdonk}},\ }\href {\doibase 10.1007/JHEP11(2014)038} {\bibfield
  {journal} {\bibinfo  {journal} {JHEP}\ }\textbf {\bibinfo {volume} {11}},\
  \bibinfo {pages} {038} (\bibinfo {year} {2014})},\ \Eprint
  {http://arxiv.org/abs/1307.1796} {arXiv:1307.1796 [hep-th]} \BibitemShut
  {NoStop}%
\bibitem [{\citenamefont {Sekino}\ and\ \citenamefont
  {Susskind}(2008)}]{Sekino:2008he}%
  \BibitemOpen
  \bibfield  {author} {\bibinfo {author} {\bibfnamefont {Y.}~\bibnamefont
  {Sekino}}\ and\ \bibinfo {author} {\bibfnamefont {L.}~\bibnamefont
  {Susskind}},\ }\href {\doibase 10.1088/1126-6708/2008/10/065} {\bibfield
  {journal} {\bibinfo  {journal} {JHEP}\ }\textbf {\bibinfo {volume} {10}},\
  \bibinfo {pages} {065} (\bibinfo {year} {2008})},\ \Eprint
  {http://arxiv.org/abs/0808.2096} {arXiv:0808.2096 [hep-th]} \BibitemShut
  {NoStop}%
\bibitem [{\citenamefont {Hayden}\ and\ \citenamefont
  {Preskill}(2007)}]{Hayden:2007cs}%
  \BibitemOpen
  \bibfield  {author} {\bibinfo {author} {\bibfnamefont {P.}~\bibnamefont
  {Hayden}}\ and\ \bibinfo {author} {\bibfnamefont {J.}~\bibnamefont
  {Preskill}},\ }\href {\doibase 10.1088/1126-6708/2007/09/120} {\bibfield
  {journal} {\bibinfo  {journal} {JHEP}\ }\textbf {\bibinfo {volume} {09}},\
  \bibinfo {pages} {120} (\bibinfo {year} {2007})},\ \Eprint
  {http://arxiv.org/abs/0708.4025} {arXiv:0708.4025 [hep-th]} \BibitemShut
  {NoStop}%
\bibitem [{\citenamefont {Verlinde}\ and\ \citenamefont
  {Verlinde}(2013{\natexlab{a}})}]{Verlinde:2013uja}%
  \BibitemOpen
  \bibfield  {author} {\bibinfo {author} {\bibfnamefont {E.}~\bibnamefont
  {Verlinde}}\ and\ \bibinfo {author} {\bibfnamefont {H.}~\bibnamefont
  {Verlinde}},\ }\href@noop {} {\  (\bibinfo {year} {2013}{\natexlab{a}})},\
  \Eprint {http://arxiv.org/abs/1306.0515} {arXiv:1306.0515 [hep-th]}
  \BibitemShut {NoStop}%
\bibitem [{\citenamefont {Verlinde}\ and\ \citenamefont
  {Verlinde}(2013{\natexlab{b}})}]{Verlinde:2013vja}%
  \BibitemOpen
  \bibfield  {author} {\bibinfo {author} {\bibfnamefont {E.}~\bibnamefont
  {Verlinde}}\ and\ \bibinfo {author} {\bibfnamefont {H.}~\bibnamefont
  {Verlinde}},\ }\href@noop {} {\  (\bibinfo {year} {2013}{\natexlab{b}})},\
  \Eprint {http://arxiv.org/abs/1306.0516} {arXiv:1306.0516 [hep-th]}
  \BibitemShut {NoStop}%
\bibitem [{\citenamefont {Polchinski}\ \emph {et~al.}(1999)\citenamefont
  {Polchinski}, \citenamefont {Susskind},\ and\ \citenamefont
  {Toumbas}}]{Polchinski:1999yd}%
  \BibitemOpen
  \bibfield  {author} {\bibinfo {author} {\bibfnamefont {J.}~\bibnamefont
  {Polchinski}}, \bibinfo {author} {\bibfnamefont {L.}~\bibnamefont
  {Susskind}}, \ and\ \bibinfo {author} {\bibfnamefont {N.}~\bibnamefont
  {Toumbas}},\ }\href {\doibase 10.1103/PhysRevD.60.084006} {\bibfield
  {journal} {\bibinfo  {journal} {Phys. Rev. D}\ }\textbf {\bibinfo {volume}
  {60}},\ \bibinfo {pages} {084006} (\bibinfo {year} {1999})},\ \Eprint
  {http://arxiv.org/abs/hep-th/9903228} {arXiv:hep-th/9903228} \BibitemShut
  {NoStop}%
\bibitem [{\citenamefont {Freivogel}\ \emph {et~al.}(2002)\citenamefont
  {Freivogel}, \citenamefont {Giddings},\ and\ \citenamefont
  {Lippert}}]{Freivogel:2002ex}%
  \BibitemOpen
  \bibfield  {author} {\bibinfo {author} {\bibfnamefont {B.}~\bibnamefont
  {Freivogel}}, \bibinfo {author} {\bibfnamefont {S.~B.}\ \bibnamefont
  {Giddings}}, \ and\ \bibinfo {author} {\bibfnamefont {M.}~\bibnamefont
  {Lippert}},\ }\href {\doibase 10.1103/PhysRevD.66.106002} {\bibfield
  {journal} {\bibinfo  {journal} {Phys. Rev. D}\ }\textbf {\bibinfo {volume}
  {66}},\ \bibinfo {pages} {106002} (\bibinfo {year} {2002})},\ \Eprint
  {http://arxiv.org/abs/hep-th/0207083} {arXiv:hep-th/0207083} \BibitemShut
  {NoStop}%
\bibitem [{\citenamefont {Hubeny}(2003)}]{Hubeny:2002dg}%
  \BibitemOpen
  \bibfield  {author} {\bibinfo {author} {\bibfnamefont {V.~E.}\ \bibnamefont
  {Hubeny}},\ }\href {\doibase 10.1142/S0218271803003992} {\bibfield  {journal}
  {\bibinfo  {journal} {Int. J. Mod. Phys. D}\ }\textbf {\bibinfo {volume}
  {12}},\ \bibinfo {pages} {1693} (\bibinfo {year} {2003})},\ \Eprint
  {http://arxiv.org/abs/hep-th/0208047} {arXiv:hep-th/0208047} \BibitemShut
  {NoStop}%
\bibitem [{\citenamefont {Susskind}\ and\ \citenamefont
  {Toumbas}(2000)}]{Susskind:1999ey}%
  \BibitemOpen
  \bibfield  {author} {\bibinfo {author} {\bibfnamefont {L.}~\bibnamefont
  {Susskind}}\ and\ \bibinfo {author} {\bibfnamefont {N.}~\bibnamefont
  {Toumbas}},\ }\href {\doibase 10.1103/PhysRevD.61.044001} {\bibfield
  {journal} {\bibinfo  {journal} {Phys. Rev. D}\ }\textbf {\bibinfo {volume}
  {61}},\ \bibinfo {pages} {044001} (\bibinfo {year} {2000})},\ \Eprint
  {http://arxiv.org/abs/hep-th/9909013} {arXiv:hep-th/9909013} \BibitemShut
  {NoStop}%
\bibitem [{\citenamefont {Giddings}\ and\ \citenamefont
  {Lippert}(2002)}]{Giddings:2001pt}%
  \BibitemOpen
  \bibfield  {author} {\bibinfo {author} {\bibfnamefont {S.~B.}\ \bibnamefont
  {Giddings}}\ and\ \bibinfo {author} {\bibfnamefont {M.}~\bibnamefont
  {Lippert}},\ }\href {\doibase 10.1103/PhysRevD.65.024006} {\bibfield
  {journal} {\bibinfo  {journal} {Phys. Rev. D}\ }\textbf {\bibinfo {volume}
  {65}},\ \bibinfo {pages} {024006} (\bibinfo {year} {2002})},\ \Eprint
  {http://arxiv.org/abs/hep-th/0103231} {arXiv:hep-th/0103231} \BibitemShut
  {NoStop}%
\bibitem [{\citenamefont {Fitzpatrick}\ \emph {et~al.}(2017)\citenamefont
  {Fitzpatrick}, \citenamefont {Kaplan}, \citenamefont {Li},\ and\
  \citenamefont {Wang}}]{Fitzpatrick:2016mtp}%
  \BibitemOpen
  \bibfield  {author} {\bibinfo {author} {\bibfnamefont {A.~L.}\ \bibnamefont
  {Fitzpatrick}}, \bibinfo {author} {\bibfnamefont {J.}~\bibnamefont {Kaplan}},
  \bibinfo {author} {\bibfnamefont {D.}~\bibnamefont {Li}}, \ and\ \bibinfo
  {author} {\bibfnamefont {J.}~\bibnamefont {Wang}},\ }\href {\doibase
  10.1007/JHEP07(2017)092} {\bibfield  {journal} {\bibinfo  {journal} {JHEP}\
  }\textbf {\bibinfo {volume} {07}},\ \bibinfo {pages} {092} (\bibinfo {year}
  {2017})},\ \Eprint {http://arxiv.org/abs/1612.06385} {arXiv:1612.06385
  [hep-th]} \BibitemShut {NoStop}%
\bibitem [{\citenamefont {Anand}\ \emph {et~al.}(2018)\citenamefont {Anand},
  \citenamefont {Chen}, \citenamefont {Fitzpatrick}, \citenamefont {Kaplan},\
  and\ \citenamefont {Li}}]{Anand:2017dav}%
  \BibitemOpen
  \bibfield  {author} {\bibinfo {author} {\bibfnamefont {N.}~\bibnamefont
  {Anand}}, \bibinfo {author} {\bibfnamefont {H.}~\bibnamefont {Chen}},
  \bibinfo {author} {\bibfnamefont {A.~L.}\ \bibnamefont {Fitzpatrick}},
  \bibinfo {author} {\bibfnamefont {J.}~\bibnamefont {Kaplan}}, \ and\ \bibinfo
  {author} {\bibfnamefont {D.}~\bibnamefont {Li}},\ }\href {\doibase
  10.1007/JHEP02(2018)012} {\bibfield  {journal} {\bibinfo  {journal} {JHEP}\
  }\textbf {\bibinfo {volume} {02}},\ \bibinfo {pages} {012} (\bibinfo {year}
  {2018})},\ \Eprint {http://arxiv.org/abs/1708.04246} {arXiv:1708.04246
  [hep-th]} \BibitemShut {NoStop}%
\bibitem [{\citenamefont {Marolf}\ and\ \citenamefont
  {Maxfield}(2020)}]{Marolf:2020xie}%
  \BibitemOpen
  \bibfield  {author} {\bibinfo {author} {\bibfnamefont {D.}~\bibnamefont
  {Marolf}}\ and\ \bibinfo {author} {\bibfnamefont {H.}~\bibnamefont
  {Maxfield}},\ }\href@noop {} {\  (\bibinfo {year} {2020})},\ \Eprint
  {http://arxiv.org/abs/2002.08950} {arXiv:2002.08950 [hep-th]} \BibitemShut
  {NoStop}%
\bibitem [{\citenamefont {Giddings}\ and\ \citenamefont
  {Turiaci}(2020)}]{Giddings:2020yes}%
  \BibitemOpen
  \bibfield  {author} {\bibinfo {author} {\bibfnamefont {S.~B.}\ \bibnamefont
  {Giddings}}\ and\ \bibinfo {author} {\bibfnamefont {G.~J.}\ \bibnamefont
  {Turiaci}},\ }\href@noop {} {\  (\bibinfo {year} {2020})},\ \Eprint
  {http://arxiv.org/abs/2004.02900} {arXiv:2004.02900 [hep-th]} \BibitemShut
  {NoStop}%
\bibitem [{\citenamefont {Balasubramanian}\ \emph {et~al.}(2008)\citenamefont
  {Balasubramanian}, \citenamefont {Czech}, \citenamefont {Hubeny},
  \citenamefont {Larjo}, \citenamefont {Rangamani},\ and\ \citenamefont
  {Simon}}]{Balasubramanian:2007qv}%
  \BibitemOpen
  \bibfield  {author} {\bibinfo {author} {\bibfnamefont {V.}~\bibnamefont
  {Balasubramanian}}, \bibinfo {author} {\bibfnamefont {B.}~\bibnamefont
  {Czech}}, \bibinfo {author} {\bibfnamefont {V.~E.}\ \bibnamefont {Hubeny}},
  \bibinfo {author} {\bibfnamefont {K.}~\bibnamefont {Larjo}}, \bibinfo
  {author} {\bibfnamefont {M.}~\bibnamefont {Rangamani}}, \ and\ \bibinfo
  {author} {\bibfnamefont {J.}~\bibnamefont {Simon}},\ }\href {\doibase
  10.1007/s10714-008-0606-8} {\bibfield  {journal} {\bibinfo  {journal} {Gen.
  Rel. Grav.}\ }\textbf {\bibinfo {volume} {40}},\ \bibinfo {pages} {1863}
  (\bibinfo {year} {2008})},\ \Eprint {http://arxiv.org/abs/hep-th/0701122}
  {arXiv:hep-th/0701122} \BibitemShut {NoStop}%
\bibitem [{\citenamefont {Giddings}\ and\ \citenamefont
  {Strominger}(1992)}]{Giddings:1992kn}%
  \BibitemOpen
  \bibfield  {author} {\bibinfo {author} {\bibfnamefont {S.~B.}\ \bibnamefont
  {Giddings}}\ and\ \bibinfo {author} {\bibfnamefont {A.}~\bibnamefont
  {Strominger}},\ }\href {\doibase 10.1103/PhysRevD.46.627} {\bibfield
  {journal} {\bibinfo  {journal} {Phys. Rev. D}\ }\textbf {\bibinfo {volume}
  {46}},\ \bibinfo {pages} {627} (\bibinfo {year} {1992})},\ \Eprint
  {http://arxiv.org/abs/hep-th/9202004} {arXiv:hep-th/9202004} \BibitemShut
  {NoStop}%
\bibitem [{\citenamefont {Balasubramanian}\ \emph {et~al.}(2020)\citenamefont
  {Balasubramanian}, \citenamefont {Kar}, \citenamefont {Parrikar},
  \citenamefont {Sarosi},\ and\ \citenamefont
  {Ugajin}}]{Balasubramanian:2020hfs}%
  \BibitemOpen
  \bibfield  {author} {\bibinfo {author} {\bibfnamefont {V.}~\bibnamefont
  {Balasubramanian}}, \bibinfo {author} {\bibfnamefont {A.}~\bibnamefont
  {Kar}}, \bibinfo {author} {\bibfnamefont {O.}~\bibnamefont {Parrikar}},
  \bibinfo {author} {\bibfnamefont {G.}~\bibnamefont {Sarosi}}, \ and\ \bibinfo
  {author} {\bibfnamefont {T.}~\bibnamefont {Ugajin}},\ }\href@noop {} {\
  (\bibinfo {year} {2020})},\ \Eprint {http://arxiv.org/abs/2003.05448}
  {arXiv:2003.05448 [hep-th]} \BibitemShut {NoStop}%
\bibitem [{\citenamefont {Marolf}\ \emph {et~al.}(2020)\citenamefont {Marolf},
  \citenamefont {Wang},\ and\ \citenamefont {Wang}}]{Marolf:2020vsi}%
  \BibitemOpen
  \bibfield  {author} {\bibinfo {author} {\bibfnamefont {D.}~\bibnamefont
  {Marolf}}, \bibinfo {author} {\bibfnamefont {S.}~\bibnamefont {Wang}}, \ and\
  \bibinfo {author} {\bibfnamefont {Z.}~\bibnamefont {Wang}},\ }\href@noop {}
  {\  (\bibinfo {year} {2020})},\ \Eprint {http://arxiv.org/abs/2006.10089}
  {arXiv:2006.10089 [hep-th]} \BibitemShut {NoStop}%
\bibitem [{\citenamefont {Dong}\ and\ \citenamefont
  {Wang}(2020)}]{Dong:2020iod}%
  \BibitemOpen
  \bibfield  {author} {\bibinfo {author} {\bibfnamefont {X.}~\bibnamefont
  {Dong}}\ and\ \bibinfo {author} {\bibfnamefont {H.}~\bibnamefont {Wang}},\
  }\href@noop {} {\  (\bibinfo {year} {2020})},\ \Eprint
  {http://arxiv.org/abs/2006.10051} {arXiv:2006.10051 [hep-th]} \BibitemShut
  {NoStop}%
\bibitem [{\citenamefont {Bianchi}\ and\ \citenamefont
  {Dona}(2019)}]{Bianchi:2019stn}%
  \BibitemOpen
  \bibfield  {author} {\bibinfo {author} {\bibfnamefont {E.}~\bibnamefont
  {Bianchi}}\ and\ \bibinfo {author} {\bibfnamefont {P.}~\bibnamefont {Dona}},\
  }\href {\doibase 10.1103/PhysRevD.100.105010} {\bibfield  {journal} {\bibinfo
   {journal} {Phys. Rev. D}\ }\textbf {\bibinfo {volume} {100}},\ \bibinfo
  {pages} {105010} (\bibinfo {year} {2019})},\ \Eprint
  {http://arxiv.org/abs/1904.08370} {arXiv:1904.08370 [hep-th]} \BibitemShut
  {NoStop}%
\bibitem [{\citenamefont {Fitzpatrick}\ \emph {et~al.}(2016)\citenamefont
  {Fitzpatrick}, \citenamefont {Kaplan}, \citenamefont {Li},\ and\
  \citenamefont {Wang}}]{Fitzpatrick:2016ive}%
  \BibitemOpen
  \bibfield  {author} {\bibinfo {author} {\bibfnamefont {A.~L.}\ \bibnamefont
  {Fitzpatrick}}, \bibinfo {author} {\bibfnamefont {J.}~\bibnamefont {Kaplan}},
  \bibinfo {author} {\bibfnamefont {D.}~\bibnamefont {Li}}, \ and\ \bibinfo
  {author} {\bibfnamefont {J.}~\bibnamefont {Wang}},\ }\href {\doibase
  10.1007/JHEP05(2016)109} {\bibfield  {journal} {\bibinfo  {journal} {JHEP}\
  }\textbf {\bibinfo {volume} {05}},\ \bibinfo {pages} {109} (\bibinfo {year}
  {2016})},\ \Eprint {http://arxiv.org/abs/1603.08925} {arXiv:1603.08925
  [hep-th]} \BibitemShut {NoStop}%
\bibitem [{\citenamefont {Casini}\ \emph {et~al.}(2014)\citenamefont {Casini},
  \citenamefont {Huerta},\ and\ \citenamefont {Rosabal}}]{Casini:2013rba}%
  \BibitemOpen
  \bibfield  {author} {\bibinfo {author} {\bibfnamefont {H.}~\bibnamefont
  {Casini}}, \bibinfo {author} {\bibfnamefont {M.}~\bibnamefont {Huerta}}, \
  and\ \bibinfo {author} {\bibfnamefont {J.~A.}\ \bibnamefont {Rosabal}},\
  }\href {\doibase 10.1103/PhysRevD.89.085012} {\bibfield  {journal} {\bibinfo
  {journal} {Phys. Rev. D}\ }\textbf {\bibinfo {volume} {89}},\ \bibinfo
  {pages} {085012} (\bibinfo {year} {2014})},\ \Eprint
  {http://arxiv.org/abs/1312.1183} {arXiv:1312.1183 [hep-th]} \BibitemShut
  {NoStop}%
\bibitem [{\citenamefont {B{\'e}ny}\ \emph {et~al.}(2007)\citenamefont
  {B{\'e}ny}, \citenamefont {Kempf},\ and\ \citenamefont
  {Kribs}}]{beny2007generalization}%
  \BibitemOpen
  \bibfield  {author} {\bibinfo {author} {\bibfnamefont {C.}~\bibnamefont
  {B{\'e}ny}}, \bibinfo {author} {\bibfnamefont {A.}~\bibnamefont {Kempf}}, \
  and\ \bibinfo {author} {\bibfnamefont {D.~W.}\ \bibnamefont {Kribs}},\
  }\href@noop {} {\bibfield  {journal} {\bibinfo  {journal} {Physical review
  letters}\ }\textbf {\bibinfo {volume} {98}},\ \bibinfo {pages} {100502}
  (\bibinfo {year} {2007})}\BibitemShut {NoStop}%
\end{thebibliography}%

\end{document}